\documentclass[prd,aps,twocolumn,preprintnumbers, showpacs, nofootinbib,superscriptaddress,notitlepage]{revtex4-1}
\usepackage{amssymb,amsthm,amsmath}
\usepackage{graphicx}   
\usepackage{color}      
\usepackage{slashed}    
\usepackage{verbatim}
\usepackage[normalem]{ulem}
\usepackage{rotating}   
\usepackage{multirow}   
\begin{document}
\title{Light quark decays of doubly heavy baryons in light front approach}

\author{ Hang Liu}
\affiliation{INPAC, Key Laboratory for Particle Astrophysics and Cosmology (MOE), Shanghai Key Laboratory for Particle Physics and Cosmology, School of Physics and Astronomy, Shanghai Jiao Tong University, Shanghai
200240, China}

\author{Zhi-Peng Xing} \email{Email:zpxing@sjtu.edu.cn}
\affiliation{Tsung-Dao Lee Institute, Shanghai Jiao Tong University, Shanghai 200240, China}

\author{Chang Yang}\email{Email:15201868391@sjtu.edu.cn}
\affiliation{INPAC, Key Laboratory for Particle Astrophysics and Cosmology (MOE), Shanghai Key Laboratory for Particle Physics and Cosmology, School of Physics and Astronomy, Shanghai Jiao Tong University, Shanghai
200240, China}
\affiliation{Tsung-Dao Lee Institute, Shanghai Jiao Tong University, Shanghai 200240, China}

\begin{abstract}
We explore the semileptonic and nonleptonic decays of  doubly heavy baryons $(\Omega_{cc}^{(*)+},\Omega_{bb}^{(*)0},\Omega_{bc}^{(*)-},\Omega_{bc}^{\prime0})$ induced by the $s\to u$ transition.  Hadronic form factors are parametrized by   transition matrix elements and are calculated in the light front quark model. With the form factors,  we make use of helicity amplitudes and analyze  semileptonic and nonleptonic decay modes of  doubly heavy baryons. Benchmark results for partial decay widths, branching fractions, forward-backward asymmetries and other phenomenological observables are derived.   We find that typical branching fractions for semileptonic decays into $\ell\bar\nu$ are at the order $10^{-7}-10^{-8}$ and the ones for nonleptonic decays are at the order $10^{-5}$, which are likely detectable such as in LHCb experiment. With the potential data accumulated in future, our results may help to shape our  understanding of the decay mechanism  in the presence of two heavy quarks. 
\end{abstract}

\maketitle

\section{Introduction}

Weak decays of quarks can play an important role in testing the standard model of particle physics, and shaping our understanding of   CP violation in our universe.  Since most quarks in nature are bounded, the study of weak decays of quarks insides a hadron also provides a platform for the exploration of strong interactions. In the past decades there have been dramatic progresses made on both experimental and theoretical sides, and the unprecedented precise experimental  measurements  and 
theoretical calculations of heavy $B$ and $D$ meson decays  lead to a stringent test of the standard model ~\cite{BESIII:2013mhi,LHCb:2014vgu,LHCb:2015gmp}.


In addition to the bottom/charm mesons and baryons that consist of one heavy quark,   baryons composed of two heavy quarks  are of special  interests and provide a unique arena  to explore the QCD dynamics in the presence of two heavy quarks. In 2017, the LHCb collaboration has firstly  observed the doubly heavy baryon $\Xi_{cc}^{++}$ via the final state $\Lambda_c^+K^-\pi^+\pi^+$~\cite{LHCb:2017iph}. Subsequently, properties including the mass and  lifetime has been precisely determined,  and a series of  decay channels were discovered in sequence~\cite{LHCb:2018zpl,LHCb:2019gqy,LHCb:2019qed,LHCb:2019ybf,LHCb:2021xba,LHCb:2021eaf,PRD:105:2022,arXiv:2208.06834}. Very recently,  a new decay mode $\Xi_{cc}^{++}\to\Xi_{c}^{\prime +}\pi^+$ was reported by the LHCb collaboration~\cite{LHCb:2022rpd}.  
Meanwhile  many properties of doubly heavy spectroscopy have been  investigated in theory~\cite{arXiv:1805.02535,arXiv:1806.09288,arXiv:2008.08026}, such as the hadron spectrum,  the ``decay constant" or the so-called pole residue, and weak and electromagnetic decays.  Focusing on  the weak decays of doubly heavy baryons, there have been intense explorations  ranging from the flavor SU(3) symmetry analyses ~\cite{Wang:2017azm}  to model-dependent determinations  of form factors and decay widths ~\cite{Hasenfratz:1980ka,Lu:2017meb,Wang:2017mqp,Yu:2017zst}.

Aside from decays where the heavy quark undergoes a weak transition, there is also a class of decays in which the heavy quarks act as a spectator while the light quark explode in a weak transition.  In particular a previous analysis which mainly focused on the light-quark decays of heavy hadrons with one heavy quark can be found in Ref.~\cite{Faller:2015oma}.  But such decay processes of double heavy baryons  are not previously explored,  and the aim of this work is to fill this blank. 

Due to the very small phase space in this class of decays, only limited channels are kinematically allowed. 
In this work, we will  explore the $s\to u$ transition of doubly heavy baryons decays, and consider   explicitly 
	\begin{itemize}
\item the spin-${1}/{2}$ to spin-${1}/{2}$ decays,
\begin{eqnarray}
\Omega_{cc}^+\to \Xi_{cc}^{++},\quad\Omega_{bb}^-\to\Xi_{bb}^0,\quad\Omega_{bc}^0\to \Xi_{bc}^+,\quad \Omega_{bc}^{\prime0}\to \Xi_{bc}^{\prime+}.\notag
\end{eqnarray}
\item the spin-${3}/{2}$ to spin-${1}/{2}$ decays,
\begin{eqnarray}
\Omega_{cc}^{*+}\to \Xi_{cc}^{++},\quad\Omega_{bb}^{*-}\to\Xi_{bb}^0,\quad\Omega_{bc}^{*0}\to \Xi_{bc}^+.\notag
\end{eqnarray}
\end{itemize}
While the small phase space will substantially suppress these decay branching fractions, and make them very difficult to be observed in experiment, the small phase space allows for solid theoretical predictions. On the one hand measurements of semileptonic decay widths can help to provide rather reliable constraints on the form factors since the recoil is small. With the knowledge on form factors, nonleptonic decays could give a direct exploration of theoretical  tools such as factorization.  On the other hand, due to the small phase space, the  helicity suppression amplitudes in certain semi-leptonic decays are uplifted by the muon mass. These effects  are absent in semi-leptonic decays with electron in the final state. Thus these class of decays induced by $s\to u$  provide a new platform for study the helicity suppression amplitudes in the experiment and offer possibilities of significant new physics (NP) contributions.

In the bottom-charm baryons, the spin of the  $bc$ system can be both 0 or 1. For convenience we use  the $\Xi_{bc}^+/\Omega_{bc}^0$ to denote the doubly heavy baryons consisting of spin-1 bc system and  the $\Xi_{bc}^{\prime+}/\Omega_{bc}^{\prime0}$ to denote the doubly heavy baryons consisting of spin-0 bc system. At this stage, it is not clear which of the two hadrons is the ground state, and thus we will consider the decays  of both hadrons.  In addition to the spin-1/2 initial state, we will  also consider the spin-3/2 doubly heavy baryons for a comparison, though whose main decay mode  might be induced by the electromagnetic transition.

 The rest of this paper is organized as follows.  In Sec.~II,  we give the theoretical framework in detail.  After the parametrization of the form factors with both spin-1/2 to spin-1/2 and spin-3/2 to spin-1/2 processes,  we will present the explicit calculation in the light-front approach.  Numerical results for   form factors and phenomenological analysis including decay widths, branching ratios and forward-backward asymmetry are given in Sec.~III. A brief summary is given in the last section.

\section{Theoretical  framework}

\begin{widetext}

\begin{figure}[htbp!] 
\includegraphics[width=0.4\columnwidth]{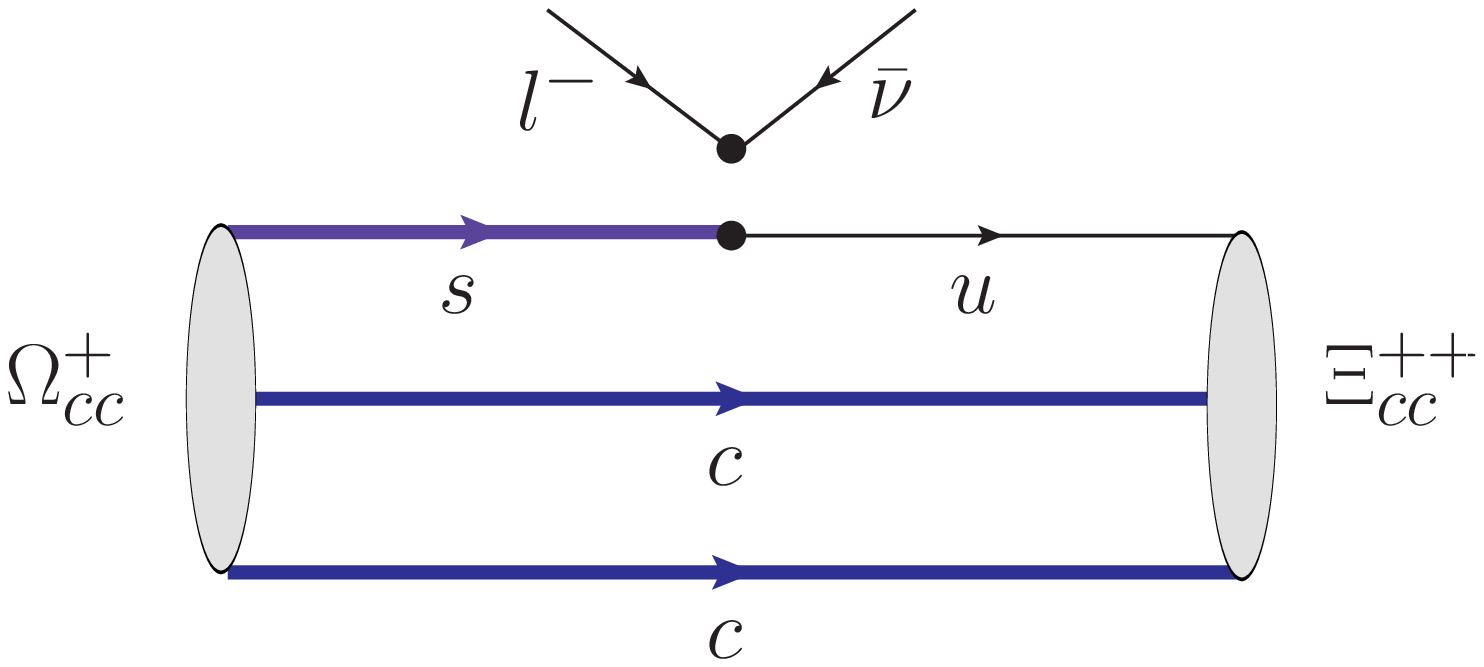}  
\includegraphics[width=0.4\columnwidth]{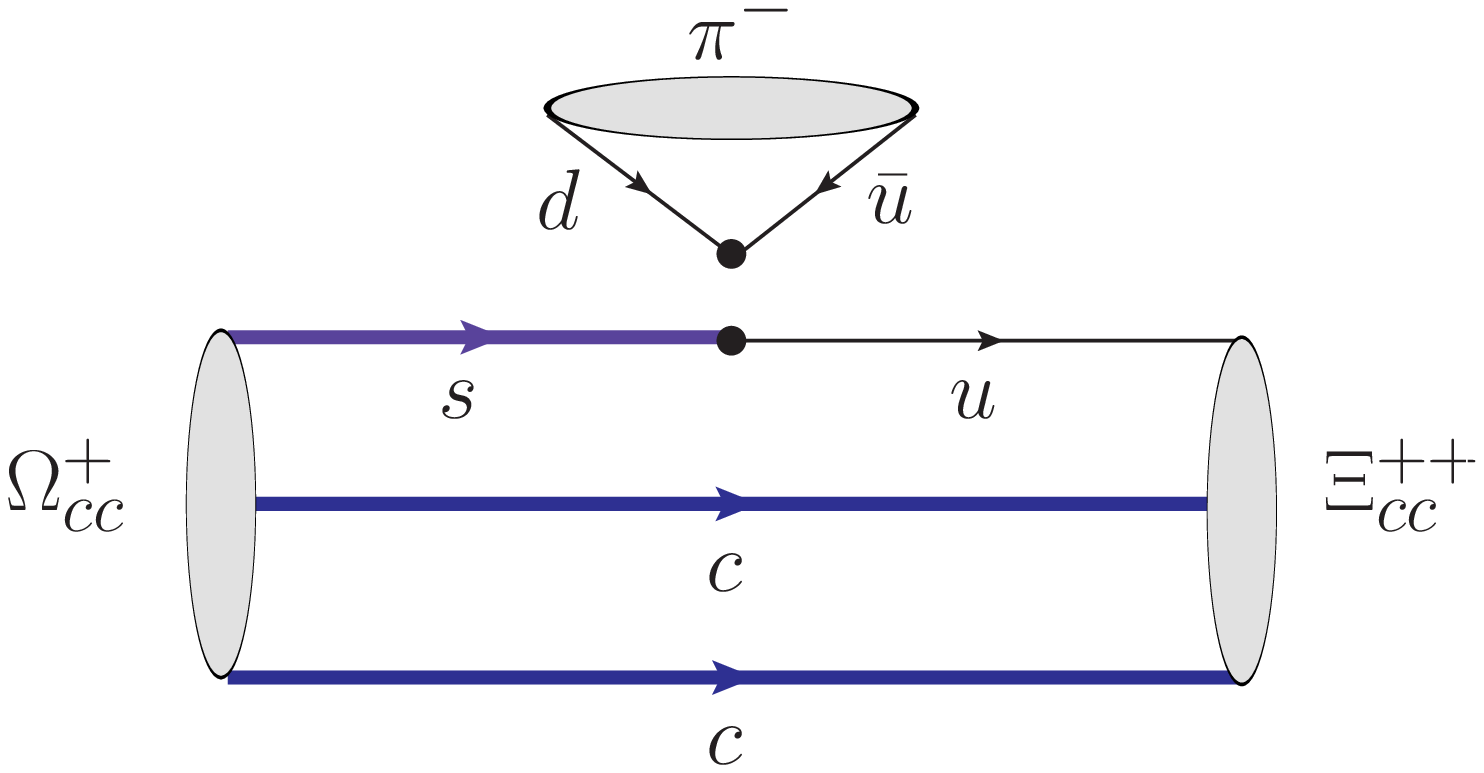}  
\caption{Feynman diagrams for    semileptonic and nonleptonic decays  of doubly heavy decay baryons. 
In these panels,    black dots correspond to the effective operators.}
\label{fd1}
\end{figure} 

\end{widetext}

Semileptonic and nonleptonic decays of the $s$ quark  are induced by the quark-level transition $s\to u \ell \bar\nu$ and $s\to u \bar u d$. The corresponding  Feynman diagrams for the processes to be investigated  are shown in Fig.~\ref{fd1}. For the nonleptonic decay processes, we only consider the factorizable contributions, which are less reliable but can give benchmark estimate of decay branching fractions.

The effective Hamiltonians for $s\to u \ell \bar \nu$ and $s\to u \bar{u} d$ are given as 
\begin{eqnarray}
&&\mathcal{H}(s\to u \ell\bar \nu)=\frac{G_F}{\sqrt{2}}\bigg\{V_{us}[\bar u\gamma_\mu(1-\gamma_5)s][\bar \ell\gamma^\mu(1-\gamma_5)\nu_\ell]\bigg\},\notag\\
&&\mathcal{H}(s\to u \bar u d)=\frac{G_F}{\sqrt{2}}V_{us}V^*_{ud}\nonumber\\
&& \times \bigg\{ C_1 [\bar u_\alpha\gamma_\mu(1-\gamma_5)s_\beta][\bar d_\beta\gamma^\mu(1-\gamma_5)u_\alpha]\nonumber\\
&& +C_2[\bar u_\alpha\gamma_\mu(1-\gamma_5)s_\alpha][\bar d_\beta\gamma^\mu(1-\gamma_5)u_\beta]\bigg\}.
\end{eqnarray}
Using the Hamiltonian, one can derive the semileptonic and color-allowed nonleptonic decay amplitudes as 
\begin{eqnarray}
&&i\mathcal{M}(\mathcal B_{Q_1Q_2}\to \mathcal{B^\prime}_{Q_1Q_2}\ell \bar\nu)=\frac{G_F}{\sqrt{2}}V_{us} \bar{l}\gamma^\mu(1-\gamma_5)\nu_\ell \nonumber\\
&&\times  \langle \mathcal{B^\prime}_{Q_1Q_2u}(P^\prime,S^\prime_z)|\bar u\gamma_\mu(1-\gamma_5)s|\mathcal B_{Q_1Q_2s}(P,S_z)\rangle,\notag\\
&&i\mathcal{M}(\mathcal B_{Q_1Q_2}\to \mathcal{B^\prime}_{Q_1Q_2}\pi^-)=\frac{G_F}{\sqrt{2}}V_{us}V^*_{ud} \nonumber\\
&&\times
a_1 \langle\pi^-(P_\pi)|\bar d\gamma^\mu(1-\gamma_5)u|0\rangle \nonumber\\
&&\times 
\langle \mathcal{B^\prime}_{Q_1Q_2u}(P^\prime,S^\prime_z)|\bar u\gamma_\mu(1-\gamma_5)s|\mathcal B_{Q_1Q_2s}(P,S_z)\rangle, \label{amp}
\end{eqnarray}
with $Q_{1,2}=b,c$, $a_1=C_1/3+C_2$.  The $C_i$ are Wilson coefficients at $m_s$ scale whose values can be taken from Ref.~\cite{Buras:1998raa}: $C_1=-0.742$,  $C_2=1.422$.  In the above the $P$ and $P^\prime$ are the momentum of doubly heavy baryons $\mathcal{B^\prime}_{Q_1Q_2}$ and $\mathcal{B}_{Q_1Q_2}$ respectively.

Hadronic parts of these processes are represented by the hadron matrix elements which can be parameterized by the form factors. For the spin-1/2 to spin-1/2 processes, the form factors are defined as
\begin{widetext}
\begin{eqnarray}
&&\langle \mathcal{B^\prime}_{Q_1Q_2u}(P^\prime,S^\prime=\frac{1}{2},S^\prime_z)|\bar u\gamma_\mu(1-\gamma_5)s |\mathcal B_{Q_1Q_2s}(P,S=\frac{1}{2},S_z)\rangle\notag\\
&=&\bar u(P^\prime,S^\prime_z)\bigg[\frac{M\gamma_\mu}{\bar M} f^{\frac{1}{2}\to\frac{1}{2}}_1(q^2)+\frac{P_\mu}{\bar M}f^{\frac{1}{2}\to\frac{1}{2}}_2(q^2)+\frac{P^\prime_\mu}{\bar M}f^{\frac{1}{2}\to\frac{1}{2}}_3(q^2)\bigg]u(P,S_z)\notag\\
&&-\bar u(P^\prime,S^\prime_z)\bigg[\frac{M\gamma_\mu}{\bar M} g^{\frac{1}{2}\to\frac{1}{2}}_1(q^2)+\frac{P_\mu}{\bar M}g^{\frac{1}{2}\to\frac{1}{2}}_2(q^2)+\frac{P^\prime_\mu}{\bar M}g^{\frac{1}{2}\to\frac{1}{2}}_3(q^2)\bigg]\gamma_5u(P,S_z),\label{ff1/2}
\end{eqnarray}
where $\bar M=M-M^\prime$. 
For the spin-3/2 to spin-1/2 process, the form factors are defined as
\begin{eqnarray}
&&\langle \mathcal{B^\prime}_{Q_1Q_2u}(P^\prime,S^\prime=\frac{1}{2},S^\prime_z)|\bar u\gamma_\mu(1-\gamma_5)s |\mathcal B_{Q_1Q_2s}(P,S=\frac{3}{2},S_z)\rangle\notag\\
&=&\bar u(P^\prime,S^\prime_z)\bigg[\frac{M\gamma_\mu P^{\prime}_\alpha}{\bar M^2} f^{\frac{3}{2}\to\frac{1}{2}}_1(q^2)+\frac{P_\mu P^{\prime}_\alpha}{\bar M^2}f^{\frac{3}{2}\to\frac{1}{2}}_2(q^2)+\frac{P^\prime_\mu P^{\prime}_\alpha}{\bar M^2}f^{\frac{3}{2}\to\frac{1}{2}}_3(q^2)+g_{\mu\alpha}f^{\frac{3}{2}\to\frac{1}{2}}_4(q^2)\bigg]\gamma_5u^\alpha(P,S_z)\notag\\
&&-\bar u(P^\prime,S^\prime_z)\bigg[\frac{M\gamma_\mu P^{\prime}_\alpha}{\bar M^2} g^{\frac{3}{2}\to\frac{1}{2}}_1(q^2)+\frac{P_\mu P^{\prime}_\alpha}{\bar M^2}g^{\frac{3}{2}\to\frac{1}{2}}_2(q^2)+\frac{P^\prime_\mu P^{\prime}_\alpha}{\bar M^2}g^{\frac{3}{2}\to\frac{1}{2}}_3(q^2)+g_{\mu\alpha}g^{\frac{3}{2}\to\frac{1}{2}}_4(q^2)\bigg]u^\alpha(P,S_z).\label{ff3/2}
\end{eqnarray}
\end{widetext}
\subsection{Light-front quark model}

 With the aid of the quark-diquark approximation, baryons are similar with the mesonic systems. In the weak transition, the two spectator quarks act as anti-quark. In doubly heavy decays,  this approximation has been widely used~\cite{Xing:2018lre,Hu:2020mxk,Zhao:2022vfr}.   Recently, a three-body vertex function in LFQM has been investigated~\cite{Ke:2019smy}. Taking the $\Lambda_b\to\Lambda_c$ and $\Sigma_b\to\Sigma_c$ transitions as the example, the authors have   demonstrated that the  three-body vertex function can give consistent results with the diquark picture, and thus have validated the diquark approximation from a certain viewpoint.

\begin{figure}[htbp!]
  \centering
\includegraphics[width=0.45\textwidth]{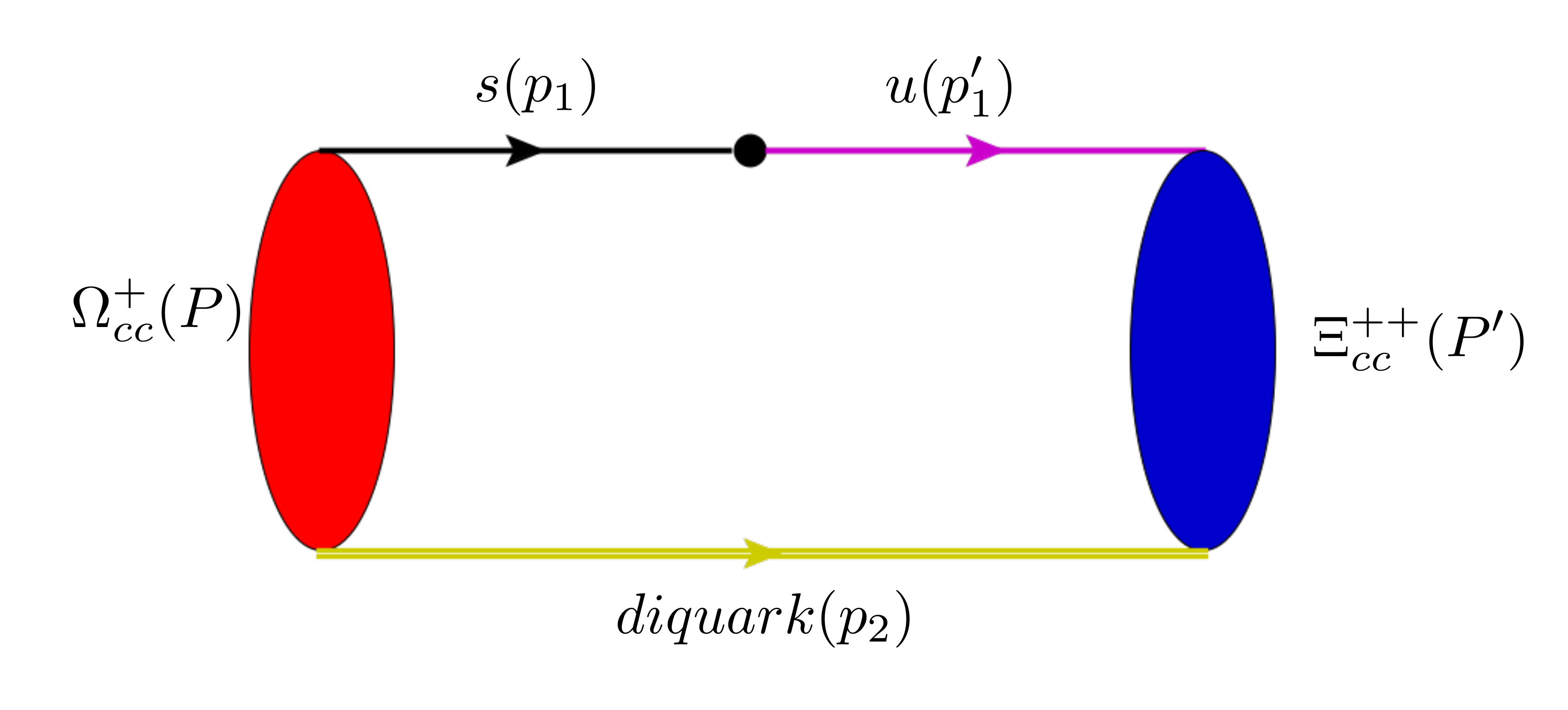}
\caption{The diquark approximation  for the baryonic transition. }
\label{diquark}
\end{figure} 


The four vector in the light front frame is represented as: $v^\mu=(v^+,v^-,v_\perp)$ with $v^\pm=v^0\pm v^3$.
In LFQM, a  baryon state can be presented by the internal quark and diquark, and  one can expand the hadron state in terms of the momenta space function and flavor-spin function. For the spin-1/2 baryon state, it is 
\begin{eqnarray}
	&& |{\cal B}(P,S,S_{z})\rangle =  \int\{d^{3}p_{1}\}\{d^{3}p_{2}\}2(2\pi)^{3}\delta^{3}(\tilde{P}-\tilde{p}_{1}-\tilde{p}_{2})\nonumber \\
	&  &\;\;\;  \times\sum_{\lambda_{1},\lambda_{2}}\Psi^{SS_{z}}(\tilde{p}_{1},\tilde{p}_{2},\lambda_{1},\lambda_{2})|q(p_{1},\lambda_{1})({\rm{di}})(p_{2},\lambda_{2})\rangle,\label{eq:state_vector}
	\end{eqnarray}
where $q$ denotes the light $s/u$ quark   and  ``$({\rm{di}})$" corresponds to the diquark shown in Fig.~\ref{diquark}. Their helicities are denoted by $\lambda_1$ and $\lambda_2$.
The momenta of the baryon, quark and diquark are $P$, $p_1$ and $p_2$, respectively.  
The momenta $\tilde{P}$, $\tilde{p}_{1}$ and $\tilde{p}_{2}$ are the three-dimensional momenta, with the notation  $\tilde{p}=(p^+,p_{\perp})$.
Obviously, the on-shell momentum has only three degrees of freedom with four components, thus the
minus component of the momentum is fixed as $p^{-}=(m^2+p_{\perp}^2)/p^{+}$.

The   wave function $\Psi$ in Eq.\eqref{eq:state_vector} can be generally decomposed as the combination of spin and momentum space: 
\begin{eqnarray}
&&\Psi^{SS_{z}}(\tilde{p}_{1},\tilde{p}_{2},\lambda_{1},\lambda_{2})=\frac{1}{\sqrt{2(p_{1}\cdot\bar{P}+m_{1}M_{0})}} \nonumber\\
&& \;\;\; \;\;\;  \;\;\; \;\;\; \times \bar{u}(p_{1},\lambda_{1})\Gamma_{S(A)} u(\bar{P},S_{z})\phi(x,k_{\perp}).\label{eq:momentum_wave_function_1/2}
\end{eqnarray}
As we mentioned above, the diquark can be a spin-0 scalar and spin-1 axial-vector. 
For the scalar diquark,  the interaction vertex $\Gamma$  is: $\Gamma_S=1$, while in a spin-1/2 baryon, the interaction vertex involving  an axial-vector diquark becomes:
\begin{align}
	\Gamma_{A} & =\frac{\gamma_{5}}{\sqrt{3}}\left(\slashed\epsilon^{*}(p_{2},\lambda_{2})-\frac{M_0+m_1+m_2}{\bar{P}\cdot p_2+m_2M_0}\epsilon^{*}(p_{2},\lambda_{2})\cdot\bar{P}\right).
	\label{eq:momentum_wave_function_1/2gamma}
\end{align}
Here $ m_1$ and $m_2$ are the masses of quark and the spectator diquark.
The $\bar P$ is the sum of the on-mass-shell momentum of the light quark $q$ and diquark. It satisfies the condition: $\bar{P}=p_1+p_2$ and $\bar{P}^2=M_{0}^2$. 
The $M_0$ is the invariant mass of $\bar P$ and is different from the baryon mass $M$. That is because the quark, the diquark and the baryon they composed can not be on their mass shells simultaneously.
The momentum $P$ and mass $M$ of baryon will satisfy the physical mass-shell condition: $M^2=P^2$.
Obviously, the momentum $\bar P$ is not equal to the $P$.

For the spin-3/2 baryon states the momenta space wave function are similar with the spin-1/2 wave function~\cite{Hu:2020mxk}:
\begin{eqnarray}
&&\Psi^{SS_{z}}(\tilde{p}_{1},\tilde{p}_{2},\lambda_{1},\lambda_{2})=\frac{1}{\sqrt{2(p_{1}\cdot\bar{P}+m_{1}M_{0})}}\nonumber\\
&& \;\;\; \;\;\;  \;\;\; \;\;\; \times\bar{u}(p_{1},\lambda_{1})\Gamma^{\alpha}_{A}(p_{2},\lambda_{2})u_{\alpha}(\bar{P},S_{z})\phi(x,k_{\perp}),\label{eq:momentum_wave_fuction_3/2}
\end{eqnarray}
In this situation, the spin of diquark can only  be $1$. The  coupling vertex $\Gamma^{\alpha}_{A}$ is then: 
 \begin{equation}
	\Gamma_{A}^{\alpha}=-\left(\epsilon^{*\alpha}(p_{2},\lambda_{2})-\frac{p_2^{\alpha}}{\bar{P}\cdot p_2+m_2M_0}\epsilon^{*}(p_{2},\lambda_{2})\cdot\bar{P}\right).
\end{equation}

The $\phi$ in Eq.~\eqref{eq:momentum_wave_function_1/2gamma} is a Gaussian-type function which is constructed as
\begin{equation}
	\phi=4\left(\frac{\pi}{\beta^{2}}\right)^{3/4}\sqrt{\frac{e_{1}e_{2}}{x_{1}x_{2}M_{0}}}\exp\left(\frac{-\vec{k}^{2}}{2\beta^{2}}\right),\label{eq:Gauss}
\end{equation}
where $e_1$ and $e_2$ represent the energy of  quark $q$ and diquark in the rest frame of $\bar{P}$.
$x_1$ and $x_2$ are the light-front momentum fractions which satisfie $0<x_2<1$ and $x_1+x_2=1$.
The internal motion of the constituent quarks are described by the internal momentum $\vec k$: 
\begin{eqnarray}
&&k_i=(k_i^-,k_i^+,k_{i\bot})=(e_i-k_{iz},e_i+k_{iz},k_{i\bot}) \nonumber\\
&& \;\;\;\; =(\frac{m_i^2+k_{i\bot}^2}{x_iM_0},x_iM_0,k_{i\bot}),\nonumber\\
&&p^+_1=x_1\bar P^+,   ~~~p^+_2=x_2 \bar P^+,  ~~~p_{1\perp}=x_1 \bar P_{\perp}+k_{1\perp}, \nonumber\\
&&  p_{2\perp}=x_2 \bar P_{\perp}+k_{2\perp},
  ~~~ k_{\perp}=-k_{1\perp}=k_{2\perp}.
\end{eqnarray}
The $\vec k$ in the Eq.~\eqref{eq:Gauss} is the internal three-momentum vector of diquark presented as $\vec k=(k_{2\bot},k_{2z})=(k_{\bot}, k_{z})$.
The parameter $\beta$ in Eq.~(\ref{eq:Gauss})  describes the momentum distributions among the constituent quarks. 
Using the definition of internal three-momentum vector, we can deduce the invariant mass square $M_0^2$ as a function of the $(x_{i},k_{i\bot})$,
 \begin{eqnarray} \label{eq:Mpz}
	  M_0^2=\frac{k_{1\perp}^2+m_1^2}{x_1}+ \frac{k_{2\perp}^2+m_2^2}{x_2}.
 \end{eqnarray}
The expressions of the energy  $e_i$ and $k_z$ can also be presented in terms of the internal variables $(x_{i},k_{i\bot})$ as
\begin{eqnarray}
&& e_i=\frac{x_iM_0}{2}+\frac{m_i^2+k_{i\perp}^2}{2x_iM_0} =\sqrt{m_i^2+k_{i\bot}^2+k_{iz}^2}, \nonumber\\
&&  k_{iz}=\frac{x_iM_0}{2}-\frac{m_i^2+k_{i\perp}^2}{2x_iM_0}.
 \end{eqnarray}
In the following,  we use the notation $x=x_2$ and  $x_1=1-x$.

\subsection{Form factors}

The hadron matrix element for the spin-1/2 to spin-1/2 processes becomes
\begin{eqnarray}
	&  & \langle{\cal B}_{Q_1Q_2u}(P^{\prime}, \frac{1}{2},S_{z}^{\prime})|\bar{u}\gamma^{\mu}(1-\gamma_{5})s|{\cal B}_{Q_1Q_2s}(P, \frac{1}{2},S_{z})\rangle\nonumber \\
	& = & \int\{d^{3}p_{2}\}\frac{\phi^{\prime}(x^{\prime},k_{\perp}^{\prime})\phi(x,k_{\perp})}{2\sqrt{p_{1}^{+}p_{1}^{\prime+}(p_{1}\cdot\bar{P}+m_{1}M_{0})(p_{1}^{\prime}\cdot\bar{P}^{\prime}+m_{1}^\prime M_{0}^{\prime})}}\nonumber \\
	&  & \times\sum_{\lambda_{2}}\bar{u}(\bar{P}^{\prime},S_{z}^{\prime})\left[\bar{\Gamma}^{\prime}(\slashed p_{1}^{\prime}+m_{1}^{\prime})\gamma^{\mu}(1-\gamma_{5})(\slashed p_{1}+m_{1})\Gamma\right] \nonumber\\
	&& \times u(\bar{P},S_{z}).\label{eq:matrix_element_onehalf}
\end{eqnarray}
For the spin-3/2 to spin-1/2 processes, the hadron matrix element is
\begin{eqnarray}
	&  & \langle{\cal B}_{Q_1Q_2u}(P^{\prime}, \frac{1}{2},S_{z}^{\prime})|\bar{u}\gamma^{\mu}(1-\gamma_{5})s|{\cal B}_{Q_1Q_2s}(P, \frac{3}{2},S_{z})\rangle\nonumber \\
		& = & \int\{d^{3}p_{2}\}\frac{\phi^{\prime}(x^{\prime},k_{\perp}^{\prime})\phi(x,k_{\perp})}{2\sqrt{p_{1}^{+}p_{1}^{\prime+}(p_{1}\cdot\bar{P}+m_{1}M_{0})(p_{1}^{\prime}\cdot\bar{P}^{\prime}+m_{1}^\prime M_{0}^{\prime})}}\nonumber \\
	&  & \times\sum_{\lambda_{2}}\bar{u}(\bar{P}^{\prime},S_{z}^{\prime})\left[\bar{\Gamma}^{\prime}_{A}(\slashed p_{1}^{\prime}+m_{1}^{\prime})\gamma^{\mu}(1-\gamma_{5})(\slashed p_{1}+m_{1})\Gamma^{\alpha}_{A}\right]\nonumber\\
	&& \times u_\alpha(\bar{P},S_{z}),\label{eq:matrix_element_threehalf}
\end{eqnarray}
and
\begin{eqnarray}
&&m_{1}=m_{s},\quad m_{1}^{\prime}=m_{u},\quad m_{2}=m_{(di)},\quad\bar{P}=p_{1}+p_{2}, \nonumber\\
&& \bar{P}^\prime=p_{1}^{\prime}+p_{2},\quad M_{0}^2=\bar{P}^{2},\quad M_{0}^{\prime2}=\bar{P}^{\prime2}.\label{parameter}
\end{eqnarray}
The momenta of $s$ quark in the initial baryon and $u$ quark in the final baryon are $p_{1}$ and $p_{1}^{\prime}$, respectively. 
The $p_2$ is the momentum of the spectator diquark. 
$P$ and $P^{\prime}$ are the four-momentum of the initial baryons ${\cal B}_{Q_1Q_2}$ and the final baryon states ${\cal B^\prime}^{+}_{Q_1Q_2}$, respectively. 
$M$ and $M^{\prime}$ are the physical masses of them.

 It is noted that  one can also define the invariant masses of the initial and final baryon states $M_{0}$ and $M^{\prime}_{0}$ with $\bar{P}^{(\prime)2}=M_0^{(\prime)2}$. The $\bar\Gamma$ in Eq.~\eqref{eq:matrix_element_onehalf} is defined as
\begin{eqnarray}
&&	\Gamma_{S}=\bar{\Gamma}^{\prime}_{S}=1,\notag\\
&&	\bar{\Gamma}^{\prime}_{A}  =\frac{1}{\sqrt{3}}\bigg(-\slashed\epsilon(p_{2},\lambda_{2}) \nonumber\\
&& \;\;\;\;\;\;\;\;\;\;\;\;\;\;\;+\frac{M_0^{\prime}+m_1^{\prime}+m_2}{\bar{P}^{\prime}\cdot p_2+m_2M_0^{\prime}}\epsilon(p_{2},\lambda_{2})\cdot\bar{P}^{\prime}\bigg)\gamma_{5}.\label{axial-vector diquarkprime}
\end{eqnarray}
Compared with the LFQM results for the hadron matrix element, the form factors defined in Eq.~\eqref{ff1/2} and Eq.~\eqref{ff3/2} can be extracted by solving the following equations as
 \begin{widetext}
\begin{eqnarray}
&&	{\rm Tr}\{(\slashed{P'}+M')\bigg[\frac{M\gamma_\mu}{\bar M} f^{\frac{1}{2}\to\frac{1}{2}}_1(q^2)+\frac{P_\mu}{\bar M}f^{\frac{1}{2}\to\frac{1}{2}}_2(q^2)+\frac{P^\prime_\mu}{\bar M}f^{\frac{1}{2}\to\frac{1}{2}}_3(q^2)\bigg](\slashed{P}+M)\{\Gamma_{i}\}_\mu\}=H^{\frac{1}{2}}_i,\notag\\
&&	{\rm Tr}\{(\slashed{P'}+M')\bigg[\frac{M\gamma_\mu}{\bar M} g^{\frac{1}{2}\to\frac{1}{2}}_1(q^2)+\frac{P_\mu}{\bar M}g^{\frac{1}{2}\to\frac{1}{2}}_2(q^2)+\frac{P^\prime_\mu}{\bar M}g^{\frac{1}{2}\to\frac{1}{2}}_3(q^2)\bigg]\gamma_5(\slashed{P}+M)\{\Gamma_{5i}\}_\mu\}=K^{\frac{1}{2}}_i,\;i=1,2,3,\notag\\
&&{\rm Tr}\{(\slashed{P'}+M')\bigg[\bigg[\frac{M\gamma_\mu P^{\prime}_\alpha}{\bar M^2} f^{\frac{3}{2}\to\frac{1}{2}}_1(q^2)+\frac{P_\mu P^{\prime}_\alpha}{\bar M^2}f^{\frac{3}{2}\to\frac{1}{2}}_2(q^2)+\frac{P^\prime_\mu P^{\prime}_\alpha}{\bar M^2}f^{\frac{3}{2}\to\frac{1}{2}}_3(q^2)+g_{\mu\alpha}f^{\frac{3}{2}\to\frac{1}{2}}_4(q^2)\bigg]\notag\\
&&\quad\quad\quad\quad\quad\quad\times\gamma_5u^\alpha(P,S_z)\bar u^\beta(P,S_z)\{\Gamma_{5j}\}_{\mu\beta}\bigg]\}=H_j^{\frac{3}{2}},\notag\\
&&{\rm Tr}\{(\slashed{P'}+M')\bigg[\bigg[\frac{M\gamma_\mu P^{\prime}_\alpha}{\bar M^2} g^{\frac{3}{2}\to\frac{1}{2}}_1(q^2)+\frac{P_\mu P^{\prime}_\alpha}{\bar M^2}g^{\frac{3}{2}\to\frac{1}{2}}_2(q^2)+\frac{P^\prime_\mu P^{\prime}_\alpha}{\bar M^2}g^{\frac{3}{2}\to\frac{1}{2}}_3(q^2)+g_{\mu\alpha}g^{\frac{3}{2}\to\frac{1}{2}}_4(q^2)\bigg]\notag\\
&&\quad\quad\quad\quad\quad\quad\times u^\alpha(P,S_z)\bar u^\beta(P,S_z)\{\Gamma_{j}\}_{\mu\beta}\bigg]\}=K_j^{\frac{3}{2}},\quad\quad\quad\quad\quad\quad\quad\quad\quad\quad\quad\quad\quad\quad\quad\quad\quad\quad j=1,2,3,4,
\end{eqnarray}
where  $H^{\frac{1}{2}}_i$, $K^{\frac{1}{2}}_i$, $H^{\frac{3}{2}}_i$ and $K^{\frac{3}{2}}_i$ are defined as
\begin{eqnarray} 
	H^{\frac{1}{2}}_i&=&\int\frac{dxd^2k_\perp}{2(2\pi)^3}\frac{\phi^\prime(x^\prime,k^\prime_\perp)\phi(x,k_\perp)}{2\sqrt{x^\prime_1x_1(p^\prime_1\cdot\bar{P}^\prime+m^\prime_1M^\prime_0)(p_1\cdot\bar{P}+m_1M_0)}}\nonumber\\
	&&\times{\rm Tr}\{(\slashed{\bar{P}}^\prime+M^\prime_0)\Gamma^\prime_{S(A)}(\slashed p^\prime_1+m^\prime_1)\gamma_\mu(\slashed p_1+m_1)\Gamma_{S(A)}(\slashed{\bar{P}}+M_0)\{\Gamma_i\}_\mu\}\notag\\
	K^{\frac{1}{2}}_i&=&\int\frac{dxd^2k_\perp}{2(2\pi)^3}\frac{\phi^\prime(x^\prime,k^\prime_\perp)\phi(x,k_\perp)}{2\sqrt{x^\prime_1x_1(p^\prime_1\cdot\bar{P}^\prime+m^\prime_1M^\prime_0)(p_1\cdot\bar{P}+m_1M_0)}}\nonumber\\
	&&\times{\rm Tr}\{(\slashed{\bar{P}}^\prime+M^\prime_0)\Gamma^\prime_{S(A)}(\slashed p^\prime_1+m^\prime_1)\gamma_\mu\gamma_5(\slashed p_1+m_1)\Gamma_{S(A)}(\slashed{\bar{P}}+M_0)\{\Gamma_{5i}\}_\mu\},\notag\\
		H^{\frac{3}{2}}_{i} & = &  \int\{d^{3}p_{2}\}\frac{\phi^{\prime}(x^{\prime},k_{\perp}^{\prime})\phi(x,k_{\perp})}{2\sqrt{p_{1}^{+}p_{1}^{\prime+}(p_{1}\cdot\bar{P}+m_{1}M_{0})(p_{1}^{\prime}\cdot\bar{P}^{\prime}+m_{1}^{\prime}M_{0}^{\prime})}}\nonumber \\
	&  &\quad \times\sum_{S_{z}^{\prime}\lambda_{2}}{\rm Tr}\Big\{ (\slashed{\bar{ P}}^\prime+M^\prime_0)\bar{\Gamma}^{\prime}_{A}(\slashed p_{1}^{\prime}+m_{1}^{\prime})\gamma_{\mu}(\slashed p_{1}+m_{1})\Gamma^\alpha_{A}u^{\alpha}(\bar{P},S_{z})\bar{u}^{\beta}(\bar{P},S_{z})\{\Gamma_{5j}\}_{\mu\beta}\Big\},\notag\\
	K^{\frac{3}{2}}_{i} & = &  \int\{d^{3}p_{2}\}\frac{\phi^{\prime}(x^{\prime},k_{\perp}^{\prime})\phi(x,k_{\perp})}{2\sqrt{p_{1}^{+}p_{1}^{\prime+}(p_{1}\cdot\bar{P}+m_{1}M_{0})(p_{1}^{\prime}\cdot\bar{P}^{\prime}+m_{1}^{\prime}M_{0}^{\prime})}}\nonumber \\
	&  &\quad \times\sum_{S_{z}^{\prime}\lambda_{2}}{\rm Tr}\Big\{ (\slashed {\bar{ P}}^\prime+M^\prime_0)\bar{\Gamma}^{\prime}_{A}(\slashed p_{1}^{\prime}+m_{1}^{\prime})\gamma_{\mu}\gamma_5(\slashed p_{1}+m_{1})\Gamma^\alpha_{A}u^{\alpha}(\bar{P},S_{z})\bar{u}^{\beta}(\bar{P},S_{z})\{\Gamma_{j}\}_{\mu\beta}\Big\}.
\end{eqnarray}
 \end{widetext}
 
The different Dirac structures $\Gamma_i$ and $\Gamma_{5i}$ are
\begin{eqnarray} 
	&&\{\Gamma_{i}\}_\mu=\{\gamma_\mu,P_{\mu},P^\prime_{\mu}\},\nonumber\\
	&& \{\Gamma_{5i}\}_\mu=\{\gamma_\mu\gamma_5,P_{\mu}\gamma_5,P^\prime_{\mu}\gamma_5\},\notag\\
	&&\{\Gamma_{j}\}_{\mu\beta}=\{\gamma_\mu P^\prime_{\beta},P_{\mu}P^\prime_{\beta},P^\prime_{\mu}P^\prime_{\beta}, g_{\mu\beta}\},\nonumber\\
	&&
	\{\Gamma_{5j}\}_{\mu\beta}=\{\gamma_\mu\gamma_5 P^\prime_{\beta},P_{\mu}P^\prime_{\beta}\gamma_5,P^\prime_{\mu}P^\prime_{\beta}\gamma_5,g_{\mu\beta}\gamma_5\}.
\end{eqnarray} 
Then the form factors are solved as as
 \begin{eqnarray}
 	&&f_1^{\frac{1}{2}\rightarrow\frac{1}{2}}=-\frac{\bar M(s_+H^{\frac{1}{2}}_1-2M^\prime H^{\frac{1}{2}}_2-2M H^{\frac{1}{2}}_3)}{4Ms_-s_+},
	\nonumber\\
	&& f_2^{\frac{1}{2}\rightarrow\frac{1}{2}}=\frac{\bar M(M^\prime s_+H^{\frac{1}{2}}_1-6M^{\prime 2}H^{\frac{1}{2}}_2+2(s_-+MM^\prime)H^{\frac{1}{2}}_3)}{2s_-s_+^2},\notag\\
	&&f_3^{\frac{1}{2}\rightarrow\frac{1}{2}}=\frac{\bar M(Ms_+H^{\frac{1}{2}}_1+2(s_-+MM^\prime)H^{\frac{1}{2}}_2-6M^2H^{\frac{1}{2}}_3)}{2s_-s_+^2}.  \nonumber\\\label{fffu}
\end{eqnarray}
The spin-3/2 to spin-1/2 form factors are solved as
\begin{widetext}
\begin{eqnarray}
	f_1^{\frac{3}{2}\rightarrow\frac{1}{2}}&=&\frac{-\bar M^2}{2s_-^2s_+^2}\big\{4{M}\big[{H^{\frac{3}{2}}_1}s_--{H^{\frac{3}{2}}_3}{M}\big]-2{H^{\frac{3}{2}}_2}\big(s_--2MM^\prime\big)+{H^{\frac{3}{2}}_4}(s_-s_+)\big\},\notag\\
	f_2^{\frac{3}{2}\rightarrow\frac{1}{2}}&=&\frac{-\bar M^2}{s_-^3s_+^2}\big\{s_-\big[H^{\frac{3}{2}}_1M\big(s_--2MM^\prime\big)+H^{\frac{3}{2}}_4\big(s_-+MM^\prime\big)s_+\big]+4H^{\frac{3}{2}}_3M^2\big(2s_+-3MM^\prime\big)\notag\\
	&&-2H^{\frac{3}{2}}_2\big[M^4-2M^3M^\prime+2M^2(6M^{\prime 2}-q^2)+2M M^\prime(q^2-M^{\prime 2})+(M^{\prime2}-q^2)^2\big]\big\},\notag\\
	f_3^{\frac{3}{2}\rightarrow\frac{1}{2}}&=&\frac{{M^2}{\bar M^2}}{s_-^3s_+^2}\big\{s_-\big(H^{\frac{3}{2}}_4s_+-2H^{\frac{3}{2}}_1 M\big)+20H^{\frac{3}{2}}_3M^2-4H^{\frac{3}{2}}_2\big(2s_+-3MM^\prime\big)\big\},\notag\\
	f_4^{\frac{3}{2}\rightarrow\frac{1}{2}}&=&\frac{1}{2s_-^2s_+}\big\{ s_-\left[{H^{\frac{3}{2}}_1}M+{H^{\frac{3}{2}}_4}s_+\right]+2{H^{\frac{3}{2}}_3}{M}^2-2{H^{\frac{3}{2}}_2}\big[s_-+MM^\prime\big]\big\},\label{ffe}
\end{eqnarray}
\end{widetext}
where $s_\pm=(M\pm M^\prime)^2-q^2$.
The expressions of form factors $g^{\frac{1}{2}\to\frac{1}{2}}_i$ and $g^{\frac{3}{2}\to\frac{1}{2}}_i$ can also be obtained by applying the transformation as
\begin{eqnarray}
	&&M^{\prime}\to -M^{\prime}, \bar M\to \bar M,  H_{j}\to K_{j},   i=1\notag\\
	&&M^{\prime}\to -M^{\prime} , \bar M\to \bar M,  H_{j}\to -K_{j},  i=2,3,4.
\end{eqnarray}

 Since the diquark can be scalar or axis-vector,   the physical baryon state is the combination of the baryon state with scalar diquark and axis-vector in LFQM. Thus the physical form factor can also be written as
 \begin{eqnarray}
&& \rm  [formfactor]^{physical}=c_S \times  [formfactor]_S \nonumber\\
&& \;\;\;\;\;\;  +c_A\times  \rm  [formfactor]_A.
 \end{eqnarray}
 The coefficients $c_S$ and $c_A$ are flavor-spin factors that can be calculated by the flavor-spin wave function of baryons in the diquark basis. 
 The flavor-spin wave functions of the spin-1/2 baryons are
  \begin{eqnarray}
&&|{\cal B}_{Q_1Q_2},S_z=\frac{1}{2}\rangle=\frac{1}{\sqrt{2}}(Q_1Q_2+Q_2Q_1)q\nonumber\\
&&\;\;\;\;\; \times \bigg(\frac{1}{\sqrt{6}}(\uparrow\downarrow\uparrow+\downarrow\uparrow\downarrow-2\uparrow\uparrow\downarrow)\bigg)=-[Q_1Q_2]_Aq,
 \end{eqnarray}
 and the flavor-spin wave function of the spin-3/2 baryons are
  \begin{eqnarray}
 &&|{\cal B}_{Q_1Q_2},S_z=\frac{3}{2}\rangle=\frac{1}{\sqrt{2}}(Q_1Q_2+Q_2Q_1)q\nonumber\\
 &&\;\;\;\;\; \times (\uparrow\uparrow\uparrow)=[Q_1Q_2]_Aq,\notag\\
 &&|{\cal B}_{Q_1Q_2},S_z=\frac{1}{2}\rangle=\frac{1}{\sqrt{2}}(Q_1Q_2+Q_2Q_1)q\nonumber\\
 &&\;\;\;\;\; \times\frac{1}{\sqrt{3}} (\uparrow\uparrow\downarrow+\uparrow\downarrow\uparrow+\downarrow\uparrow\uparrow)=[Q_1Q_2]_Aq. 
   \end{eqnarray}
   
 For the doubly heavy baryons ${\cal B}_{bc}$, there is another set of the flavor-spin wave functions in which the $\rm bc$ system is scalar. Their flavor-spin wave functions are
   \begin{eqnarray}
&&|{\cal B}_{bc},S_z=\frac{1}{2}\rangle=\frac{1}{\sqrt{2}}(bc-bc)q\nonumber\\
&&\;\;\;\;\; \times\bigg(\frac{1}{\sqrt{2}}(\uparrow\downarrow\uparrow-\downarrow\uparrow\downarrow)\bigg)=[bc]_S q,
 \end{eqnarray}
  where the flavor wave functions of the diquark basis are
    \begin{eqnarray}
&&[Q_1Q_2]_A=\frac{1}{\sqrt{2}}(Q_1Q_2+Q_2Q_1),\nonumber\\
&& [Q_1Q_2]_S=\frac{1}{\sqrt{2}}(Q_1Q_2-Q_2Q_1).
 \end{eqnarray}
 Thus the overlapping factors are extracted and shown in Table~.\ref{ol}.

 \begin{table}[!htb]
\caption{The spin  factors for the doubly heavy baryon decay induced by $s\to u$..}
\label{ol} %
\begin{tabular}{|c|c|c|c|c|c|c|c|c|}
\hline \hline
channel & $\quad c_S\quad$  & $\quad c_A\quad$   \tabularnewline
\hline
${\cal B}_{Q_1Q_2s}\to{\cal B}_{Q_1Q_2u}$&0&1\tabularnewline
\hline
$\Omega_{bc}^{\prime0}\to\Omega_{bc}^{\prime+}$&1&0\tabularnewline
\hline
${\cal B}^*_{Q_1Q_2s}\to{\cal B}_{Q_1Q_2u}$&0&$-1$\tabularnewline
\hline
\end{tabular}
\end{table}
 
\section{Numerical results}

After deriving the analytic expression in  the LFQM,  we now give the numerical results. In the calculation, the masses of quarks are taken from Refs.~\cite{Li:2010bb,Verma:2011yw,Shi:2016gqt}
\begin{eqnarray}
	 && m_u=m_d= 0.25~{\rm GeV}, \;\; m_s=0.37~{\rm GeV}, \nonumber\\
	 && m_c=1.4~{\rm GeV},\;\;m_b=4.8~{\rm GeV}.\label{eq:mass_quark}
\end{eqnarray}
The diquark masses can be approximatively given as 
\begin{eqnarray}
	m_{[bb]}=2m_b,\quad m_{[cc]}=2m_{c},\quad m_{[bc]}=m_b+m_c.\label{eq:mass_diquark}
\end{eqnarray}
The baryon masses, lifetime, and input parameter $\beta_{[Q_1Q_2]q}$ are shown in Table.~\ref{mass_beta}.

\begin{widetext}

\begin{table}[!htb]
\caption{Doubly heavy baryon masses~\cite{Ghalenovi:2022dok,Workman:2022ynf}, lifetime~\cite{Cheng:2018mwu,Kiselev:2001fw} and input parameter $\beta_{[Q_1Q_2]q}$. The mass of $\Xi_{cc}^{++}$ is the experimental data and other parameters are theorical prediction.}
\label{mass_beta} %
\begin{tabular}{|c|c|c|c|c|c|c|c|c|c|c|c|c|c|}
\hline \hline
Baryon & $\Xi_{cc}^{++}$  & $\Omega_{cc}^+$  & $\Omega_{cc}^{*+}$  & $\Xi_{bc}^{(\prime)+}$  & $\Xi_{bc}^{*+}$  & $\Omega_{bc}^{(\prime) 0}$  & $\Omega_{bc}^{*0}$  & $\Xi_{bb}^0$ &$\Omega_{bb}^-$ & $\Omega_{bb}^{*-}$   \tabularnewline
\hline
Mass $({\rm GeV})$&3.622&3.798&3.831&6.958&6.991&7.137&7.170&10.322&10.500&10.533\tabularnewline
\hline
Life time (fs)&256&180&-&244&-&220&-&370&800&-\tabularnewline
\hline
$ \beta_{[cc]u}=0.470$&\multicolumn{2}{|c|}{$ \beta_{[cc]s}=0.535$}&\multicolumn{2}{|c|}{$ \beta_{[bc]u}=0.562$}&\multicolumn{2}{|c|}{$ \beta_{[bc]s}=0.623$}& \multicolumn{2}{|c|}{$ \beta_{[bb]u}=0.562$}&\multicolumn{2}{|c|}{$ \beta_{[bb]s}=0.623$}\\\hline
\end{tabular}
\end{table}
\end{widetext}

\subsection{Form factors}

From the parametrization of form factors in Eq.~\eqref{ff1/2} and Eq.~\eqref{ff3/2}, one can notice  that the form factors  are the functions of $q^2$. To access  the $q^2$ dependence of the form factors,  we take the following parametrization scheme called the single pole model as
 \begin{eqnarray}
	F(q^2)=\frac{F(0)}{1-\frac{q^2}{m_{\rm pole}^2}},\label{fit}
\end{eqnarray}
where $F(0)$ is the numerical results of form factor at $q^2=0$. In this formula, the $m_{\rm pole}$ is set as $m_{\rm pole}=0.494{\rm GeV}$ for the form factor $g_i$ and $m_{\rm pole}=0.845{\rm GeV}$ for $f_i$. It can be understood by inserting a set of hadron complete states in the transition matrix element as
 \begin{eqnarray}
&& \langle B^\prime|\bar u\gamma^\mu(1-\gamma_5)s|B\rangle= \nonumber\\
 &=&\sum_\lambda\int \frac{d^4q}{(2\pi)^4}\frac{1}{q^2-m_{\lambda}^2}\langle B^\prime \lambda|B \rangle\langle 0|\bar u\gamma^\mu(1-\gamma_5)s|\lambda\rangle\notag\\
&=&\int\frac{d^4k}{(2\pi)^4}\bigg[\frac{f_{K^*}k^\mu}{k^2-m_{K^*}^2}\langle B^\prime K^*|B \rangle-\frac{f_{K}k^\mu}{k^2-m_{K}^2}\langle B^\prime K|B \rangle\bigg] \nonumber\\
&& +\quad\cdot\cdot\cdot\notag\\
&=&\frac{f_{K^*}q^\mu}{q^2-m_{K^*}^2}{\mathcal M}(B\to B^\prime K^*)-\frac{f_{K}q^\mu}{q^2-m_{K}^2}{\mathcal M}(B\to B^\prime K)\nonumber\\
&& +\quad\cdot\cdot\cdot,
\end{eqnarray}
with
 \begin{eqnarray}
\langle 0|\bar u\gamma^\mu s|K^*(q)\rangle&=&f_{K^*}q^\mu, \nonumber\\
\langle 0|\bar u\gamma^\mu \gamma_5 s|K(q)\rangle &=&f_{K}q^\mu.
\end{eqnarray}
It can be easily seen that the $q^2$ dependence behavior of the transition matrix element will have a pole at $q^2=m_K^2$ for the axis-vector current and $q^2=m^2_{K^*}$ for the vector current. Therefore it is reasonable to set   $m_{\rm pole}=m_K$ for the form factors $g_i$ and $m_{\rm pole}=m_{K^*}$ for $f_i$.
Numerical results for the form factors are collected in Table.~\ref{ffnr} and  Table.~\ref{ffnr3}.

 \begin{table}[!htb]
\caption{Numerical results for the form factors of spin-1/2 to spin-1/2 transitions. The F$^*(0)$ indicates the form factors simplified in the heavy quark limit in Eq.\eqref{fit1}.}
\label{ffnr} %
\begin{tabular}{|c|c|c|c|c|c|c|c|c|c|c|c|}
\hline \hline
channel &F & F(0) &F$^*(0)$  &F& F(0) &F$^*(0)$  \\
\hline
\multirow{3}{*}{$\Omega_{cc}^+\to \Xi_{cc}^{++}$}&$ f_1^{\frac{1}{2}\to\frac{1}{2}}$&-1.53&-1.56&$ g_1^{\frac{1}{2}\to\frac{1}{2}}$&-0.013&0\cr\cline{2-7}
&$ f_2^{\frac{1}{2}\to\frac{1}{2}}$&1.68&1.22&$ g_2^{\frac{1}{2}\to\frac{1}{2}}$&-3.95&-4.41
\cr\cline{2-7}
&$ f_3^{\frac{1}{2}\to\frac{1}{2}}$&-0.091&0.34&$ g_3^{\frac{1}{2}\to\frac{1}{2}}$&3.57&4.41\tabularnewline
\hline
\multirow{3}{*}{$\Omega_{bb}^-\to\Xi_{bb}^0$} &$ f_1^{\frac{1}{2}\to\frac{1}{2}}$&0.47&0.48&$ g_1^{\frac{1}{2}\to\frac{1}{2}}$&-0.0048&0\cr\cline{2-7}
 &$ f_2^{\frac{1}{2}\to\frac{1}{2}}$&0.89&0.22&$ g_2^{\frac{1}{2}\to\frac{1}{2}}$&-16.34&-44.61\cr\cline{2-7}
  &$ f_3^{\frac{1}{2}\to\frac{1}{2}}$&-1.36&-0.70&$ g_3^{\frac{1}{2}\to\frac{1}{2}}$&15.92&44.61\tabularnewline
\hline
\multirow{3}{*}{$\Omega_{bc}^0\to \Xi_{bc}^+$} &$ f_1^{\frac{1}{2}\to\frac{1}{2}}$&0.30&0.31&$ g_1^{\frac{1}{2}\to\frac{1}{2}}$&-0.007&0\cr\cline{2-7}
 &$ f_2^{\frac{1}{2}\to\frac{1}{2}}$&0.99&0.30&$ g_2^{\frac{1}{2}\to\frac{1}{2}}$&-11.78&-29.94\cr\cline{2-7}
  &$ f_3^{\frac{1}{2}\to\frac{1}{2}}$&-1.28&-0.61&$ g_3^{\frac{1}{2}\to\frac{1}{2}}$&11.37&29.94\tabularnewline
\hline
\multirow{3}{*}{$\Omega_{bc}^{\prime0}\to \Xi_{bc}^{\prime+}$} &$ f_1^{\frac{1}{2}\to\frac{1}{2}}$&0.46&0.47&$ g_1^{\frac{1}{2}\to\frac{1}{2}}$&0.02&0\cr\cline{2-7}
 &$ f_2^{\frac{1}{2}\to\frac{1}{2}}$&0.91&0.22&$ g_2^{\frac{1}{2}\to\frac{1}{2}}$&35.34&89.81\cr\cline{2-7}
  &$ f_3^{\frac{1}{2}\to\frac{1}{2}}$&-1.36&-0.69&$ g_3^{\frac{1}{2}\to\frac{1}{2}}$&-34.11&-89.81\tabularnewline
\hline
\end{tabular}
\end{table}

 \begin{table}[!htb]
\caption{Numerical results for the form factors in spin-3/2 to spin-1/2 doubly heavy baryon decays induced by $s\to u$. The F$^*(0)$ indicates the form factors simplified in the heavy quark limit in Eq.\eqref{fit2}.}
\label{ffnr3} %
\begin{tabular}{|c|c|c|c|c|c|c|c|c|c|c|c|}
\hline \hline
channel &form factor & F(0) &F$^*(0)$  &form factor& F(0) &F$^*(0)$   \\
\hline
\multirow{4}{*}{$\Omega_{cc}^{*+}\to \Xi_{cc}^{++}$}&$ f_1^{\frac{3}{2}\to\frac{1}{2}}$&0.0034&0&$ g_1^{\frac{3}{2}\to\frac{1}{2}}$&-0.33&-0.87\cr\cline{2-7}
&$ f_2^{\frac{3}{2}\to\frac{1}{2}}$&0.06&0.08&$ g_2^{\frac{3}{2}\to\frac{1}{2}}$&0.30&1.92
\cr\cline{2-7}
&$ f_3^{\frac{3}{2}\to\frac{1}{2}}$&-0.06&-0.08&$ g_3^{\frac{3}{2}\to\frac{1}{2}}$&-0.018&-1.05\cr\cline{2-7}
&$ f_4^{\frac{3}{2}\to\frac{1}{2}}$&-2.19&-2.17&$ g_4^{\frac{3}{2}\to\frac{1}{2}}$&-1.13&-1.66\tabularnewline
\hline
\multirow{4}{*}{$\Omega_{bb}^{*-} \to\Xi_{bb}^0$}&$ f_1^{\frac{3}{2}\to\frac{1}{2}}$&0.0012&0&$ g_1^{\frac{3}{2}\to\frac{1}{2}}$&-1.77&-2.44\cr\cline{2-7}
&$ f_2^{\frac{3}{2}\to\frac{1}{2}}$&0.035&0.056&$ g_2^{\frac{3}{2}\to\frac{1}{2}}$&1.23&3.26
\cr\cline{2-7}
&$ f_3^{\frac{3}{2}\to\frac{1}{2}}$&-0.035&-0.056&$ g_3^{\frac{3}{2}\to\frac{1}{2}}$&0.53&-0.82\cr\cline{2-7}
&$ f_4^{\frac{3}{2}\to\frac{1}{2}}$&-5.93&-5.91&$ g_4^{\frac{3}{2}\to\frac{1}{2}}$&-1.54&-2.21\tabularnewline
\hline
\multirow{4}{*}{$\Omega_{bc}^{*0}\to \Xi_{bc}^+$}&$ f_1^{\frac{3}{2}\to\frac{1}{2}}$&0.002&0&$ g_1^{\frac{3}{2}\to\frac{1}{2}}$&-1.71&-2.38\cr\cline{2-7}
&$ f_2^{\frac{3}{2}\to\frac{1}{2}}$&0.039&0.06&$ g_2^{\frac{3}{2}\to\frac{1}{2}}$&1.15&3.20
\cr\cline{2-7}
&$ f_3^{\frac{3}{2}\to\frac{1}{2}}$&-0.039&-0.06&$ g_3^{\frac{3}{2}\to\frac{1}{2}}$&0.54&-0.83\cr\cline{2-7}
&$ f_4^{\frac{3}{2}\to\frac{1}{2}}$&-3.93&-3.91&$ g_4^{\frac{3}{2}\to\frac{1}{2}}$&-1.51&-2.19\tabularnewline
\hline
\end{tabular}
\end{table}

\subsection{Heavy quark limit}

The transition form factors for the doubly heavy baryon decays can be greatly simplified in heavy quark limit. In this limit, the initial baryon mass $M$ and the final baryon mass $M^\prime$ become
 \begin{eqnarray}
M\approx M^\prime \approx m_Q,
\end{eqnarray}
where the $m_Q$ is the mass of heavy diquark. Then the formula of the form factors in Eq.~\eqref{fffu} and Eq.~\eqref{ffe} can be expanded in $1/m_Q$. In the first step, the dimensionless functions $\hat{H_i}$ and $\hat{K_i}$ are introduced as
 \begin{eqnarray}
&&\hat{H_1}^{\frac{1}{2}}=\frac{H_1^{\frac{1}{2}}}{(M+M^\prime)\bar M},\quad \hat{H_2}^{\frac{1}{2}}=\frac{H_2^{\frac{1}{2}}}{(M+M^\prime)^2\bar M},\notag\\
&& \hat{H_3}^{\frac{1}{2}}=\frac{H_3^{\frac{1}{2}}}{(M+M^\prime)^2\bar M}, \quad \hat{K_1}^{\frac{1}{2}}=\frac{K_1^{\frac{1}{2}}}{(M+M^\prime)\bar M},\notag\\
&& \hat{K_2}^{\frac{1}{2}}=\frac{K_2^{\frac{1}{2}}}{\bar M^3},\quad \hat{K_3}^{\frac{1}{2}}=\frac{K_3^{\frac{1}{2}}}{\bar M^3}\label{fit1}.
\end{eqnarray}
Using these dimensionless function, one can simplify the form factors at $q^2=0$ in the heavy quark limit as
 \begin{eqnarray}
&&f_1^{\frac{1}{2}\rightarrow\frac{1}{2}}=\frac{1}{2}(-\hat{H_1}^{\frac{1}{2}}+\hat{H_2}^{\frac{1}{2}}+\hat{H_3}^{\frac{1}{2}})+{\cal O}(1/m^3_Q),\notag\\
&& f_2^{\frac{1}{2}\rightarrow\frac{1}{2}}=\frac{1}{4}(\hat{H_1}^{\frac{1}{2}}+\hat{H_2}^{\frac{1}{2}}-3\hat{H_3}^{\frac{1}{2}})+{\cal O}(1/m^3_Q),\notag\\
&&f_3^{\frac{1}{2}\rightarrow\frac{1}{2}}=\frac{1}{4}(\hat{H_1}^{\frac{1}{2}}-3\hat{H_2}^{\frac{1}{2}}+\hat{H_3}^{\frac{1}{2}})+{\cal O}(1/m^3_Q),\notag\\
&& g_1^{\frac{1}{2}\rightarrow\frac{1}{2}}=0, \notag\\
&&g_2^{\frac{1}{2}\rightarrow\frac{1}{2}}=-g_3^{\frac{1}{2}\rightarrow\frac{1}{2}}=\frac{1}{4}(\hat{K_1}^{\frac{1}{2}}+3\hat{K_2}^{\frac{1}{2}}-3\hat{K_3}^{\frac{1}{2}})\notag\\
&&\;\;\;\; +{\cal O}(1/m^3_Q).
\end{eqnarray}

For the spin-$3/2$ to spin-$1/2$ processes form factors, the dimensionless functions $\hat{H_i}$ and $\hat{K_i}$ are defined as
 \begin{eqnarray}
&&\hat{H_1}^{\frac{3}{2}}=\frac{H_1^{\frac{3}{2}}}{(M+M^\prime)\bar M^2},\quad \hat{H_2}^{\frac{3}{2}}=\frac{H_2^{\frac{3}{2}}}{\bar M^4},\notag\\
&& \hat{H_3}^{\frac{3}{2}}=\frac{H_3^{\frac{3}{2}}}{\bar M^4},\quad\quad \hat{H_4}^{\frac{3}{2}}=\frac{H_4^{\frac{3}{2}}}{\bar M^2}, \notag\\
&&\hat{K_1}^{\frac{3}{2}}=\frac{K_1^{\frac{3}{2}}}{(M+M^\prime)\bar M^2},\quad \hat{K_2}^{\frac{3}{2}}=\frac{K_2^{\frac{3}{2}}}{(M+M^\prime)^2\bar M^2},\notag\\
&& \hat{K_3}^{\frac{3}{2}}=\frac{K_3^{\frac{3}{2}}}{(M+M^\prime)^2\bar M^2},\quad \hat{K_4}^{\frac{3}{2}}=\frac{K_4^{\frac{3}{2}}}{(M+M^\prime)^2}.\label{fit2}
\end{eqnarray}
Then the  form factors at $q^2=0$ can be expressed in the heavy quark limit as
\begin{eqnarray}
&& f_1^{\frac{3}{2}\rightarrow\frac{1}{2}}=0,\notag\\
&& f_2^{\frac{3}{2}\rightarrow\frac{1}{2}}=-f_3^{\frac{3}{2}\rightarrow\frac{1}{2}}=
\notag\\
&&\;\;\;\; \;\;\;=
 \frac{1}{4}(\hat{H_1}^{\frac{3}{2}}-5\hat{H_2}^{\frac{3}{2}}+5\hat{H_3}^{\frac{3}{2}}-\hat{H_4}^{\frac{3}{2}})+{\cal O}(1/m^3_Q),\notag\\
&& f_4^{\frac{3}{2}\rightarrow\frac{1}{2}}=\frac{1}{4}(\hat{H_1}^{\frac{3}{2}}+\hat{H_2}^{\frac{3}{2}}-\hat{H_3}^{\frac{3}{2}}+2\hat{H_4}^{\frac{3}{2}})+{\cal O}(1/m^3_Q),\notag\\
&&g_1^{\frac{3}{2}\rightarrow\frac{1}{2}}=\frac{1}{2}(-2\hat{K_1}^{\frac{3}{2}}+3\hat{K_2}^{\frac{3}{2}}+\hat{K_3}^{\frac{3}{2}}-\hat{K_4}^{\frac{3}{2}}),\notag\\
&& g_2^{\frac{3}{2}\rightarrow\frac{1}{2}}=\frac{3}{4}(\hat{K_1}^{\frac{3}{2}}-3\hat{K_2}^{\frac{3}{2}}+\hat{K_3}^{\frac{3}{2}}+\hat{K_4}^{\frac{3}{2}})+{\cal O}(1/m^3_Q),
\notag\\
&&g_3^{\frac{3}{2}\rightarrow\frac{1}{2}}=\frac{1}{4}(\hat{K_1}^{\frac{3}{2}}+3\hat{K_2}^{\frac{3}{2}}-5\hat{K_3}^{\frac{3}{2}}-\hat{K_4}^{\frac{3}{2}})+{\cal O}(1/m^3_Q),\notag\\
&& g_4^{\frac{3}{2}\rightarrow\frac{1}{2}}=\frac{1}{4}(-\hat{K_1}^{\frac{3}{2}}+3\hat{K_2}^{\frac{3}{2}}-\hat{K_3}^{\frac{3}{2}}-2\hat{K_4}^{\frac{3}{2}})+{\cal O}(1/m_Q^3). \nonumber\\
\end{eqnarray}
Using these simplified formulas, one can   estimate the numerical results of form factors in heavy quark limit. The numerical results are also shown in Table.~\ref{ffnr} and  Table.~\ref{ffnr3}.

\subsection{Semileptonic decays}

  In this subsection, the semileptonic decays are studied with the help of helicity amplitudes. 
 The full decay amplitude can be decomposed in terms of hadron helicity amplitude and lepton helicity amplitude as
 \begin{eqnarray}
	i\mathcal{M}(\mathcal B_{Q_1Q_2}\to \mathcal{B^\prime}_{Q_1Q_2}\ell \nu)=\frac{G_F}{\sqrt{2}}V_{us} H^{S,\lambda}_{\lambda^\prime,\lambda_W}L^{\lambda_W}_{\lambda_\ell},
	\notag\\
	H^{S,\lambda}_{\lambda^\prime,\lambda_W}=HV^{S,\lambda}_{\lambda^\prime,\lambda_W}-HA^{S,\lambda}_{\lambda^\prime,\lambda_W}.
\end{eqnarray}
The helicity amplitudes $HV,\;HA$ and $L$ are defined as
 \begin{eqnarray}
&&HV^{S,\lambda}_{\lambda^\prime,\lambda_W}=\langle \mathcal{B^\prime}(\lambda^\prime)|\bar u\gamma_\mu s|\mathcal B(S,\lambda)\rangle\epsilon_W^{*\mu}(\lambda_W), \nonumber\\
&&  HA^{S,\lambda}_{\lambda^\prime,\lambda_W}=\langle \mathcal{B^\prime}(\lambda^\prime)|\bar u\gamma_\mu \gamma_5 s|\mathcal B(S,\lambda)\rangle\epsilon_W^{*\mu}(\lambda_W),\notag\\
&&L^{\lambda_W}_{\lambda_\ell}=\bar{l}\gamma_\mu(1-\gamma_5)\nu_l(\lambda_\ell)\epsilon_W^{\mu}(\lambda_W),
\end{eqnarray}
where $\epsilon_W^{\mu}$ is the polarization vector of the virtual propagator and $\lambda_W$ means the polarization of the virtual propagator. The $\lambda$, $\lambda^\prime$, and $\lambda_\ell$ are the helicities of the initial baryon, the final baryon and the lepton. $S$ denotes the spin of the initial baryon states.

Then the differential decay width of semileptonic decays is
\begin{eqnarray}
&&\frac{d^2\Gamma}{dq^2d\cos\theta}=\frac{G_F^2|V_{CKM}^2|}{2 }\frac{\sqrt{s_-s_+}}{512\pi^3M^3}(1-\hat{m_\ell}^2) \nonumber\\
&& \times \bigg[L_1+L_2\cos\theta+L_3\cos2\theta\bigg].\label{angle}
\end{eqnarray}
Here the coefficients $L_i$ can be expressed by the helicity amplitudes $H^{S,\lambda}_{\lambda^\prime,\lambda_W}$.
For the spin-1/2 to spin-1/2 processes, the coefficients are
\begin{widetext}
\begin{eqnarray}
	L_1&=&\frac{q^2}{2}(1-\hat{m_\ell}^2) \bigg[(3+\hat{m_\ell}^2)(|H_{\frac{1}{2},1}^{\frac{1}{2},\frac{1}{2}}|^2+|H_{-\frac{1}{2},-1}^{\frac{1}{2},-\frac{1}{2}}|^2)+2(1+\hat{m_\ell}^2)(|H_{-\frac{1}{2},0}^{\frac{1}{2},\frac{1}{2}}|^2+|H_{\frac{1}{2},0}^{\frac{1}{2},-\frac{1}{2}}|^2)\notag\\
	&&+4\hat{m_\ell}^2 (|H_{-\frac{1}{2},t}^{\frac{1}{2},\frac{1}{2}}|^2+|H_{\frac{1}{2},t}^{\frac{1}{2},-\frac{1}{2}}|^2)\bigg],\notag\\
	L_2&=&2q^2(1-\hat{m_\ell}^2)\bigg[\big(|H_{\frac{1}{2},1}^{\frac{1}{2},\frac{1}{2}}|^2-|H_{-\frac{1}{2},-1}^{\frac{1}{2},-\frac{1}{2}}|^2\big)+2\hat{m_\ell}^2[\mathcal{R}_e(H_{-\frac{1}{2},t}^{\frac{1}{2},\frac{1}{2}}H_{-\frac{1}{2},0}^{*\frac{1}{2},\frac{1}{2}} )+\mathcal{R}_e(H_{\frac{1}{2},t}^{\frac{1}{2},-\frac{1}{2}}H_{\frac{1}{2},0}^{*\frac{1}{2},-\frac{1}{2}} )]\bigg],\notag\\	
	L_3&=&\frac{q^2}{2}(1-\hat{m_\ell}^2)^2\bigg[(|H_{\frac{1}{2},1}^{\frac{1}{2},\frac{1}{2}}|^2+|H_{-\frac{1}{2},-1}^{\frac{1}{2},-\frac{1}{2}}|^2)-2(|H_{-\frac{1}{2},0}^{\frac{1}{2},\frac{1}{2}}|^2+|H_{\frac{1}{2},0}^{\frac{1}{2},-\frac{1}{2}}|^2)\bigg].\label{22L}
\end{eqnarray}

For the spin-3/2 to spin-1/2 processes, the coefficients can be written as

\begin{eqnarray}
	L_1&=&\frac{q^2}{4}(1-\hat{m_\ell}^2) \bigg[(3+\hat{m_\ell}^2)(|H_{-\frac{1}{2},1}^{\frac{3}{2},\frac{3}{2}}|^2+|H_{\frac{1}{2},-1}^{\frac{3}{2},-\frac{3}{2}}|^2+|H_{\frac{1}{2},1}^{\frac{3}{2},\frac{1}{2}}|^2+|H_{-\frac{1}{2},-1}^{\frac{3}{2},-\frac{1}{2}}|^2)+2(1+\hat{m_\ell}^2)(|H_{-\frac{1}{2},0}^{\frac{3}{2},\frac{1}{2}}|^2+|H_{\frac{1}{2},0}^{\frac{3}{2},-\frac{1}{2}}|^2)\notag\\
	&&+4\hat{m_\ell}^2 (|H_{-\frac{1}{2},t}^{\frac{3}{2},\frac{1}{2}}|^2+|H_{\frac{1}{2},t}^{\frac{3}{2},-\frac{1}{2}}|^2)\bigg],\notag\\
	L_2&=&q^2(1-\hat{m_\ell}^2)\bigg[\big(|H_{-\frac{1}{2},1}^{\frac{3}{2},\frac{3}{2}}|^2-|H_{\frac{1}{2},-1}^{\frac{3}{2},-\frac{3}{2}}|^2+|H_{\frac{1}{2},1}^{\frac{3}{2},\frac{1}{2}}|^2-|H_{-\frac{1}{2},-1}^{\frac{3}{2},-\frac{1}{2}}|^2\big)+2\hat{m_\ell}^2[\mathcal{R}_e(H_{-\frac{1}{2},t}^{\frac{3}{2},\frac{1}{2}}H_{-\frac{1}{2},0}^{*\frac{3}{2},\frac{1}{2}} )+\mathcal{R}_e(H_{\frac{1}{2},t}^{\frac{3}{2},-\frac{1}{2}}H_{\frac{1}{2},0}^{*\frac{3}{2},-\frac{1}{2}} )]\bigg],\notag\\
	L_3&=&\frac{q^2}{4}(1-\hat{m_\ell}^2)^2\bigg[(|H_{-\frac{1}{2},1}^{\frac{3}{2},\frac{3}{2}}|^2+|H_{\frac{1}{2},-1}^{\frac{3}{2},-\frac{3}{2}}|^2+|H_{\frac{1}{2},1}^{\frac{3}{2},\frac{1}{2}}|^2+|H_{-\frac{1}{2},-1}^{\frac{3}{2},-\frac{1}{2}}|^2)-2(|H_{-\frac{1}{2},0}^{\frac{3}{2},\frac{1}{2}}|^2+|H_{\frac{1}{2},0}^{\frac{3}{2},-\frac{1}{2}}|^2)\bigg],\label{32L}
\end{eqnarray}
\end{widetext}
where $\hat{m_\ell}=m_\ell/\sqrt{q^2}$.   The detailed  expressions of helicity amplitudes can be found  in Appendix.~\ref{A}.

The $q^2$ distribution differential decay widths can be deduced by integrating out the angle $\theta$. For the spin-1/2 to spin-1/2 processes, the differential decay widths are
\begin{eqnarray}
	\frac{d\Gamma_L}{dq^2}&=&\frac{G_F^2|V_{CKM}^2|}{2 }\frac{\sqrt{s_-s_+}(1-\hat{m_\ell}^2)}{384\pi^3M^3}q^2(1-\hat{m_\ell}^2)\notag\\
	&&\bigg[(\hat{m_{\ell}}^2+2)(|H_{-\frac{1}{2},0}^{\frac{1}{2},\frac{1}{2}}|^2+|H_{\frac{1}{2},0}^{\frac{1}{2},-\frac{1}{2}}|^2)\notag\\
	&&+3\hat{m_\ell}^2(|H_{-\frac{1}{2},t}^{\frac{1}{2},\frac{1}{2}}|^2+|H_{\frac{1}{2},t}^{\frac{1}{2},-\frac{1}{2}}|^2)\bigg],\notag\\
	\frac{d\Gamma_T}{dq^2}&=&\frac{G_F^2|V_{CKM}^2|}{2 }\frac{\sqrt{s_-s_+}(1-\hat{m_\ell}^2)}{384\pi^3M^3}q^2(1-\hat{m_\ell}^2)\notag\\
	&&(\hat{m_\ell}^2+2)\bigg(|H_{\frac{1}{2},1}^{\frac{1}{2},\frac{1}{2}}|^2+|H_{-\frac{1}{2},-1}^{\frac{1}{2},-\frac{1}{2}}|^2\bigg),\notag\\
	\frac{d\Gamma}{dq^2}&=&\frac{G_F^2|V_{CKM}^2|}{2 }\frac{\sqrt{s_-s_+}(1-\hat{m_\ell}^2)}{512\pi^3M^3}(2L_1-\frac{2}{3}L_3). \nonumber\\
	\label{gamma}
\end{eqnarray}
For the spin-3/2 to spin-1/2 processes, the differential decay widths can be obtained as
\begin{eqnarray}
	\frac{d\Gamma_L}{dq^2}&=&\frac{G_F^2|V_{CKM}^2|}{2 }\frac{\sqrt{s_-s_+}(1-\hat{m_\ell}^2)}{768\pi^3M^3}q^2(1-\hat{m_\ell}^2)\notag\\
	&&\bigg[(\hat{m_{\ell}}^2+2)(|H_{-\frac{1}{2},0}^{\frac{3}{2},\frac{1}{2}}|^2+|H_{\frac{1}{2},0}^{\frac{3}{2},-\frac{1}{2}}|^2)\notag\\
	&&+3\hat{m_\ell}^2(|H_{-\frac{1}{2},t}^{\frac{3}{2},\frac{1}{2}}|^2+|H_{\frac{1}{2},t}^{\frac{3}{2},-\frac{1}{2}}|^2)\bigg],\notag\\
	\frac{d\Gamma_T}{dq^2}&=&\frac{G_F^2|V_{CKM}^2|}{2 }\frac{\sqrt{s_-s_+}(1-\hat{m_\ell}^2)^2}{768\pi^3M^3}q^2(\hat{m_\ell}^2+2)\notag\\
	&&\bigg(|H_{-\frac{1}{2},1}^{\frac{3}{2},\frac{3}{2}}|^2+|H_{\frac{1}{2},-1}^{\frac{3}{2},-\frac{3}{2}}|^2+|H_{\frac{1}{2},1}^{\frac{3}{2},\frac{1}{2}}|^2+|H_{-\frac{1}{2},-1}^{\frac{3}{2},-\frac{1}{2}}|^2\bigg),\notag\\
	\frac{d\Gamma}{dq^2}&=&\frac{G_F^2|V_{CKM}^2|}{2 }\frac{\sqrt{s_-s_+}(1-\hat{m_\ell}^2)}{512\pi^3M^3}(2L_1-\frac{2}{3}L_3).\label{gamma3/2}
\end{eqnarray}

\begin{table*}[!htb]
\caption{Numerical results of decay widths and branching fractions in doubly heavy baryon semileptonic decays using the form factor F($q^2$) in Eq.~\eqref{fit}.}
\label{semileptonq2} %
\begin{tabular}{|c|c|c|c|c|c|c|c|c|}
\hline \hline
channel & $\Gamma(\times10^{-19}\rm{GeV})$   & $\Gamma_L/\Gamma_T$& ${\cal B}$($\%$)&channel & $\Gamma(\times10^{-19}\rm{GeV})$   & $\Gamma_L/\Gamma_T$&${\cal B}$($\%$)& ${\mathcal R}_{\Xi_{QQ}}$    \tabularnewline
\hline
$\Omega_{cc}^+\to \Xi_{cc}^{++}e^-\bar \nu_e$&4.62&0.56&$1.26\times10^{-5}$&$\Omega_{cc}^+\to \Xi_{cc}^{++}\mu^-\bar\nu_\mu$&1.45&2.41&$3.95\times10^{-6}$&0.31\tabularnewline
\hline
$\Omega_{bb}^-\to\Xi_{bb}^0e^-\bar \nu_e$&1.72&0.70&$2.09\times10^{-5}$&$\Omega_{bb}^-\to\Xi_{bb}^0\mu^-\bar \nu_\mu$&1.13&4.69&$1.36\times10^{-5}$&0.65\tabularnewline
\hline
$\Omega_{bc}^0\to \Xi_{bc}^+e^-\bar \nu_e$&3.20&2.64&$1.07\times10^{-5}$&$\Omega_{bc}^0\to \Xi_{bc}^+\mu^-\bar \nu_\mu$&1.82&8.87&$6.08\times10^{-6}$&0.57\tabularnewline
\hline
$\Omega_{bc}^{\prime0}\to \Xi_{bc}^{\prime+}e^-\bar \nu_e$&11.93&0.79&$3.98\times10^{-5}$&$\Omega_{bc}^{\prime0}\to \Xi_{bc}^{\prime+}\mu^-\bar \nu_\mu$&3.84&1.66&$1.28\times10^{-5}$&0.32\tabularnewline
\hline
$\Omega_{cc}^{*+}\to \Xi_{cc}^{++}e^-\bar \nu_e$&$12.93$&0.36&-&$\Omega_{cc}^{*+}\to \Xi_{cc}^{++}\mu^-\bar \nu_\mu$&$4.96$&0.41&-&0.38\tabularnewline
\hline
$\Omega_{bb}^{*-}\to\Xi_{bb}^0e^-\bar \nu_e$&$23.26$&0.41&-&$\Omega_{bb}^{*-}\to\Xi_{bb}^0\mu^-\bar \nu_\mu$&$9.12$&0.45&-&0.39\tabularnewline
\hline
$\Omega_{bc}^{*0}\to \Xi_{bc}^+e^-\bar \nu_e$&21.68&0.35&-&$\Omega_{bc}^{*0}\to \Xi_{bc}^+\mu^-\bar \nu_\mu$&8.70&0.41&-&0.40\tabularnewline
\hline
\end{tabular}
\end{table*}

\begin{table*}[!htb]
\caption{Numerical results of decay widths and branching fractions in doubly heavy baryon semileptonic decays using  constant form factors $F(q^2)=F(0)$.}
\label{semilepton0} %
\begin{tabular}{|c|c|c|c|c|c|c|c|c|}
\hline \hline
channel & $\Gamma(\times10^{-19}\rm{GeV})$   & $\Gamma_L/\Gamma_T$& ${\cal B}$($\%$)&channel & $\Gamma(\times10^{-19}\rm{GeV})$   & $\Gamma_L/\Gamma_T$&${\cal B}$($\%$)& ${\mathcal R}_{\Xi_{QQ}}$    \tabularnewline
\hline
$\Omega_{cc}^+\to \Xi_{cc}^{++}e^-\bar \nu_e$&4.35&0.57&$1.19\times10^{-5}$&$\Omega_{cc}^+\to \Xi_{cc}^{++}\mu^-\bar\nu_\mu$&1.33&2.50&$3.62\times10^{-6}$&0.31\tabularnewline
\hline
$\Omega_{bb}^-\to\Xi_{bb}^0e^-\bar \nu_e$&1.54&0.73&$1.86\times10^{-5}$&$\Omega_{bb}^-\to\Xi_{bb}^0\mu^-\bar \nu_\mu$&1.02&5.24&$1.24\times10^{-5}$&0.66\tabularnewline
\hline
$\Omega_{bc}^0\to \Xi_{bc}^+e^-\bar \nu_e$&2.98&2.92&$0.99\times10^{-5}$&$\Omega_{bc}^0\to \Xi_{bc}^+\mu^-\bar \nu_\mu$&1.67&10.07&$5.58\times10^{-6}$&0.56\tabularnewline
\hline
$\Omega_{bc}^{\prime0}\to \Xi_{bc}^{\prime+}e^-\bar \nu_e$&10.50&0.84&$3.50\times10^{-5}$&$\Omega_{bc}^{\prime0}\to \Xi_{bc}^{\prime+}\mu^-\bar \nu_\mu$&3.34&1.84&$1.11\times10^{-5}$&0.32\tabularnewline
\hline
$\Omega_{cc}^{*+}\to \Xi_{cc}^{++}e^-\bar \nu_e$&$10.22$&0.35&-&$\Omega_{cc}^{*+}\to \Xi_{cc}^{++}\mu^-\bar \nu_\mu$&$3.73$&0.40&-&0.37\tabularnewline
\hline
$\Omega_{bb}^{*-}\to\Xi_{bb}^0e^-\bar \nu_e$&$18.31$&0.40&-&$\Omega_{bb}^{*-}\to\Xi_{bb}^0\mu^-\bar \nu_\mu$&$6.85$&0.45&-&0.37\tabularnewline
\hline
$\Omega_{bc}^{*0}\to \Xi_{bc}^+e^-\bar \nu_e$&16.97&0.34&-&$\Omega_{bc}^{*0}\to \Xi_{bc}^+\mu^-\bar \nu_\mu$&6.51&0.40&-&0.38\tabularnewline
\hline
\end{tabular}
\end{table*}

After integrating  out the $q^2$, one can give the partial decay widths and the branching fractions which are presented in Table~\ref{semileptonq2} and in Table~\ref{semilepton0}. Most of the partial decays are at the order ${\cal O}(10^{-19})$ GeV, which corresponds to a small branching fraction.

\begin{figure*}[htbp!]
	\begin{minipage}[t]{0.4\linewidth}
		\centering
		\includegraphics[width=1\columnwidth]{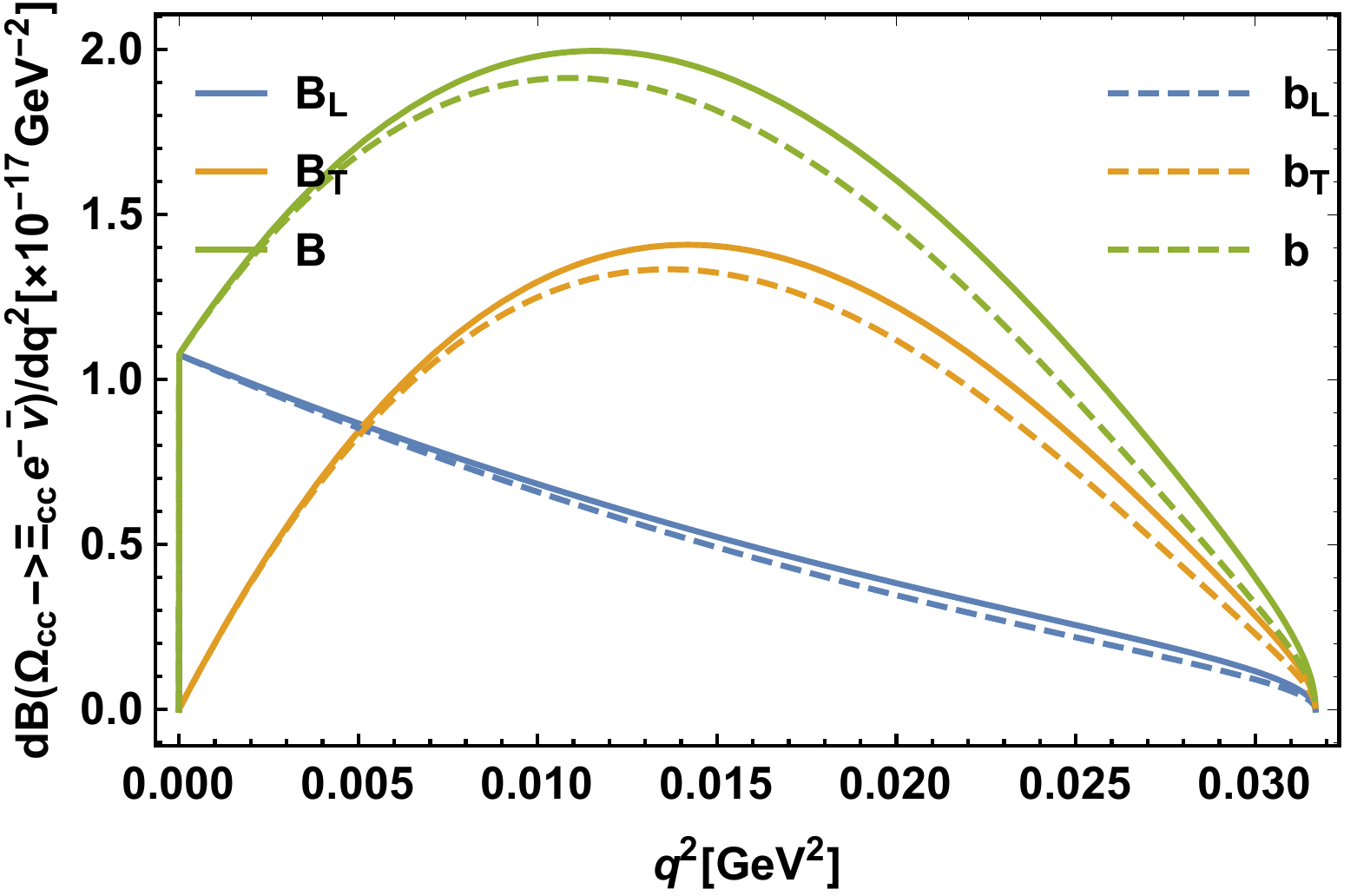}
	\end{minipage}
	\begin{minipage}[t]{0.4\linewidth}
		\centering
		\includegraphics[width=1\columnwidth]{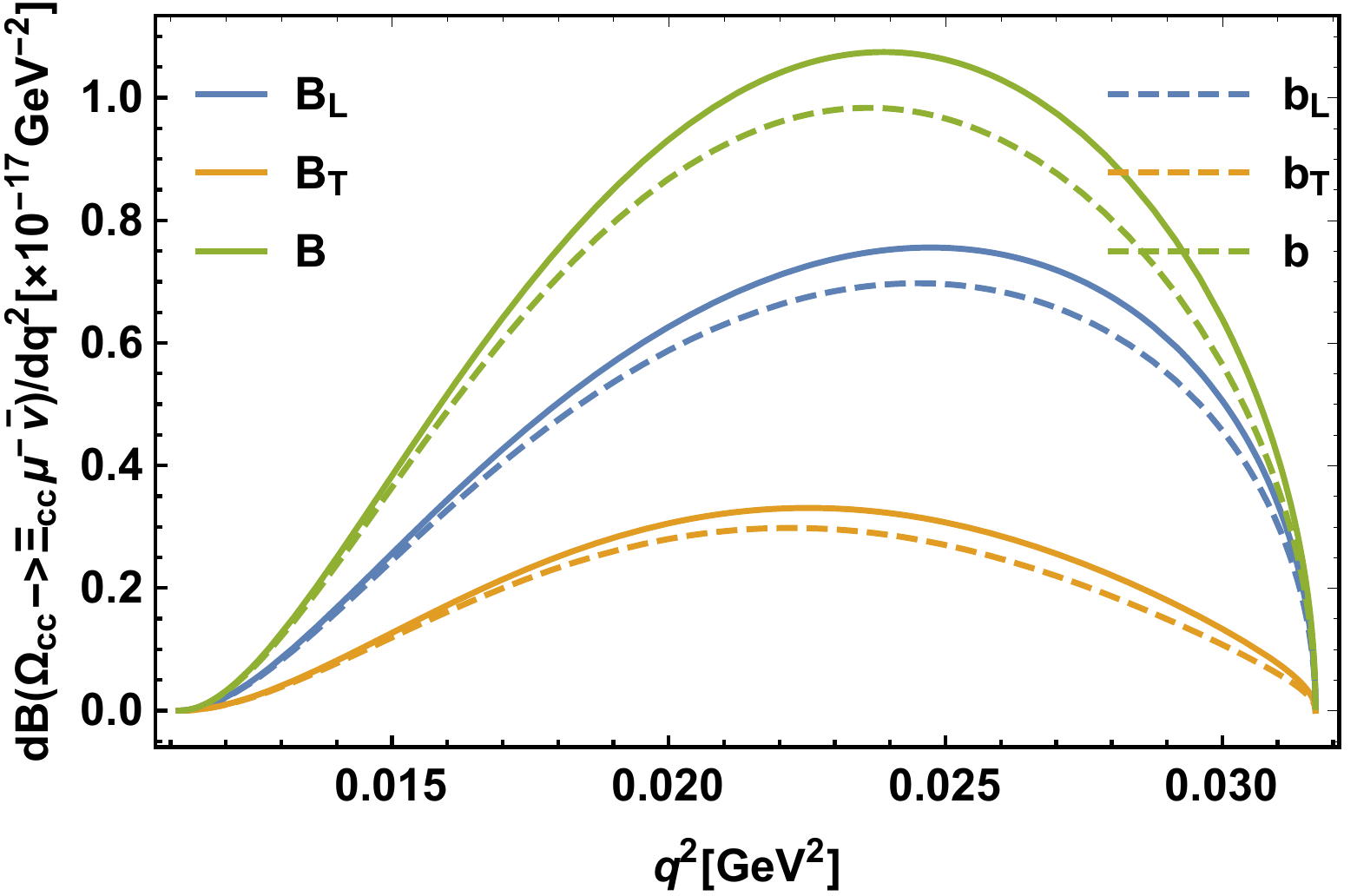}
	\end{minipage}
	\begin{minipage}[t]{0.4\linewidth}
		\centering
		\includegraphics[width=1\columnwidth]{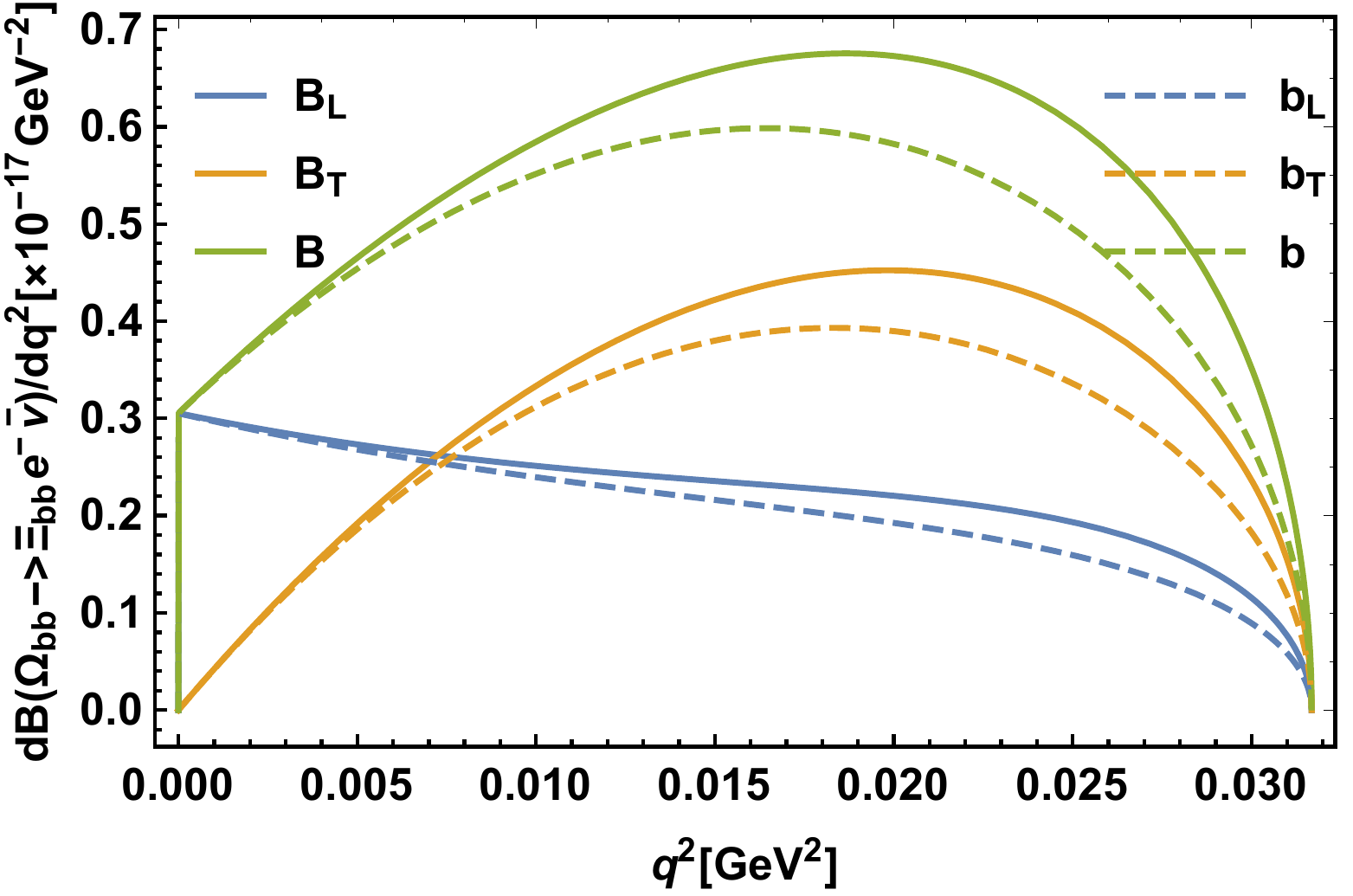}
	\end{minipage}
	\begin{minipage}[t]{0.4\linewidth}
		\centering
		\includegraphics[width=1\columnwidth]{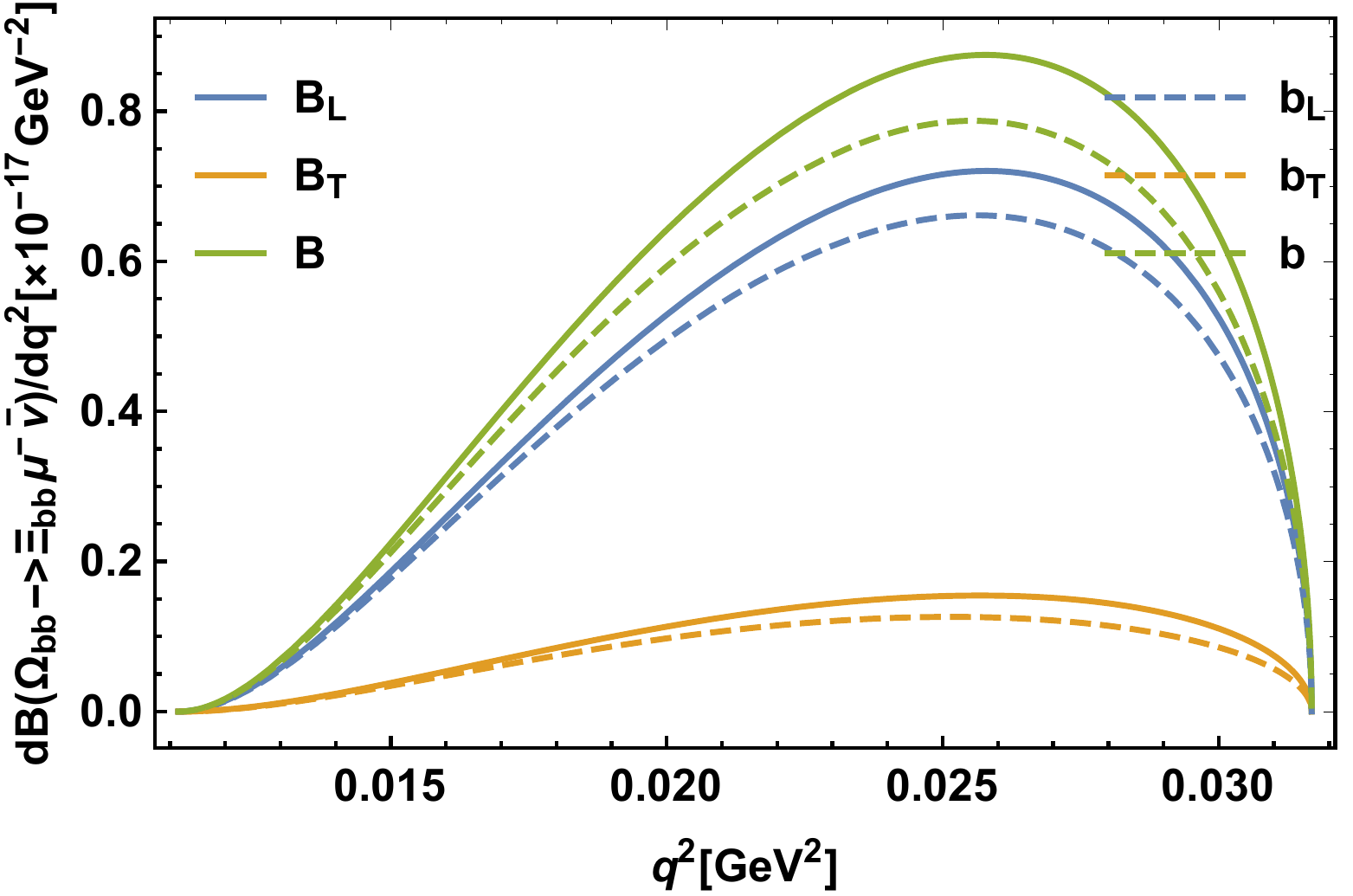}
	\end{minipage}
	\begin{minipage}[t]{0.4\linewidth}
	    \centering
	    \includegraphics[width=1\columnwidth]{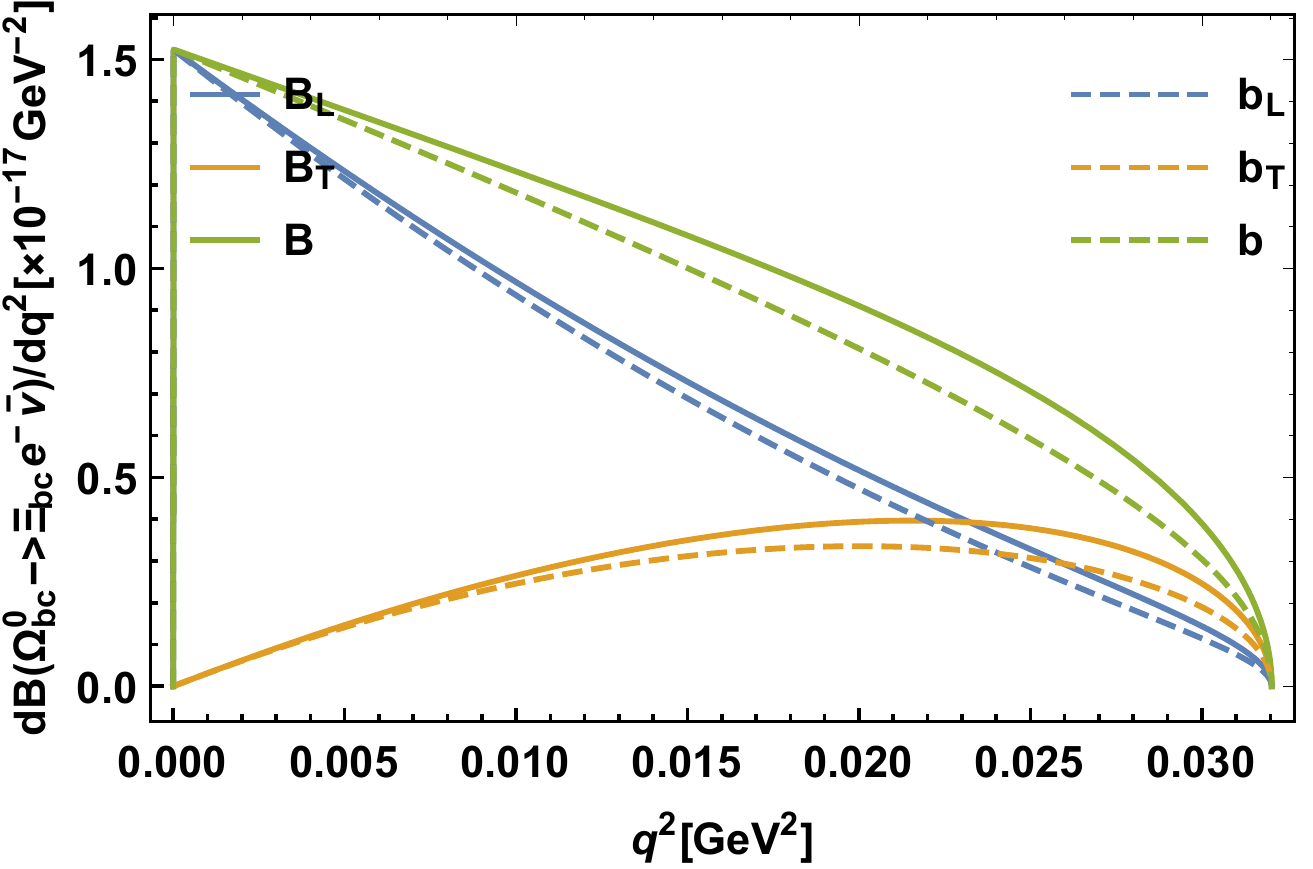}
    \end{minipage}
	\begin{minipage}[t]{0.4\linewidth}
	    \centering
	    \includegraphics[width=1\columnwidth]{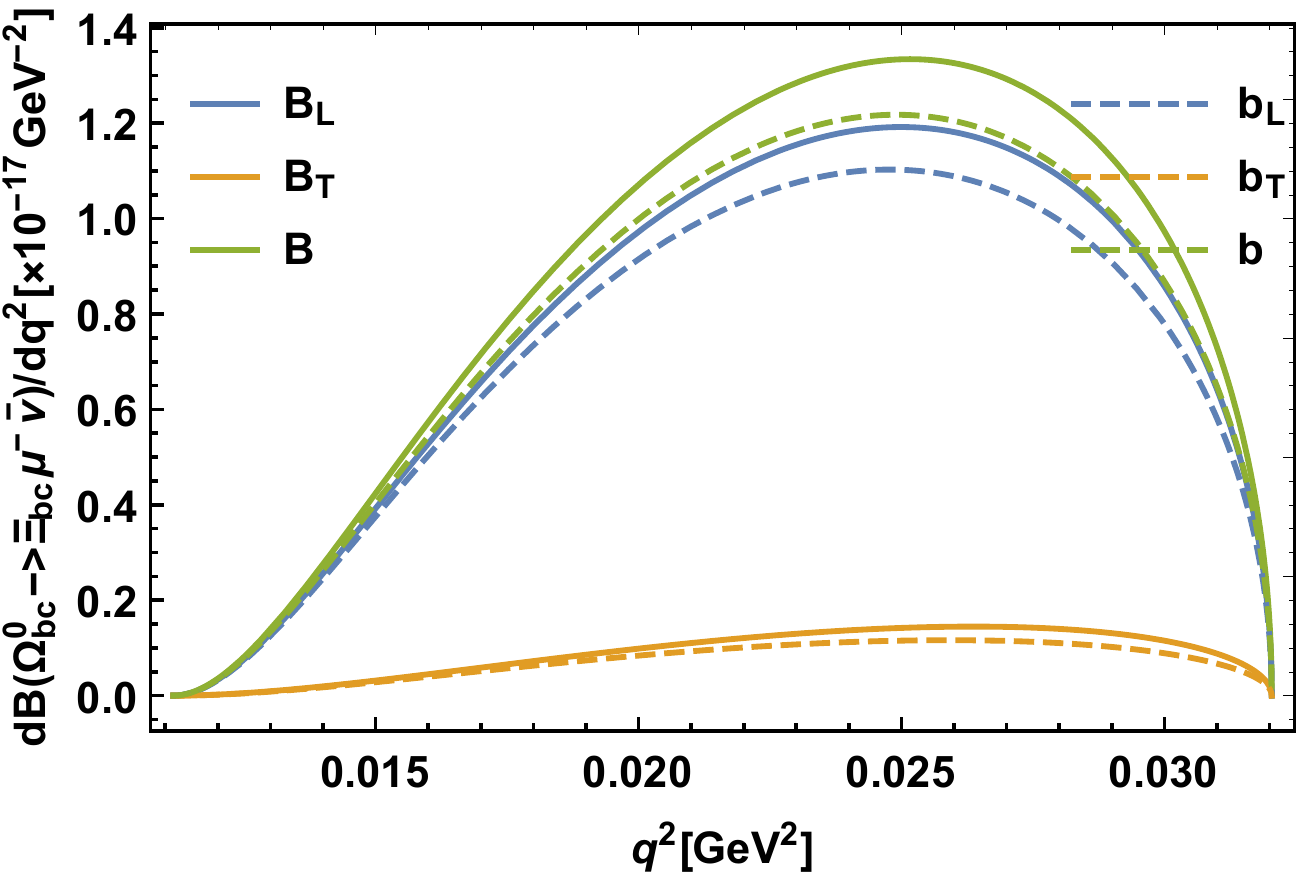}
    \end{minipage}
	\begin{minipage}[t]{0.4\linewidth}
	    \centering
     	\includegraphics[width=1\columnwidth]{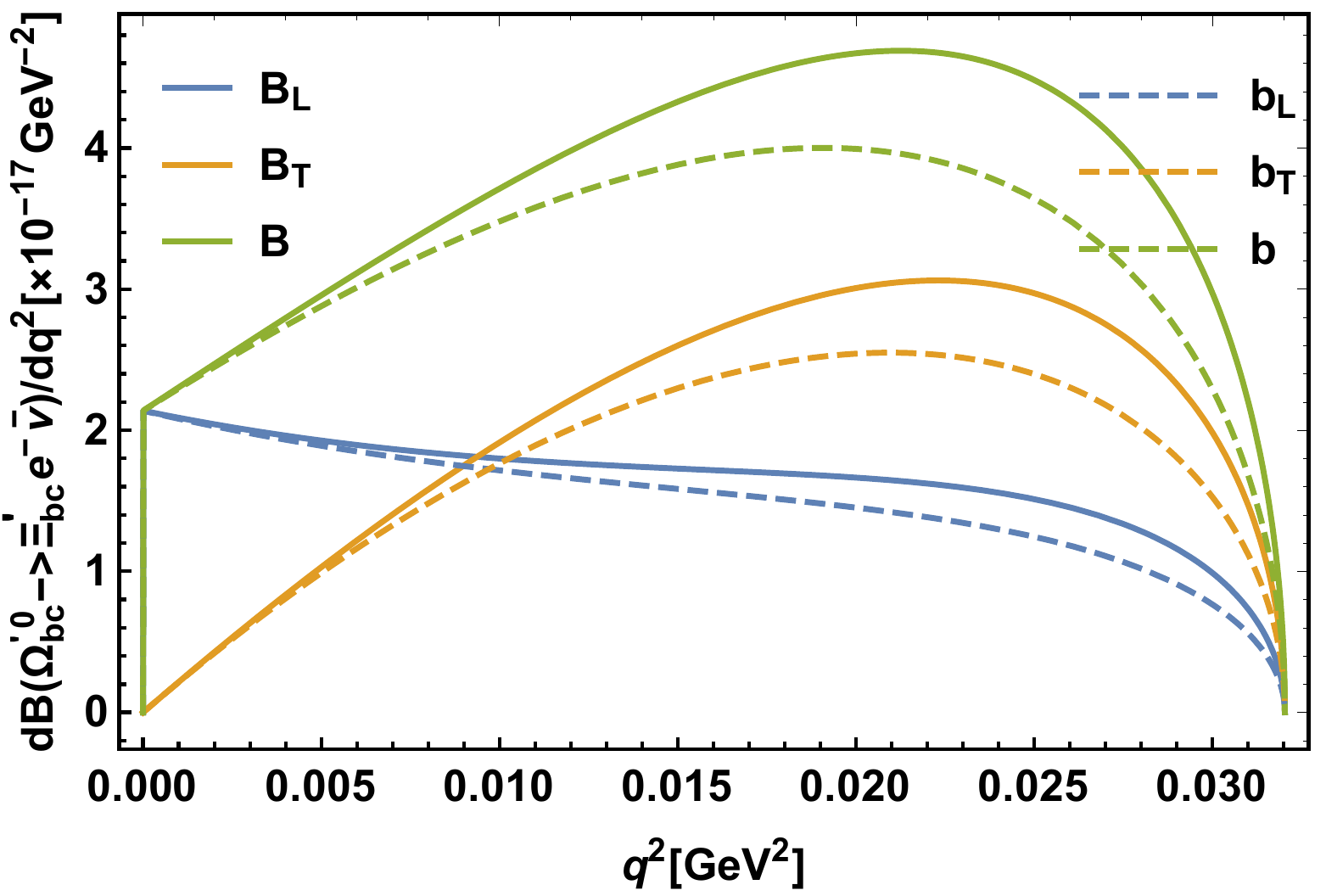}
    \end{minipage}
	\begin{minipage}[t]{0.4\linewidth}
	    \centering
    	\includegraphics[width=1\columnwidth]{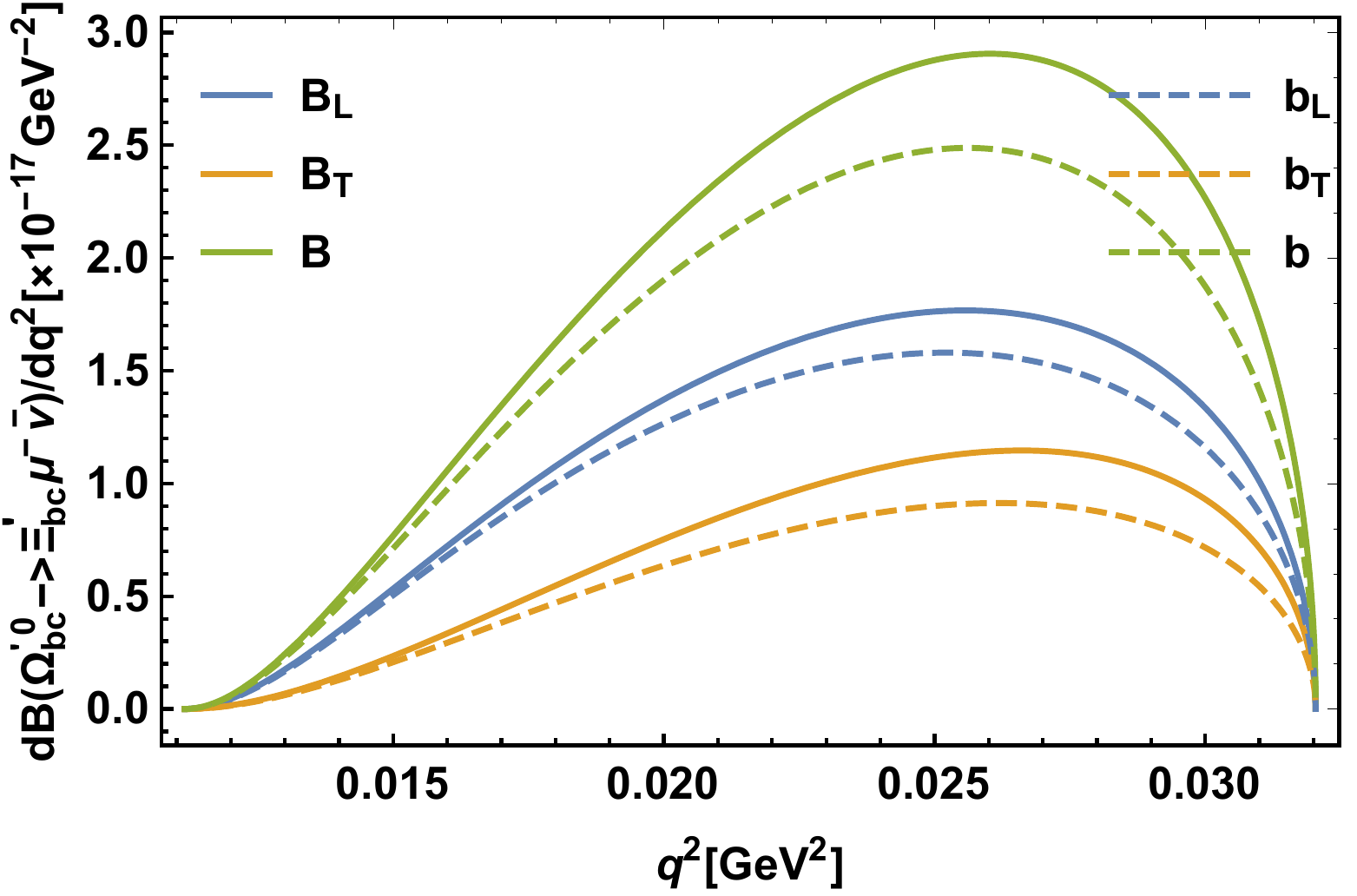}
    \end{minipage}
	\caption{  The differential branching fraction dB/d$q^2$(db/d$q^2$)
		of spin-1/2 to spin-1/2 doubly heavy baryon decay processes. The $B_L$($b_L$) and
		$B_T$($b_T$) represent the contribution of longitudinal polarisation and transverse polarisation in branching fractions in Eq.~\eqref{gamma}.}
\label{fd2}
\end{figure*}

\begin{figure*}[htbp!]
	\begin{minipage}[t]{0.4\linewidth}
	\centering
	\includegraphics[width=1\columnwidth]{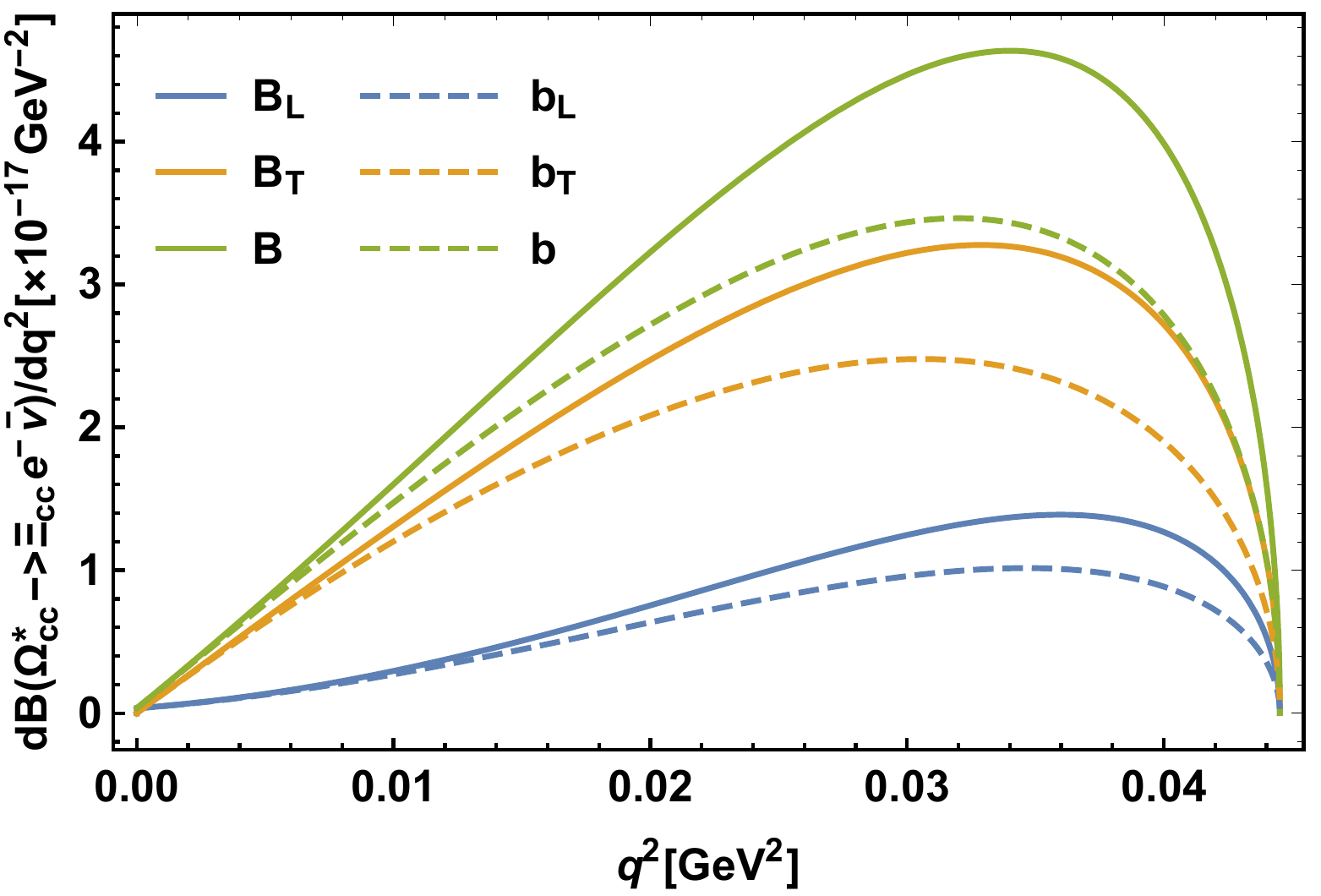}
    \end{minipage}
	\begin{minipage}[t]{0.4\linewidth}
	\centering
	\includegraphics[width=1\columnwidth]{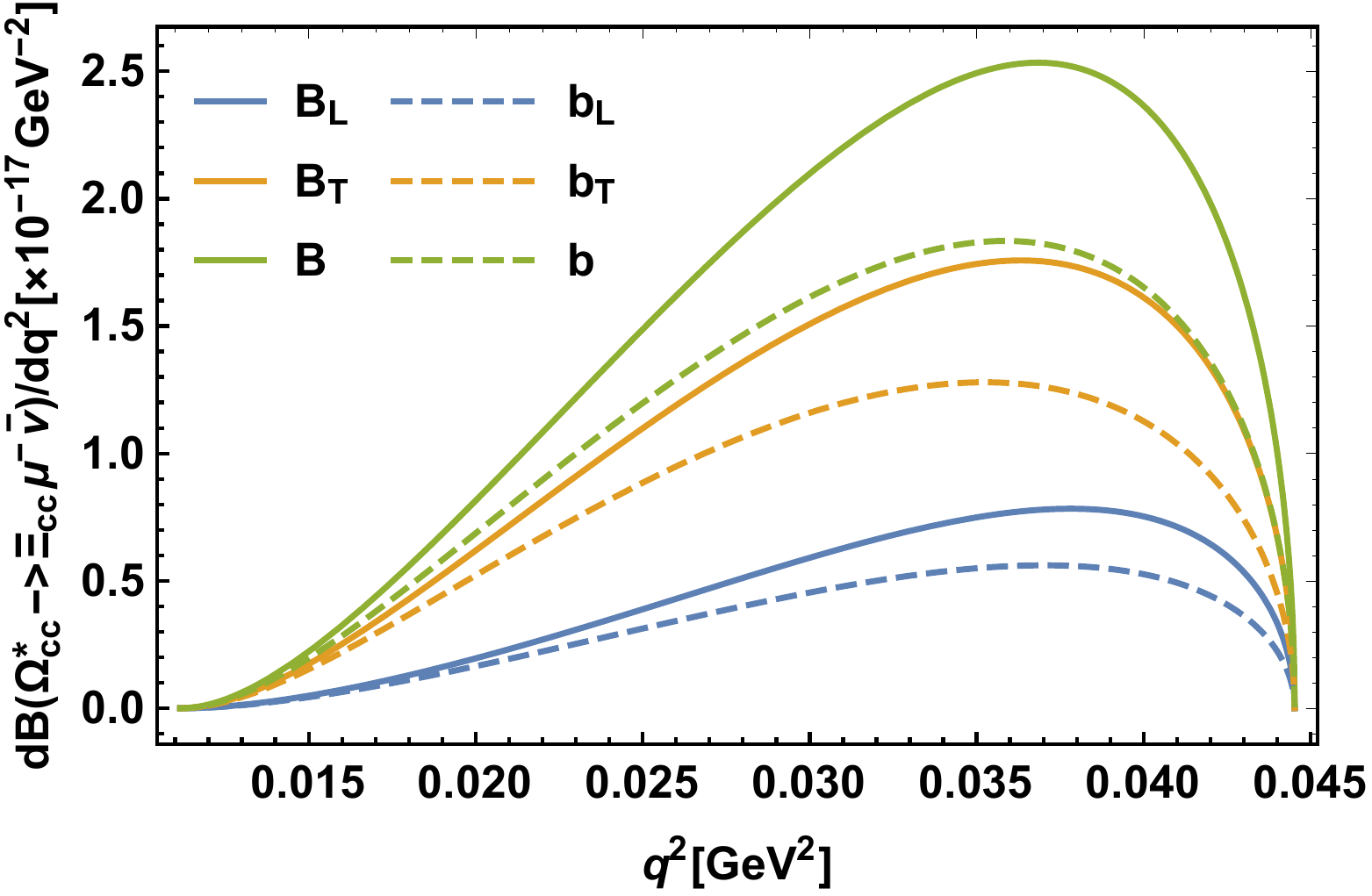}
	\end{minipage}
    \begin{minipage}[t]{0.4\linewidth}
	\centering
	\includegraphics[width=1\columnwidth]{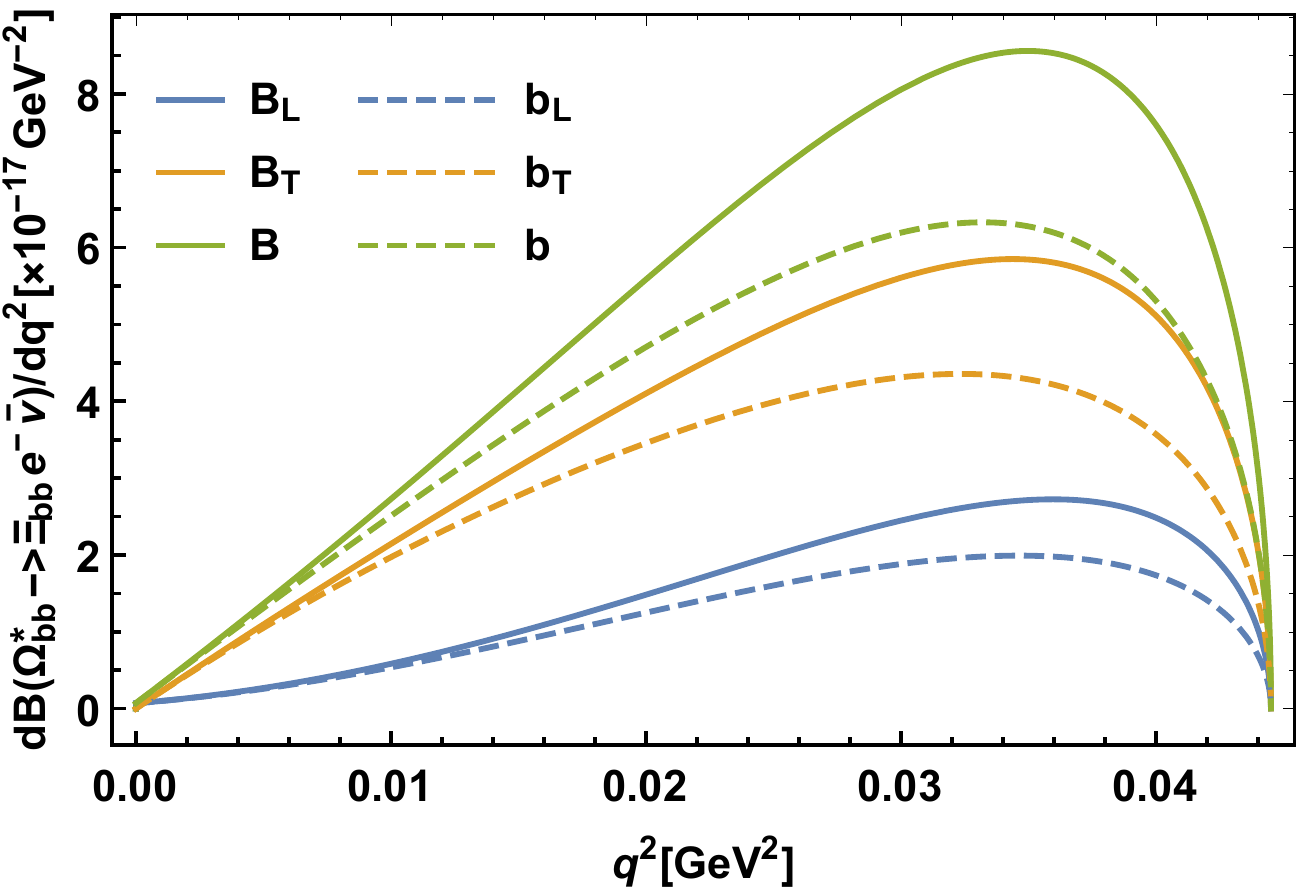}
    \end{minipage}
    \begin{minipage}[t]{0.4\linewidth}
	\centering
	\includegraphics[width=1\columnwidth]{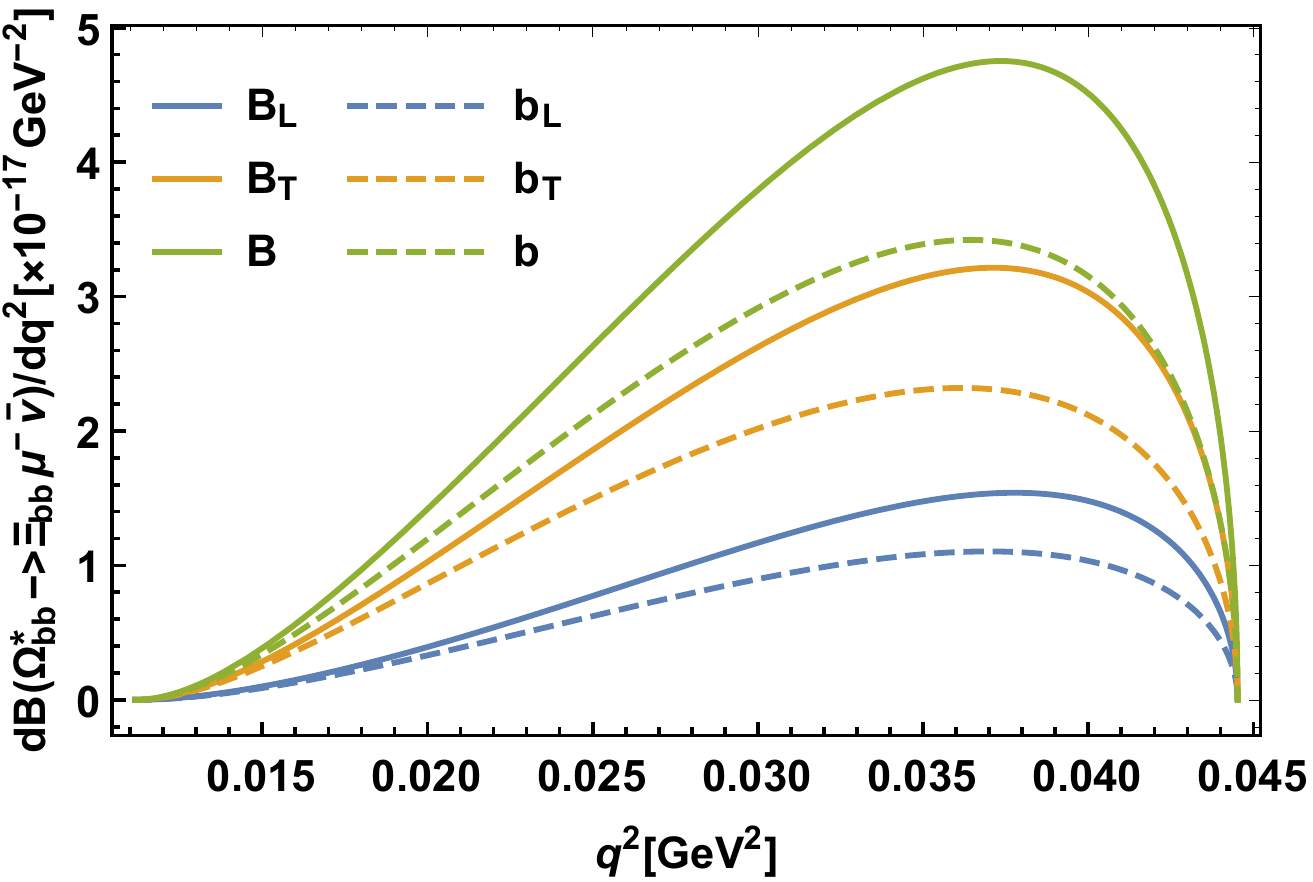}
    \end{minipage}
    \begin{minipage}[t]{0.4\linewidth}
	\centering
	\includegraphics[width=1\columnwidth]{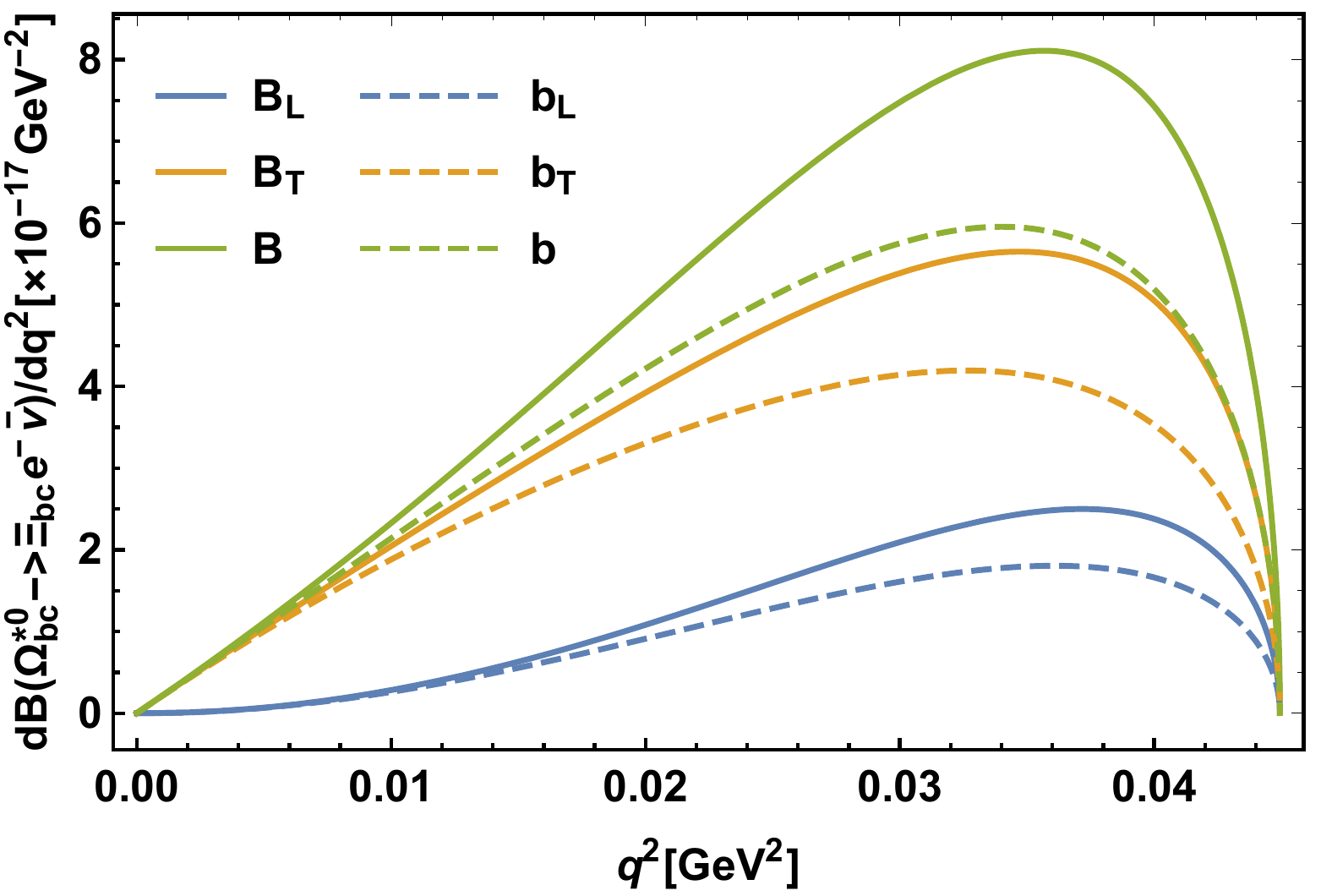}
    \end{minipage}
    \begin{minipage}[t]{0.4\linewidth}
	\centering
	\includegraphics[width=1\columnwidth]{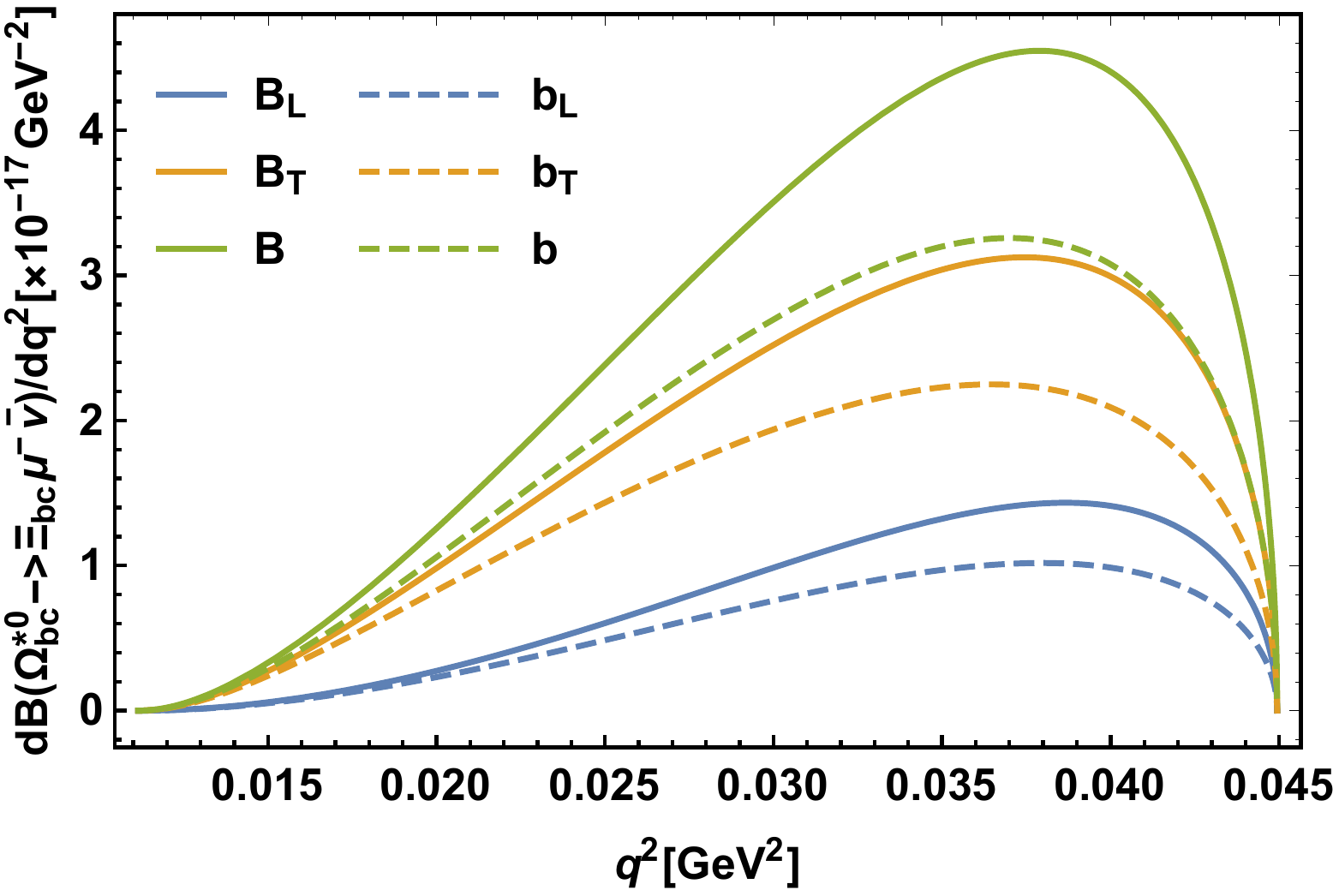}
    \end{minipage}
	\caption{The differential branching fraction dB/d$q^{2}$(db/d$q^{2}$)
		of spin-3/2 to spin-1/2 doubly heavy baryon decay processes. The $B_L$($b_L$) and
		$B_T$($b_T$) represent the contribution of longitudinal polarisation and transverse polarisation in branching fractions in Eq.~\eqref{gamma3/2}.}
	\label{fd4}
\end{figure*}

For these   $s\to u$ decay processes, one can find that the phase space is small. Thus in our analysis, we also give the results for decay widths by setting the form factors $F(q^2)=F(0)$. Using the Eq.~\eqref{gamma} and Eq.~\eqref{gamma3/2}, we study the $q^2$ distributions of $(B,B_L,B_T)$ and $(b,b_L,b_T)$ which   correspond  to the decay widths $(\Gamma,\Gamma_L,\Gamma_T)$. $B$ represents the results from the form factors $F(q^2)$ in Eq.~\eqref{fit} and $b$ is obtained by $F(q^2)=F(0)$. The distributions are shown in Fig.\ref{fd2} and Fig.\ref{fd4}.

One can see that the $q^2$ distributions of decay widths are much different for  different lepton types $(\ell=\mu,e)$. For the $\ell=e$ case,  branching ratios $B_L(b_L)$ are obviously enhanced in the low-$q^2$ region. As shown in Eq.~\eqref{A1} and Eq.~\eqref{A2}, because the helicity amplitudes $H^{S\lambda}_{\lambda^\prime0}$ and $H^{S \lambda }_{\lambda^\prime,t}$ contain the $1/\sqrt{q^2}$ term, the branching ratio $B_L(b_L)$ which includes the $H_0$ and $H_t$ is enhanced in the $q^2\to m_\ell^2$ region. However, this effect does not exist in the $\ell=\mu$ case. The difference comes from the phase space integral which includes $(1-{\hat{m_\ell}}^2)$ term. This term will suppress the differential decay width when $q^2\to m_\ell^2$.  For the $\ell=\mu$, the phase space factor has a more severe impact  since the mass of $\mu$ is much larger than $m_e$. Especially in these $s\to u$ semileptonic decays the phase space is very small and $q^2$ is comparable to the $m_\mu$ for these processes.

\begin{figure*}[htbp!]
	\begin{minipage}[t]{0.4\linewidth}
	    \centering
	    \includegraphics[width=1\columnwidth]{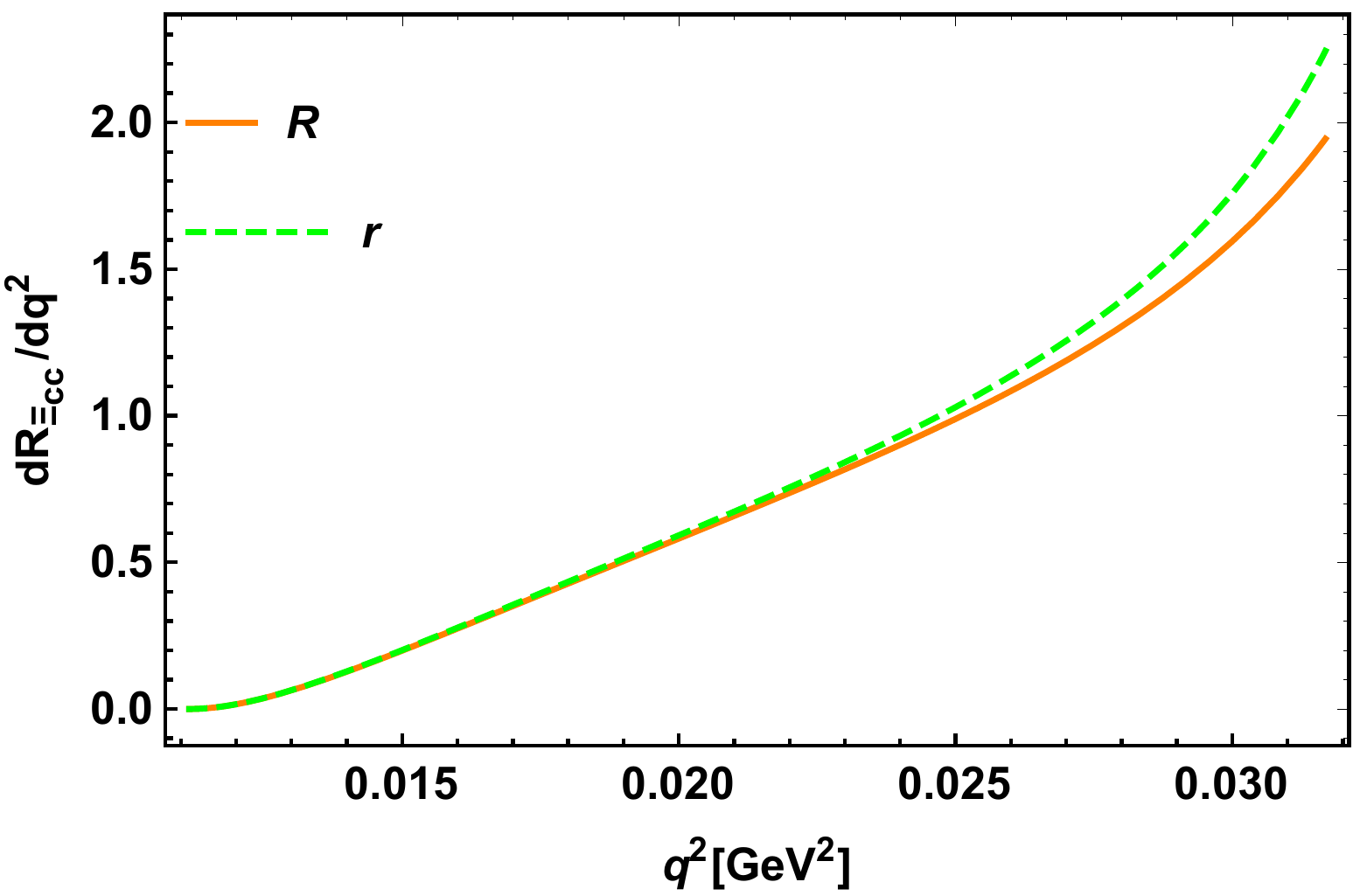}
    \end{minipage}
	\begin{minipage}[t]{0.4\linewidth}
	    \centering
	    \includegraphics[width=1\columnwidth]{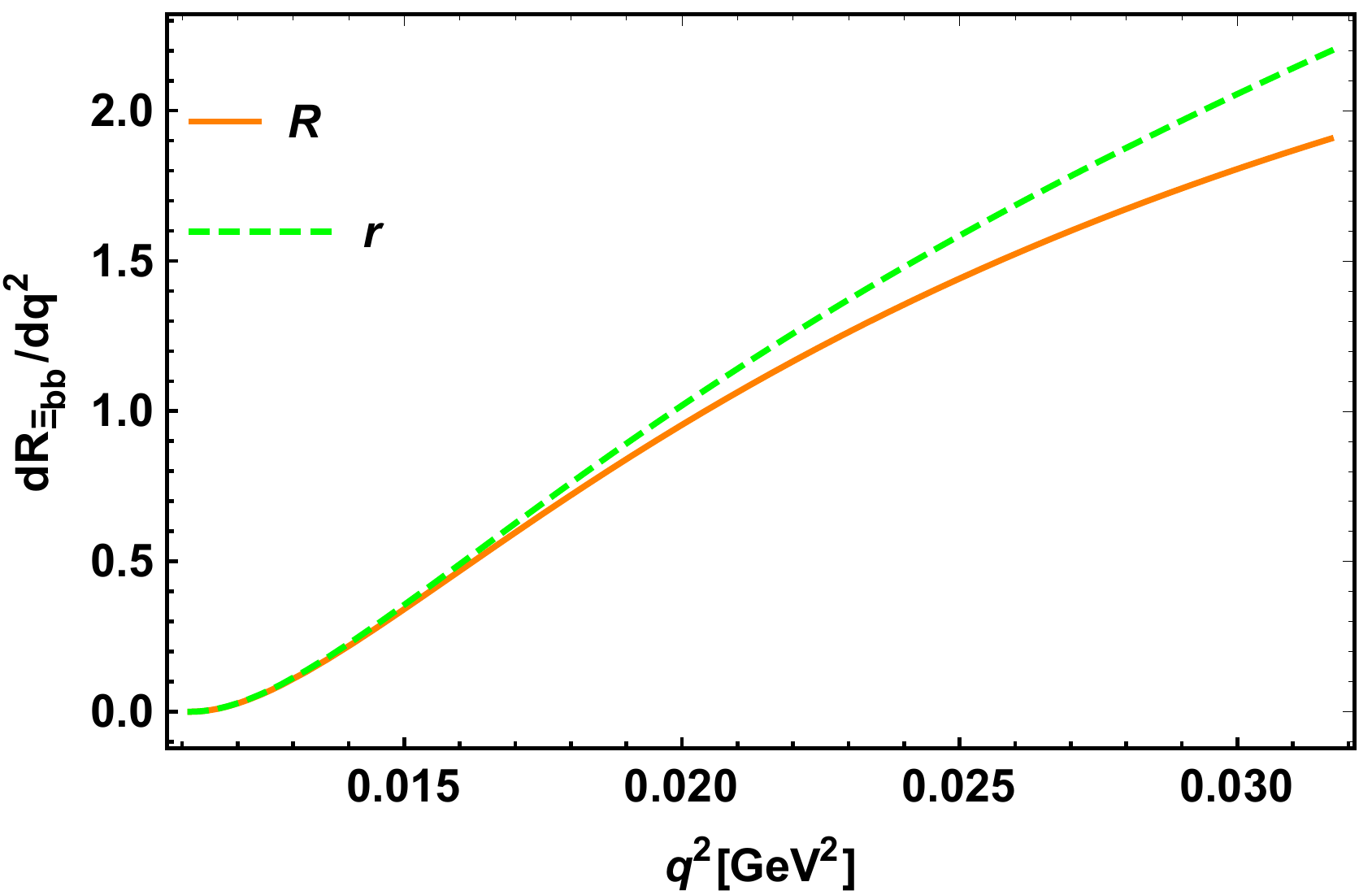}
    \end{minipage}
	\begin{minipage}[t]{0.4\linewidth}
	    \centering
     	\includegraphics[width=1\columnwidth]{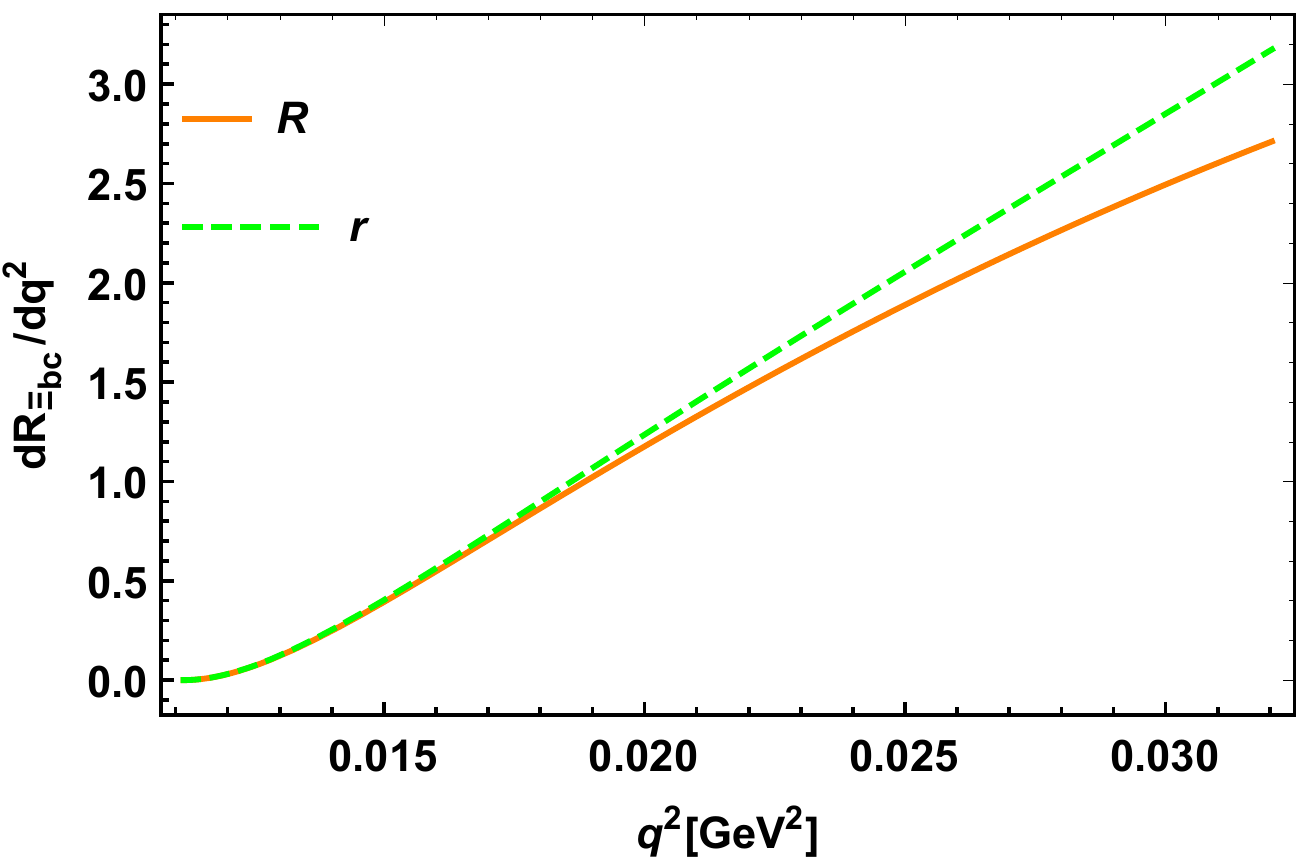}
    \end{minipage}
	\begin{minipage}[t]{0.4\linewidth}
	    \centering
    	\includegraphics[width=1\columnwidth]{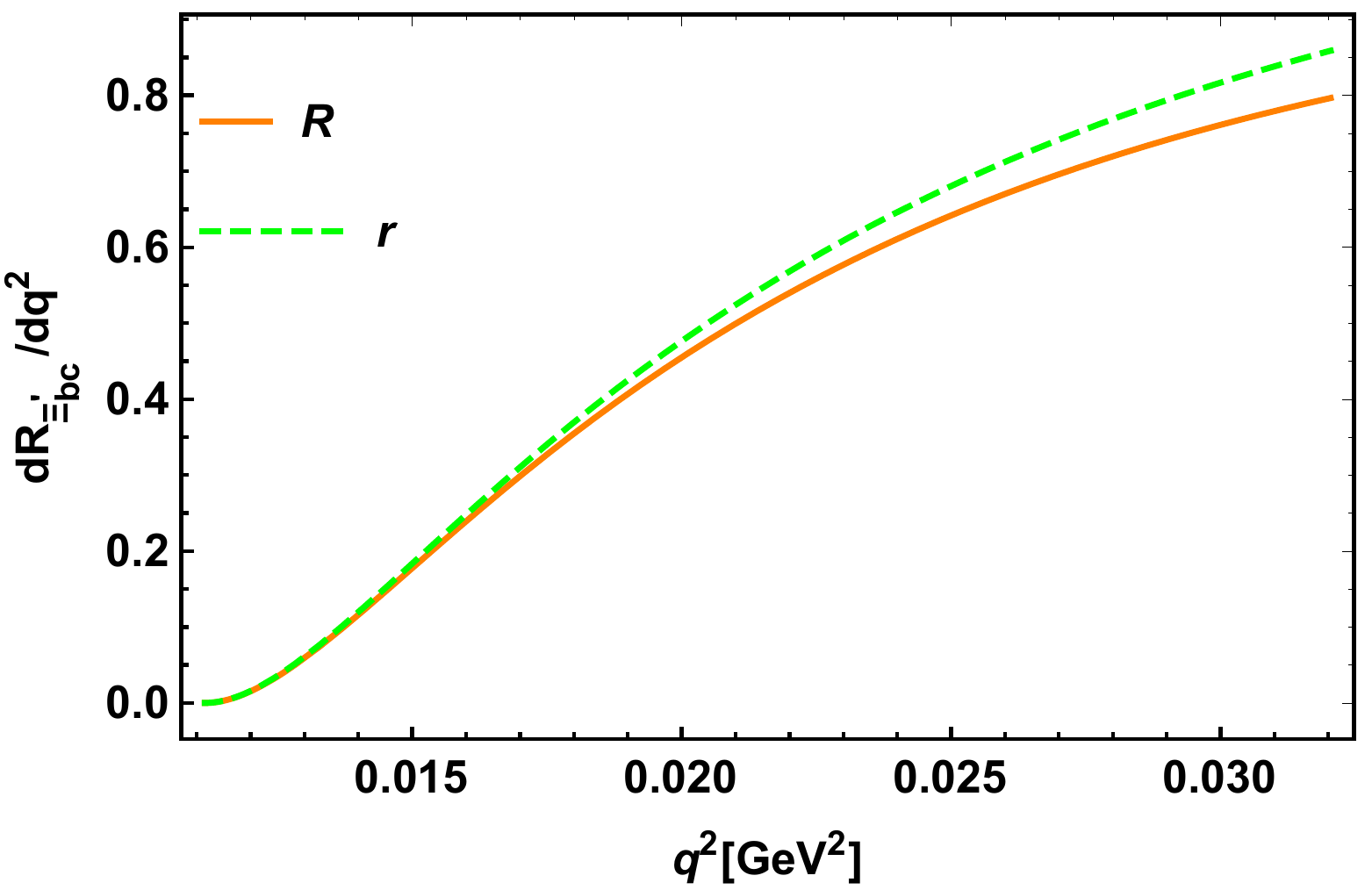}
    \end{minipage}
	\caption{The d$\mathcal R_{\Xi_{Q Q}}/d q^2$ of spin-1/2 to spin-1/2 doubly heavy baryon decay processes. The R and r represent the results using F($q^2$) in Eq.~\eqref{fit} and $F(q^2)=F(0)$ in Eq.~\eqref{47} respectively.}
\label{fd3}
\end{figure*}

\begin{figure*}[htbp!]
	\begin{minipage}[t]{0.4\linewidth}
	\centering
	\includegraphics[width=1\columnwidth]{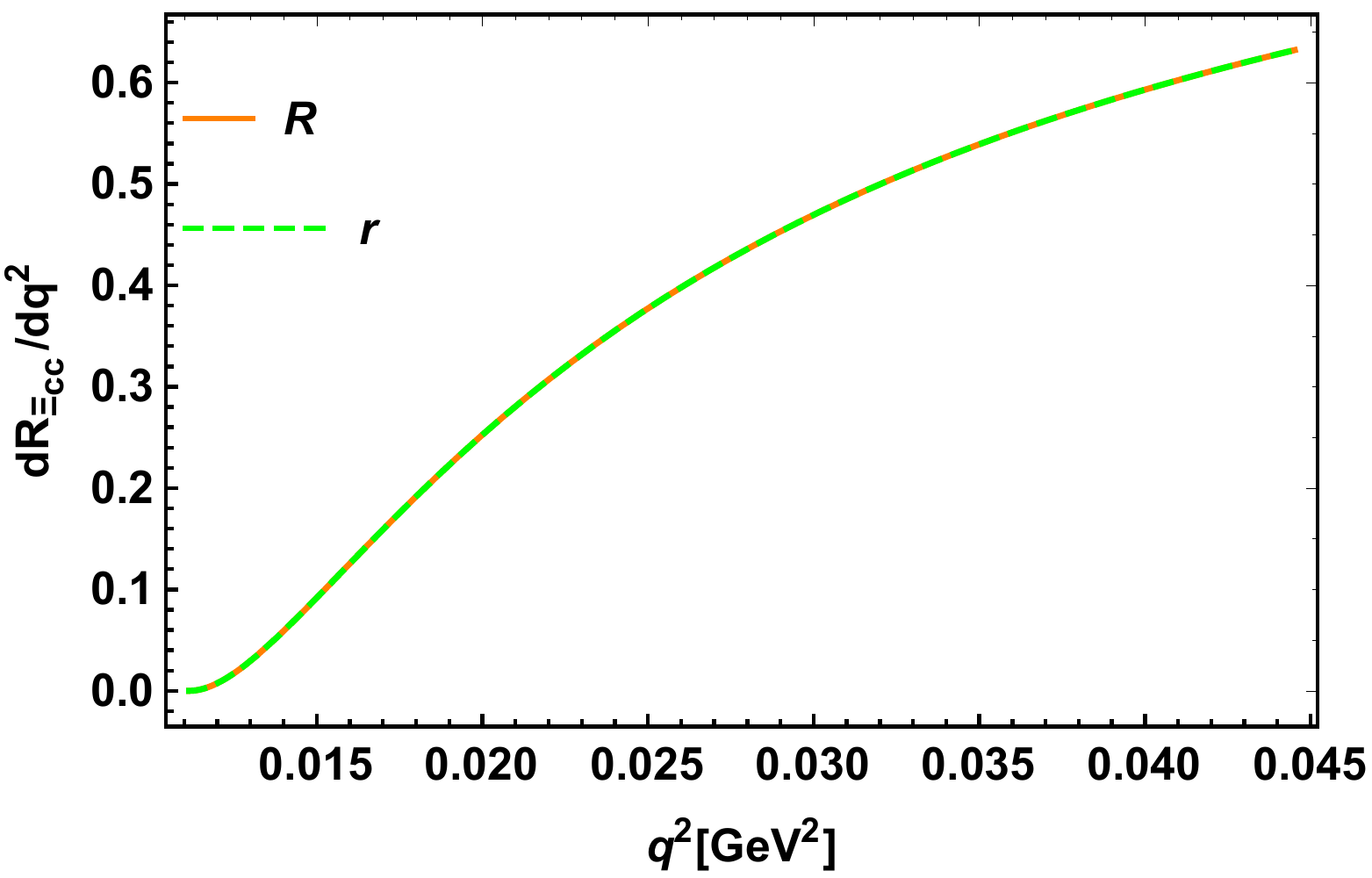}
    \end{minipage}
	\begin{minipage}[t]{0.4\linewidth}
	\centering
	\includegraphics[width=1\columnwidth]{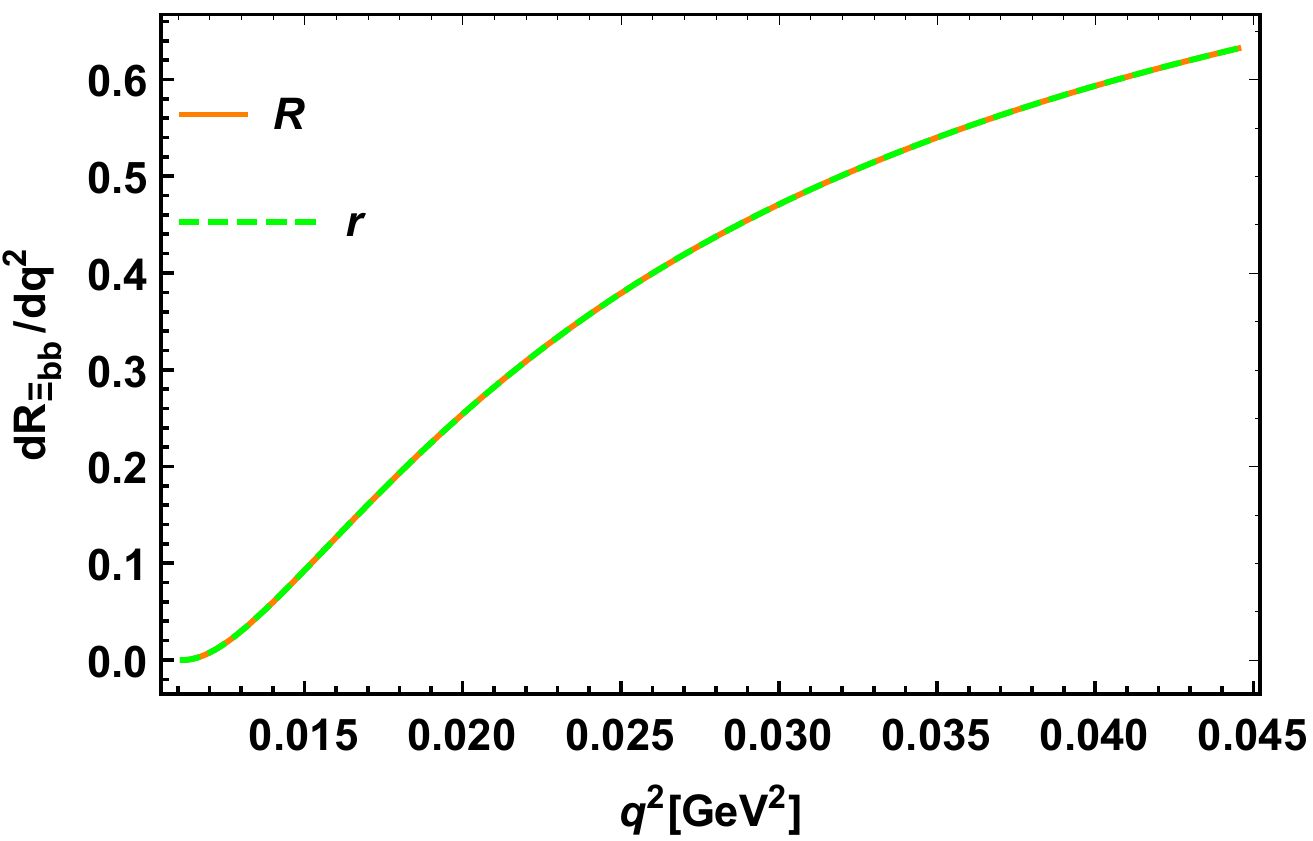}
	\end{minipage}
    \begin{minipage}[t]{0.4\linewidth}
	\centering
	\includegraphics[width=1\columnwidth]{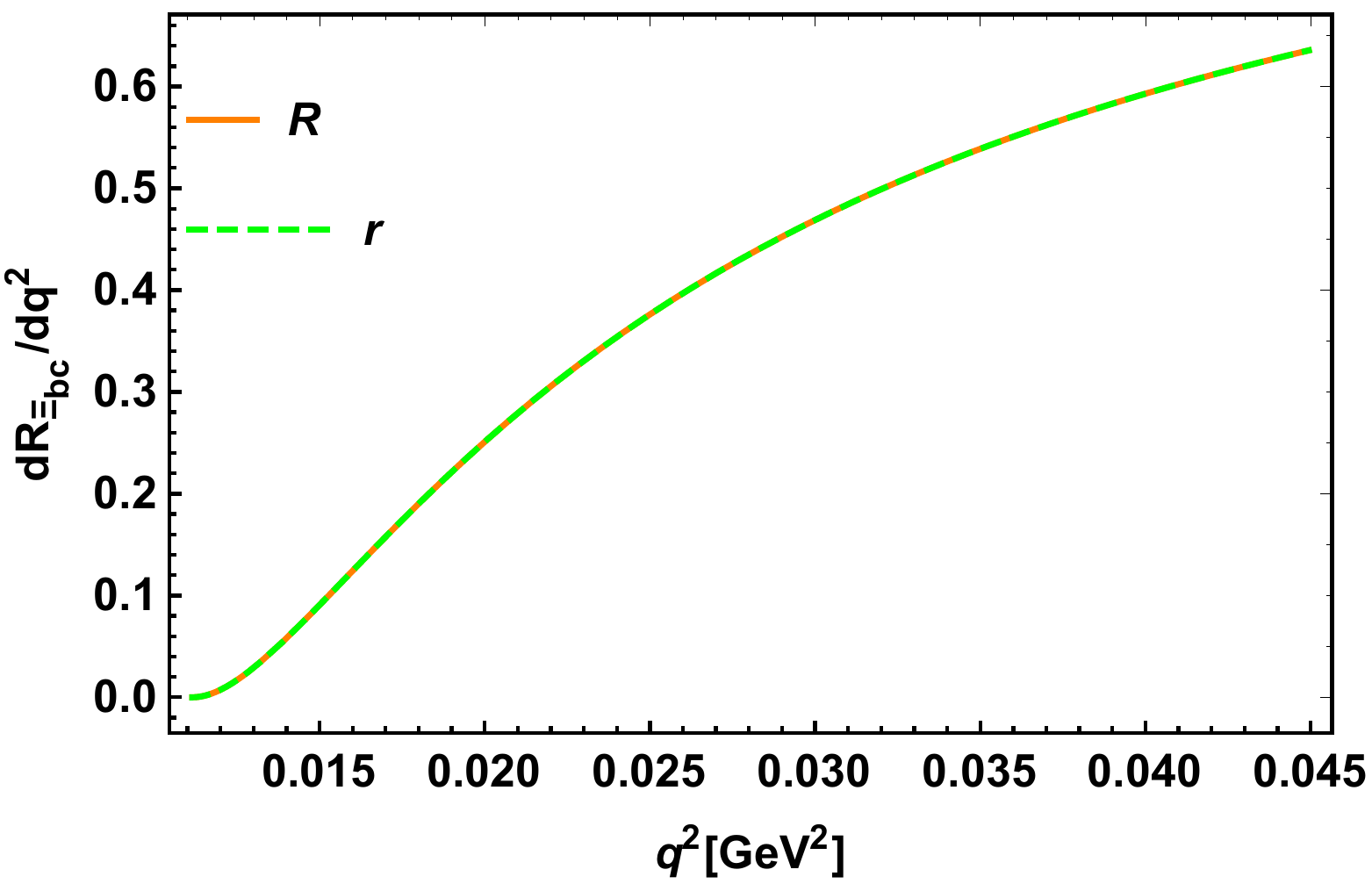}
    \end{minipage}
	\caption{ The d$\mathcal R_{\Xi_{Q Q}}/d q^2$ of spin-3/2 to spin-1/2 doubly heavy baryon decay processes. The R and r represent the results from F($q^2$) in Eq.~\eqref{fit} and constant form factor $F(0)$ respectively.}
	\label{fd5}
\end{figure*}

Differences in decay widths  with two different final lepton types $(\mu,e)$  can be used as a platform to explore the lepton flavor universality. Therefore one can define the value ${\mathcal R}_{\Xi_{QQ}}$ as
\begin{eqnarray}
{\mathcal R}_{\Xi_{QQ}}=\frac{\Gamma(\Omega_{QQ}\to \Xi_{QQ} \mu\bar{\nu_\mu})}{\Gamma(\Omega_{QQ}\to \Xi_{QQ} e\bar{\nu_e}))}.
\end{eqnarray}
The values of these ratios are shown in Table.~\ref{semileptonq2} and Table~\ref{semilepton0}. Since these semileptonic decay processes in our work are sensitive to the lepton mass, the value of the ratio ${\mathcal R}_{\Xi_{QQ}}$ will be far from the 1. Namely, in these processes the muon mass can uplift the helicity suppression of certain semi-leptonic decay amplitudes which are unobservable in decays with electron in the final state.   In the presence of new physics that is sensitive to the lepton flavor, these  ratios can serve as  a   platform for searching new physics effects in future.  
For more information on the lepton universality, one can also define the differential ratio $\frac{d{\mathcal R}_{\Xi_{QQ}}}{dq^2}$ as
\begin{eqnarray}
\frac{d{\mathcal R}_{\Xi_{QQ}}}{dq^2}=\frac{d\Gamma(\Omega_{QQ}\to \Xi_{QQ} \mu\bar{\nu_\mu})/dq^2}{d\Gamma(\Omega_{QQ}\to \Xi_{QQ} e\bar{\nu_e}))/dq^2}.\label{47}
\end{eqnarray}
The behaviors of  the differential ratio $\frac{d{\mathcal R}_{\Xi_{QQ}}}{dq^2}$ are shown in Fig.~\ref{fd3} and Fig.~\ref{fd5}.



For further studying these processes, one can also explore the $\theta$ distribution through Eq.~\eqref{angle}.  The  normalized forward-backward asymmetry can be defined as
\begin{eqnarray}
	\frac{dA_{FB}}{dq^2}&=&\frac{\big[\int^1_0-\int^0_{-1}\big]d\cos\theta\frac{d^2\Gamma}{dq^2 d\cos\theta_\Lambda}}{\big[\int^1_0-\int^0_{-1}\big]d\cos\theta\frac{d\Gamma}{ d\cos\theta_\Lambda}}\propto L_2.
\end{eqnarray}
The normalized forward-backward asymmetries of these processes are shown in Fig.~\ref{fd6} and Fig.~\ref{fd7}. 

One can find that the behavior of $A_{FB}$ is proportional to the coefficient $L_2$ in Eq.~\eqref{angle}. As the helicity amplitude shown in Eq.~\eqref{A1}, the first part of $L_2$ in Eq.~\eqref{22L} and Eq.~\eqref{32L} are the interference terms between the vector helicity amplitude $HV$ and the axis-vector helicity amplitude $HA$. If we set $\hat{m_\ell}\to0$, the coefficient $L_2$ should be proportional to $f_1\times g_1$ and the $A_{FB}$ should be positive. However in Fig.~\ref{fd2} and Fig.~\ref{fd3} the $A_{FB}$ is not always positive in the three channels $\Omega_{cc}^+\to\Xi_{cc}^{++}\mu\bar\nu$, $\Omega_{bc}^0\to\Xi_{bc}^{+}\mu\bar\nu$ and $\Omega_{cc}^{*+}\to\Xi_{cc}^{++}\mu\bar\nu$. We can see that when $m_\ell=m_\mu$, the assumption $\hat{m_\ell}\to0$ is inappropriate and the second part which is proportional to $\hat{m_\ell}^2$ in $L_2$ should be taken into count. The second part in $L_2$ is the interference term of different polarisation helicity amplitude which can induce a negative contribution which depends on the form factor in hadron transition matrix element. As we mentioned above when the $m_\ell=m_\mu$, 
the muon mass can uplift the helicity suppression of certain semi-leptonic decay amplitudes.
Therefore the forward-backward asymmetry $A_{FB}$ may show some diffferent behavior with muon in the final state and it would be a new method to study the hadron transition matrix element and form factors in the future.

\subsection{Nonleptonic decays}

Nonleptonic decays with pion emission can be estimated  with the helicity amplitude method. In Eq.~\eqref{amp}, the local matrix element  $\langle  \pi^-(P_\pi)|\bar d\gamma^{\mu}(1-\gamma_5)u| 0\rangle$  can be expressed by the decay constant $f_\pi$ as
\begin{equation}
	\langle  \pi^-(P_\pi)|\bar d\gamma^{\mu}(1-\gamma_5)u| 0\rangle=if_\pi P_\pi^\mu,
\end{equation}
where $f_\pi=0.130\rm GeV$.
Then the amplitude of nonleptonic decays becomes
\begin{eqnarray}
&&i\mathcal{M}(\mathcal B_{Q_1Q_2}\to \mathcal{B^\prime}_{Q_1Q_2}\pi^-)=\frac{iG_F}{\sqrt{2}}f_\pi V_{us}V^*_{ud}a_1\notag\\
&&\;\;\;\;\;\;\langle \mathcal{B^\prime}_{Q_1Q_2}(P^\prime,S_z^\prime)|\bar u\slashed P_\pi(1-\gamma_5)s|\mathcal B_{Q_1Q_2}(P,S_z)\rangle  ,\notag\\
&&Q_{1,2}=b,c.
\end{eqnarray}
The helicity amplitudes in the nonleptonic decay processes are defined as
\begin{eqnarray}
&&HV^{S}_{\lambda,\lambda^\prime}=\langle \mathcal{B^\prime}_{Q_1Q_2}(P^\prime,\lambda^\prime)|\bar u\slashed P_\pi s|\mathcal B_{Q_1Q_2}(P,\lambda)\rangle,\notag\\
&& HA^{S}_{\lambda,\lambda^\prime}=\langle \mathcal{B^\prime}_{Q_1Q_2}(P^\prime,\lambda^\prime)|\bar u\slashed P_\pi \gamma_5 s|\mathcal B_{Q_1Q_2}(P,\lambda)\rangle.
\end{eqnarray}
With the help of helicity amplitude, the total decay width of the spin-1/2 to spin-1/2 and the spin-3/2 to spin-1/2 processes can be expressed as
\begin{eqnarray}
	&&\Gamma^{\frac{1}{2}\to\frac{1}{2}}={\cal P}\frac{\sqrt{s_+s_-}}{32\pi M^3}\bigg(|H_{\frac{1}{2},-\frac{1}{2}}^{\frac{1}{2}}|^2+|H_{-\frac{1}{2},\frac{1}{2}}^{\frac{1}{2}}|^2\bigg),\notag\\
	&&\Gamma^{\frac{3}{2}\to\frac{1}{2}}={\cal P}\frac{\sqrt{s_+s_-}}{64\pi M^3}\bigg(|H_{\frac{1}{2},-\frac{1}{2}}^{\frac{3}{2}}|^2+|H_{-\frac{1}{2},\frac{1}{2}}^{\frac{3}{2}}|^2\bigg),\notag\\
	&&{\cal P}=\frac{G_F^2}{2}|V_{us}V^*_{ud}|^2 a_1^2f_\pi^2.
\end{eqnarray}
where $H^{S}_{\lambda,\lambda^\prime}=HV^{S}_{\lambda,\lambda^\prime}-HA^{S}_{\lambda,\lambda^\prime}$. The expressions of the nonleptonic helicity amplitudes are shown in Appendix.~\ref{A}.

Using the formulas above,  we give   numerical results of doubly heavy hadron nonleptionic two-body decays in Table.~\ref{f0} and Table.~\ref{fq}.  Comparing the two numerical results derived by the form factor in Eq.~\eqref{fit} and the constant form factors, we find that the effects induced by the different form factors are about $10\%$. 

\begin{table}[!htb]
\caption{Numerical results of decay width and branching fraction in doubly heavy baryon nonleptonic decays using F($q^2$).}\label{fq} %
\begin{tabular}{|c|c|c|c|c|c|c|c|c|}
\hline \hline
channel & $\Gamma(\times10^{-17}\rm{GeV})$   & ${\cal B}$($\%$)   \tabularnewline
\hline
$\Omega_{cc}^+\to \Xi_{cc}^{++}\pi^-$&4.21&$1.15\times10^{-3}$\tabularnewline
\hline
$\Omega_{bb}^-\to\Xi_{bb}^0\pi^-$&3.72&$4.51\times10^{-3}$\tabularnewline
\hline
$\Omega_{bc}^0\to \Xi_{bc}^+\pi^-$&6.68&$2.23\times10^{-3}$\tabularnewline
\hline
$\Omega_{bc}^{\prime0}\to \Xi_{bc}^{\prime+}\pi^-$&7.59&$2.53\times10^{-3}$\tabularnewline
\hline
$\Omega_{cc}^{*+}\to \Xi_{cc}^{++}\pi^-$&$2.85\times10^{-2}$&-\tabularnewline
\hline
$\Omega_{bb}^{*-}\to\Xi_{bb}^0\pi^-$&$8.78\times10^{-2}$&-\tabularnewline
\hline
$\Omega_{bc}^{*0}\to \Xi_{bc}^+\pi^-$&$1.30\times10^{-2}$&-\tabularnewline
\hline
\end{tabular}
\end{table}

\begin{table}[!htb]
\caption{Numerical results of decay width and branching fraction in doubly heavy baryon nonleptonic decays using  F(0).}\label{f0} %
\begin{tabular}{|c|c|c|c|c|c|c|c|c|}
\hline \hline
channel & $\Gamma(\times10^{-17}\rm{GeV})$   & ${\cal B}$($\%$)   \tabularnewline
\hline
$\Omega_{cc}^+\to \Xi_{cc}^{++}\pi^-$&3.98&$1.09\times10^{-3}$\tabularnewline
\hline
$\Omega_{bb}^-\to\Xi_{bb}^0\pi^-$&3.52&$4.26\times10^{-3}$\tabularnewline
\hline
$\Omega_{bc}^0\to \Xi_{bc}^+\pi^-$&6.68&$2.23\times10^{-3}$\tabularnewline
\hline
$\Omega_{bc}^{\prime0}\to \Xi_{bc}^{\prime+}\pi^-$&7.18&$2.39\times10^{-3}$\tabularnewline
\hline
$\Omega_{cc}^{*+}\to \Xi_{cc}^{++}\pi^-$&$2.41\times10^{-2}$&-\tabularnewline
\hline
$\Omega_{bb}^{*-}\to\Xi_{bb}^0\pi^-$&$7.43\times10^{-2}$&-\tabularnewline
\hline
$\Omega_{bc}^{*0}\to \Xi_{bc}^+\pi^-$&$1.10\times10^{-2}$&-\tabularnewline
\hline
\end{tabular}
\end{table}

\section{Summary}
The weak decays of doubly heavy baryons induced by $s\to u$ are studied in this work. Not only the spin-1/2 to spin-1/2 processes but also spin-3/2 to spin-1/2 processes are considered. Using the light-front-quark model,  we use the diquark picture in which the two spectator heavy quarks can be seen as a scalar or an axis vector diquark and the baryon state can be treated as the meson state.  Using the form factors defined in the hadron matrix element, we estimate the decay width branching fractions of semileptonic and two-body nonleptonic decays and give their phenomenological applications. 

To obtain the phenomenological results, the helicity amplitude are used. With the help of the lifetime of doubly heavy baryons shown in Table.~\ref{mass_beta}, we have predicted the  branching fraction in Table~\ref{semileptonq2}, Table~\ref{semilepton0}, Table.~\ref{fq} and Table.~\ref{f0}. The $q^2$ distribution and angular distribution are also studied which is shown in Fig.~\ref{fd2}, Fig.~\ref{fd4} and Fig.~\ref{fd6}, Fig.~\ref{fd7}. Since the integral region in phase space is very small which is comparable to the $m_\mu^2$, there are many strange phenomenological results in these channel such as branching fraction, forward-backward asymmetry and ${\mathcal R}_{\Xi_{QQ}}$.  The decay processes induced by $s\to u$ with $\mu$ leptons in the final state offer possibilities of significant NP contributions not present in processes with electron.  Therefore these phenomenological results are excited and they may become a new method to search new physics in the future.

\section*{Acknowledgements}

We thank Prof. Wei Wang  for useful discussions. 
This work was supported in part by NSFC under Grant Nos.12147147,  11735010, 11911530088, U2032102,  12125503.

\appendix
\section{Helicity amplitude}\label{A}
The helicity amplitude can be expressed by the form factors defined in hadron matrix element. Let $(M^2-M^{\prime2})\pm q^2 =\hat{M}_q^{\pm}$ and for the spin-1/2 to spin-1/2 semileptonic and nonleptonic decay processes, the helicity amplitudes are
    \begin{eqnarray}
	&&HV_{\frac{1}{2},+1}^{\frac{1}{2},\frac{1}{2}}=\frac{-iM\sqrt{2s_-}}{\bar M}f^{\frac{1}{2}\to\frac{1}{2}}_1,\notag\\
	&&HV_{\frac{1}{2},0}^{\frac{1}{2},-\frac{1}{2}}=\frac{-i\sqrt{s_-}}{2\bar{M}\sqrt{q^2}}[2M(M+M^\prime)f^{\frac{1}{2}\to\frac{1}{2}}_1\notag\\
	&&\qquad\qquad\quad+s_+(f^{\frac{1}{2}\to\frac{1}{2}}_2+f^{\frac{1}{2}\to\frac{1}{2}}_3)],\notag\\
	&&HV_{\frac{1}{2},t}^{\frac{1}{2},-\frac{1}{2}}=\frac{-i\sqrt{s_+}}{2\bar{M}\sqrt{q^2}}[2M\bar{M}f^{\frac{1}{2}\to\frac{1}{2}}_1+\hat{M}_q^{+}f^{\frac{1}{2}\to\frac{1}{2}}_2+\hat{M}_q^{-}f^{\frac{1}{2}\to\frac{1}{2}}_3],\notag\\
	&&HV_{-\frac{1}{2},\frac{1}{2}}^{\frac{1}{2}}=\frac{-i\sqrt{s_+}m_\pi}{2\bar{M}\sqrt{q^2}}[2M\bar{M}f^{\frac{1}{2}\to\frac{1}{2}}_1+\hat{M}_q^{+}f^{\frac{1}{2}\to\frac{1}{2}}_2+\hat{M}_q^{-}f^{\frac{1}{2}\to\frac{1}{2}}_3],\notag\\
	&&HV_{-\lambda^\prime,-\lambda_W}^{\frac{1}{2},-\lambda}=HV_{\lambda^\prime,\lambda_W}^{\frac{1}{2},\lambda},\notag\\ &&HV_{-\lambda,-\lambda^\prime}^{\frac{1}{2}}=HV_{\lambda,\lambda^\prime}^{\frac{1}{2}}\label{A1}\;.
\end{eqnarray}
 and 
\begin{eqnarray}
	&&HA_{\frac{1}{2},+1}^{\frac{1}{2},\frac{1}{2}}=\frac{-iM\sqrt{2s_+}}{\bar M}g^{\frac{1}{2}\to\frac{1}{2}}_1,\notag\\
	&&HA_{\frac{1}{2},0}^{\frac{1}{2},-\frac{1}{2}}=\frac{-i\sqrt{s_+}}{2\bar{M}\sqrt{q^2}}[2M\bar{M}g^{\frac{1}{2}\to\frac{1}{2}}_1-s_-(g^{\frac{1}{2}\to\frac{1}{2}}_2+g^{\frac{1}{2}\to\frac{1}{2}}_3)],\notag
	\end{eqnarray}
	\begin{eqnarray}
	&&HA_{\frac{1}{2},t}^{\frac{1}{2},-\frac{1}{2}}=\frac{-i\sqrt{s_-}}{2\bar{M}\sqrt{q^2}}[2M(M+M^\prime)g^{\frac{1}{2}\to\frac{1}{2}}_1\notag\\
	&&\qquad\qquad\quad-\hat{M}_q^{+}g^{\frac{1}{2}\to\frac{1}{2}}_2-\hat{M}_q^{-}g^{\frac{1}{2}\to\frac{1}{2}}_3],\notag\\
	&&HA_{-\frac{1}{2},\frac{1}{2}}^{\frac{1}{2}}=\frac{-i\sqrt{s_-}m_\pi}{2\bar{M}\sqrt{q^2}}[2M(M+M^\prime)g^{\frac{1}{2}\to\frac{1}{2}}_1\notag\\
	&&\qquad\qquad\quad-\hat{M}_q^{+}g^{\frac{1}{2}\to\frac{1}{2}}_2-\hat{M}_q^{-}g^{\frac{1}{2}\to\frac{1}{2}}_3],\notag\\
	&&HA_{-\lambda^\prime,-\lambda_W}^{\frac{1}{2},-\lambda}=-HA_{\lambda^\prime,\lambda_W}^{\frac{1}{2},\lambda},\notag\\
	&&HA_{-\lambda,-\lambda^\prime}^{\frac{1}{2}}=-HA_{\lambda,\lambda^\prime}^{\frac{1}{2}}\label{A2}\;.
\end{eqnarray}

For the spin-3/2 to spin-1/2 semileptonic and nonleptonic decay processes, the helicity amplitudes are
\begin{eqnarray}
    &&HV_{\frac{1}{2},-1}^{\frac{3}{2},-\frac{3}{2}}=-i\sqrt{s_-}f^{\frac{3}{2}\to\frac{1}{2}}_4,\notag\\
    && HV_{\frac{1}{2},+1}^{\frac{3}{2},\frac{1}{2}}=\frac{-i\sqrt{s_-}}{\sqrt{3}\bar{M}^2}[s_+f^{\frac{3}{2}\to\frac{1}{2}}_1+\bar{M}^2f^{\frac{3}{2}\to\frac{1}{2}}_4],\notag\\
    &&HV_{\frac{1}{2},0}^{\frac{3}{2},-\frac{1}{2}}=\frac{-i\sqrt{s_-}}{2M\bar{M}^2\sqrt{6q^2}}[2M\bar{M}s_+f^{\frac{3}{2}\to\frac{1}{2}}_1\notag\\
    &&\qquad\qquad\quad-s_+s_-(f^{\frac{3}{2}\to\frac{1}{2}}_2+f^{\frac{3}{2}\to\frac{1}{2}}_3)+2\bar{M}^2\hat{M}_q^{+}f^{\frac{3}{2}\to\frac{1}{2}}_4],\notag\\
    &&HV_{\frac{1}{2},t}^{\frac{3}{2},-\frac{1}{2}}=\frac{-is_-\sqrt{s_+}}{2M\bar{M}^2\sqrt{6q^2}}[2M(M+M^\prime)f^{\frac{3}{2}\to\frac{1}{2}}_1\notag\\
    &&\qquad\qquad\quad-\hat{M}_q^{+}f^{\frac{3}{2}\to\frac{1}{2}}_2-\hat{M}_q^{-}f^{\frac{3}{2}\to\frac{1}{2}}_3+2\bar{M}^2f^{\frac{3}{2}\to\frac{1}{2}}_4],\notag
\end{eqnarray}
\begin{eqnarray}
    &&HV_{-\frac{1}{2},\frac{1}{2}}^{\frac{3}{2}}=\frac{-is_-\sqrt{s_+}m_\pi}{2M\bar{M}^2\sqrt{6q^2}}[2M(M+M^\prime)f^{\frac{3}{2}\to\frac{1}{2}}_1\notag\\
    &&\qquad\qquad\quad-\hat{M}_q^{+}f^{\frac{3}{2}\to\frac{1}{2}}_2-\hat{M}_q^{-}f^{\frac{3}{2}\to\frac{1}{2}}_3+2\bar{M}^2f^{\frac{3}{2}\to\frac{1}{2}}_4],\notag\\
   	&&HV_{-\lambda^\prime,-\lambda_W}^{\frac{3}{2},-\lambda}=-HV_{\lambda^\prime,\lambda_W}^{\frac{3}{2},\lambda},\notag\\
   	&&HV_{-\lambda,-\lambda^\prime}^{\frac{3}{2}}=-HV_{\lambda,\lambda^\prime}^{\frac{3}{2}}\;.
\end{eqnarray}
 
\begin{eqnarray}
   &&HA_{\frac{1}{2},-1}^{\frac{3}{2},-\frac{3}{2}}=i\sqrt{s_+}g^{\frac{3}{2}\to\frac{1}{2}}_4,\notag\\
   &&HA_{\frac{1}{2},+1}^{\frac{3}{2},\frac{1}{2}}=\frac{-i\sqrt{s_+}}{\sqrt{3}\bar{M}^2}[s_-g^{\frac{3}{2}\to\frac{1}{2}}_1-\bar{M}^2g^{\frac{3}{2}\to\frac{1}{2}}_4],\notag\\
   &&HA_{\frac{1}{2},0}^{\frac{3}{2},-\frac{1}{2}}=\frac{-i\sqrt{s_+}}{2M\bar{M}^2\sqrt{6q^2}}[2M(M+M^\prime)s_-g^{\frac{3}{2}\to\frac{1}{2}}_1\notag\\
   &&\qquad\qquad\quad+s_+s_-(g^{\frac{3}{2}\to\frac{1}{2}}_2+g^{\frac{3}{2}\to\frac{1}{2}}_3)-2\bar{M}^2\hat{M}_q^{+}g^{\frac{3}{2}\to\frac{1}{2}}_4],\notag\\
   &&HA_{\frac{1}{2},t}^{\frac{3}{2},-\frac{1}{2}}=\frac{-is_+\sqrt{s_-}}{2M\bar{M}^2\sqrt{6q^2}}[2M\bar{M}g^{\frac{3}{2}\to\frac{1}{2}}_1\notag\\
   &&\qquad\qquad\quad+\hat{M}_q^{+}g^{\frac{3}{2}\to\frac{1}{2}}_2+\hat{M}_q^{-}g^{\frac{3}{2}\to\frac{1}{2}}_3-2\bar{M}^2g^{\frac{3}{2}\to\frac{1}{2}}_4],\notag\\
   &&HA_{-\frac{1}{2},\frac{1}{2}}^{\frac{3}{2}}=\frac{-is_+\sqrt{s_-}m_\pi}{2M\bar{M}^2\sqrt{6q^2}}[2M\bar{M}g^{\frac{3}{2}\to\frac{1}{2}}_1\notag\\
   &&\qquad\qquad\quad+\hat{M}_q^{+}g^{\frac{3}{2}\to\frac{1}{2}}_2+\hat{M}_q^{-}g^{\frac{3}{2}\to\frac{1}{2}}_3-2\bar{M}^2g^{\frac{3}{2}\to\frac{1}{2}}_4],\notag\\
   &&HA_{-\lambda^\prime,-\lambda_W}^{\frac{3}{2},-\lambda}=HA_{\lambda^\prime,\lambda_W}^{\frac{3}{2},\lambda},\notag\\ &&HA_{-\lambda,-\lambda^\prime}^{\frac{3}{2}}=HA_{\lambda,\lambda^\prime}^{\frac{3}{2}}\;.
\end{eqnarray} 
 

\begin{figure*}[htbp!]
	\begin{minipage}[t]{0.4\linewidth}
		\centering
		\includegraphics[width=1\columnwidth]{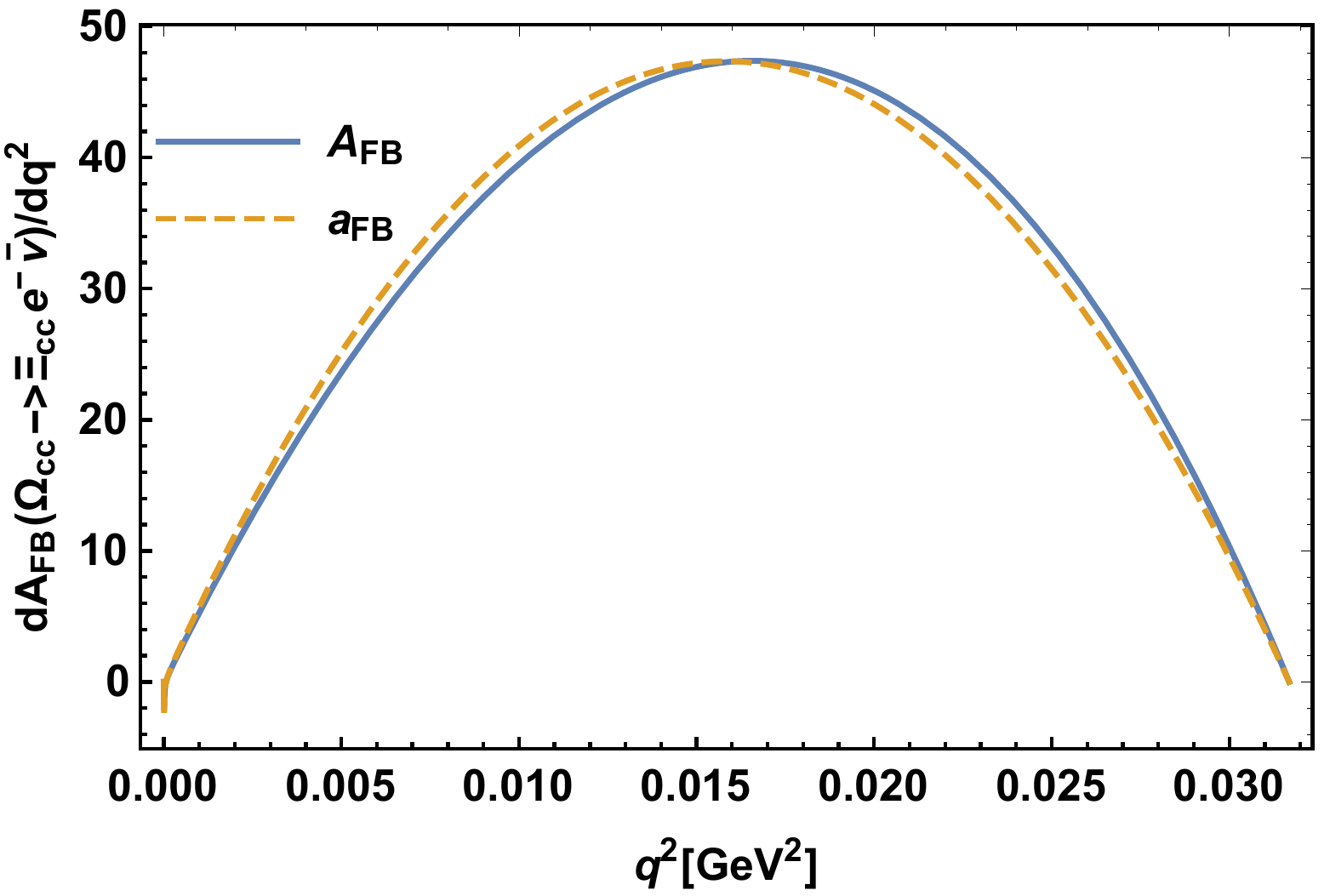}
	\end{minipage}
	\begin{minipage}[t]{0.4\linewidth}
		\centering
		\includegraphics[width=1\columnwidth]{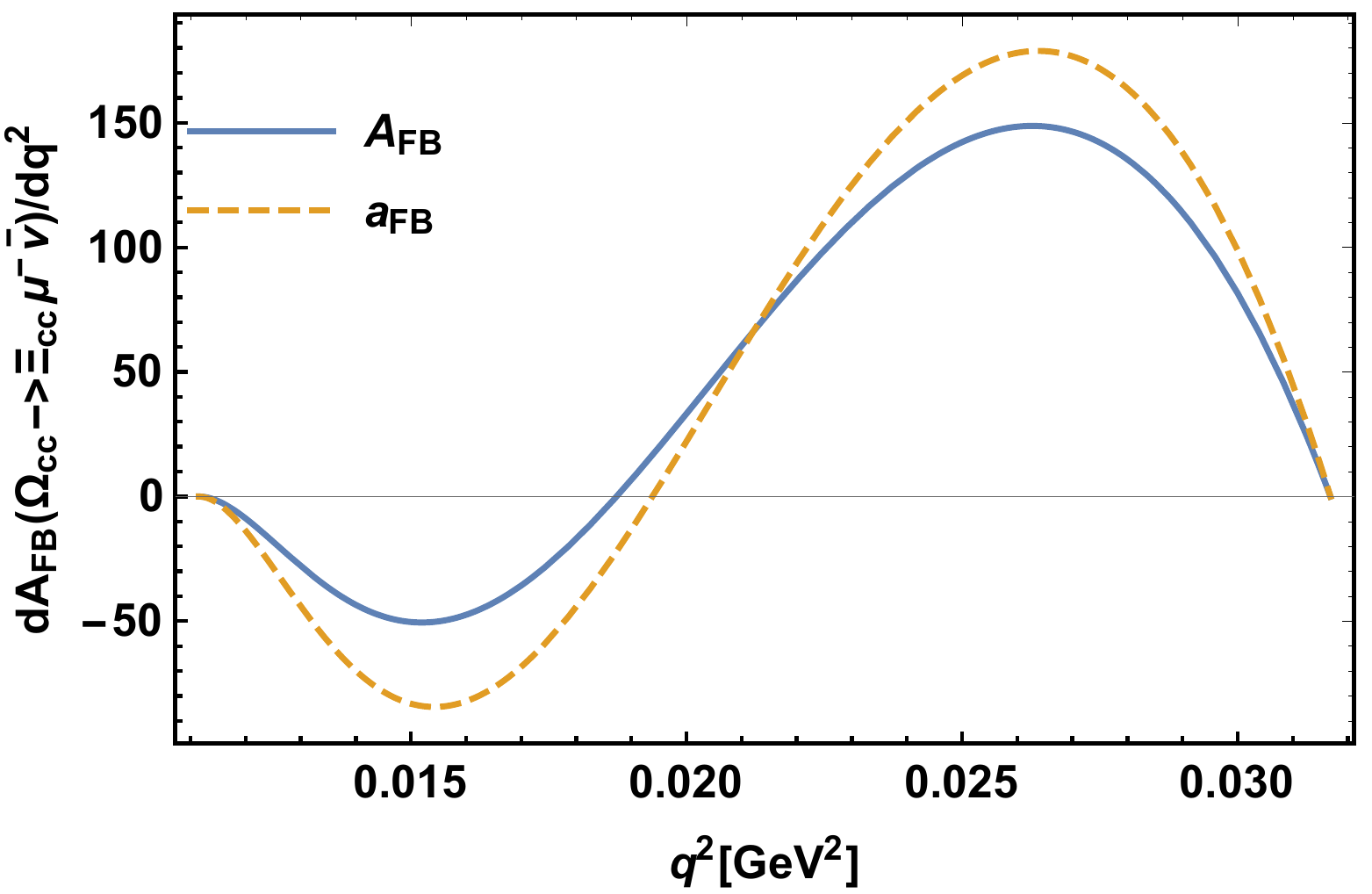}
	\end{minipage}
	\begin{minipage}[t]{0.4\linewidth}
		\centering
		\includegraphics[width=1\columnwidth]{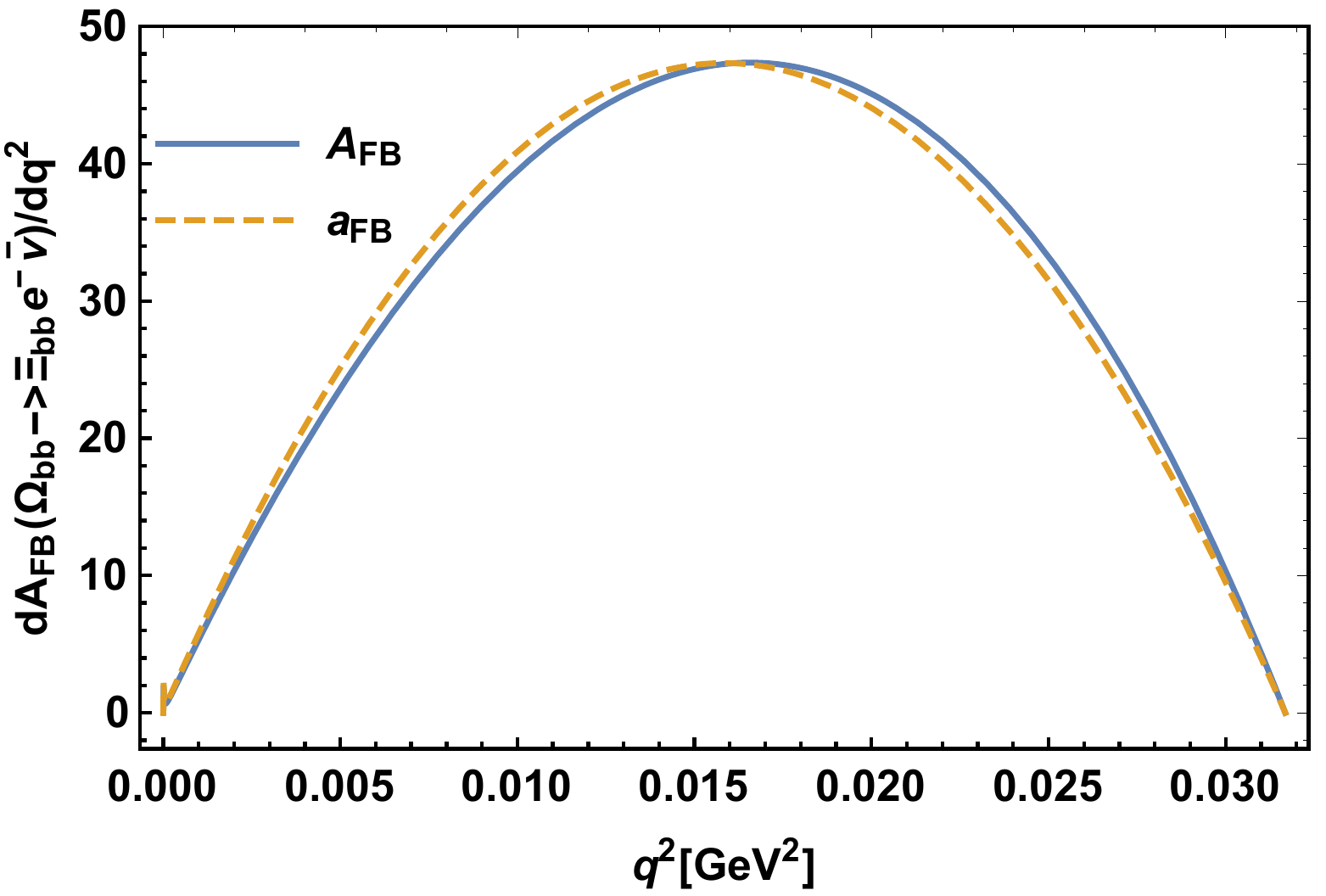}
	\end{minipage}
	\begin{minipage}[t]{0.4\linewidth}
		\centering
		\includegraphics[width=1\columnwidth]{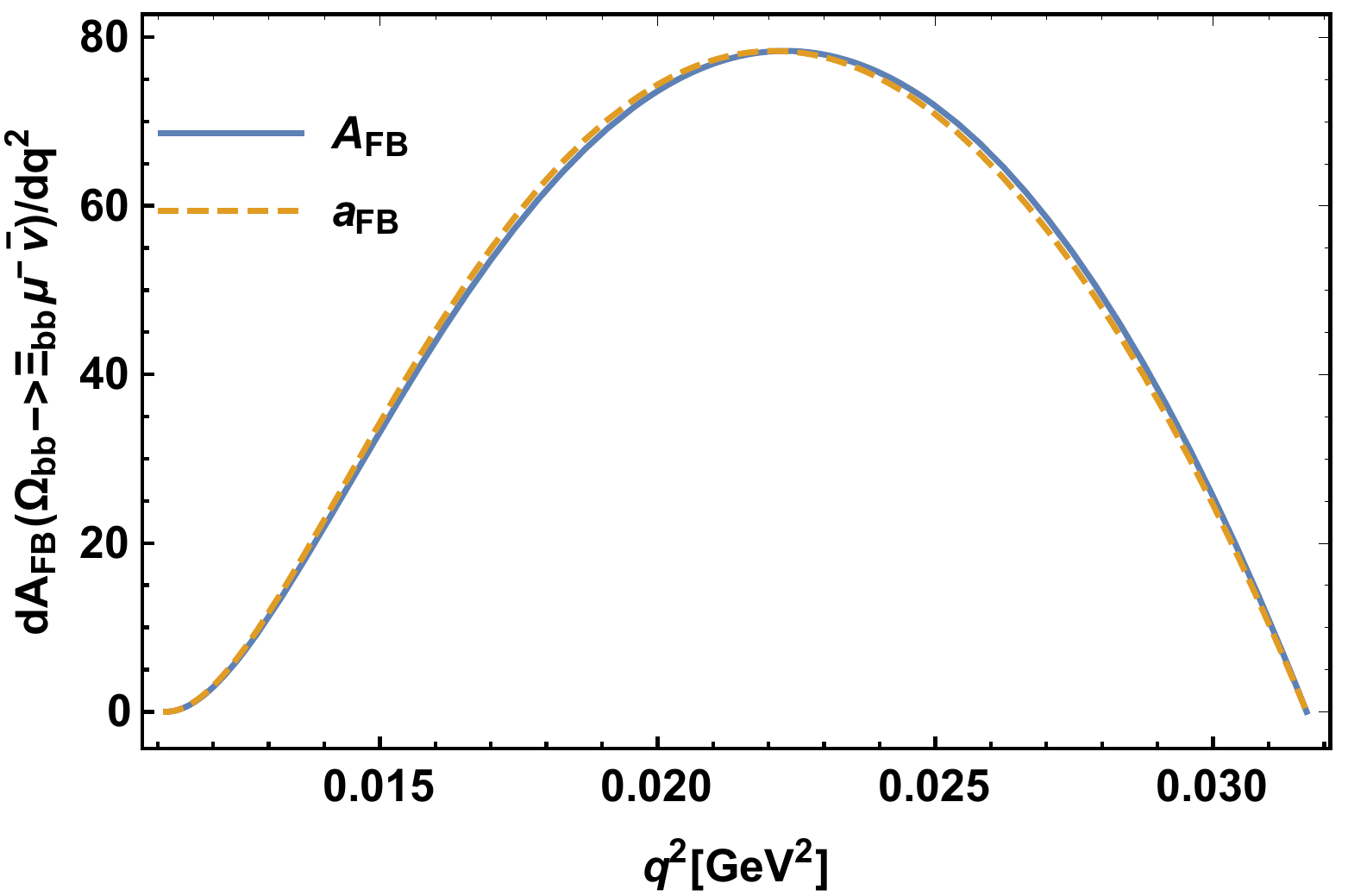}
	\end{minipage}
	\begin{minipage}[t]{0.4\linewidth}
		\centering
		\includegraphics[width=1\columnwidth]{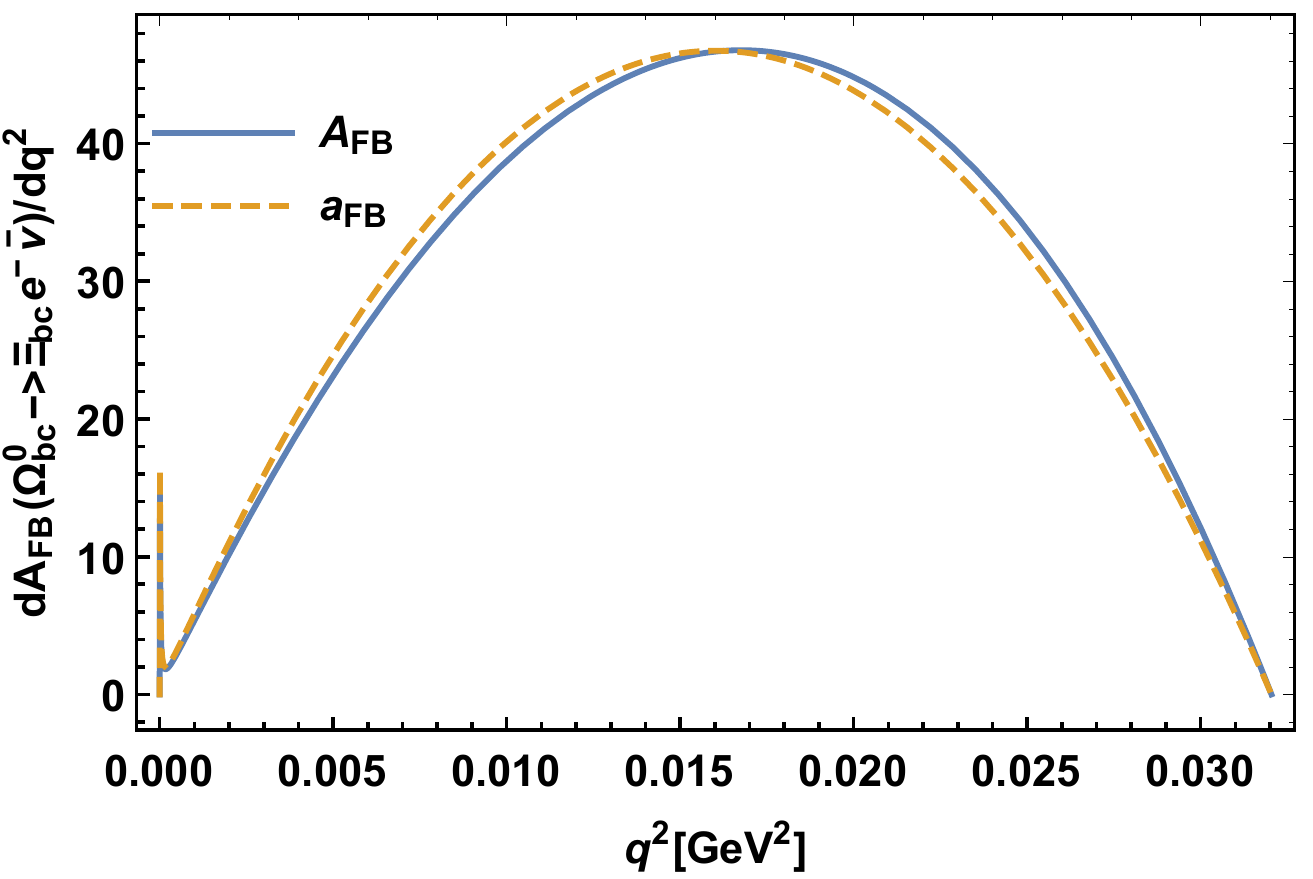}
	\end{minipage}
	\begin{minipage}[t]{0.4\linewidth}
		\centering
		\includegraphics[width=1\columnwidth]{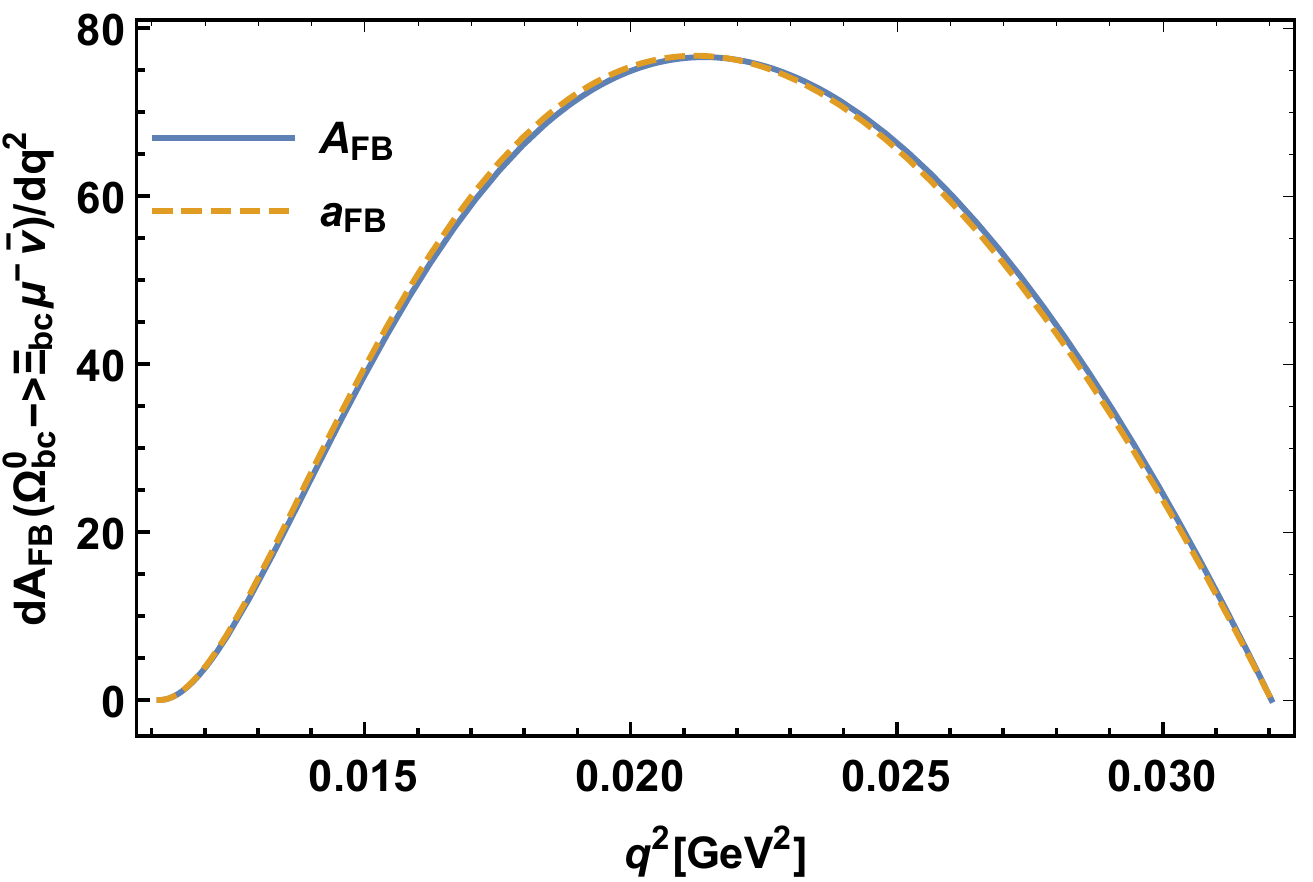}
	\end{minipage}
	\begin{minipage}[t]{0.4\linewidth}
		\centering
		\includegraphics[width=1\columnwidth]{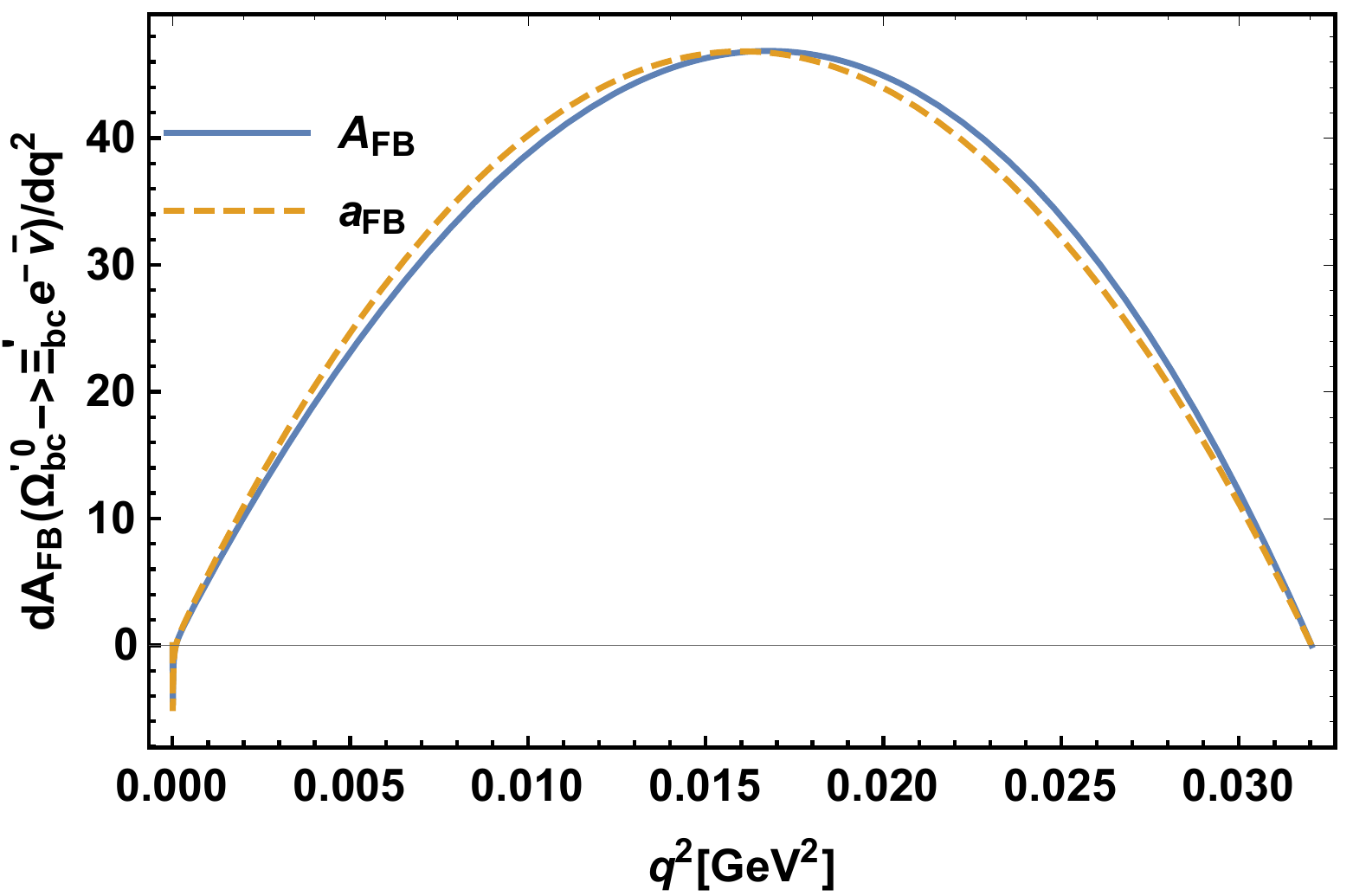}
	\end{minipage}
	\begin{minipage}[t]{0.4\linewidth}
		\centering
		\includegraphics[width=1\columnwidth]{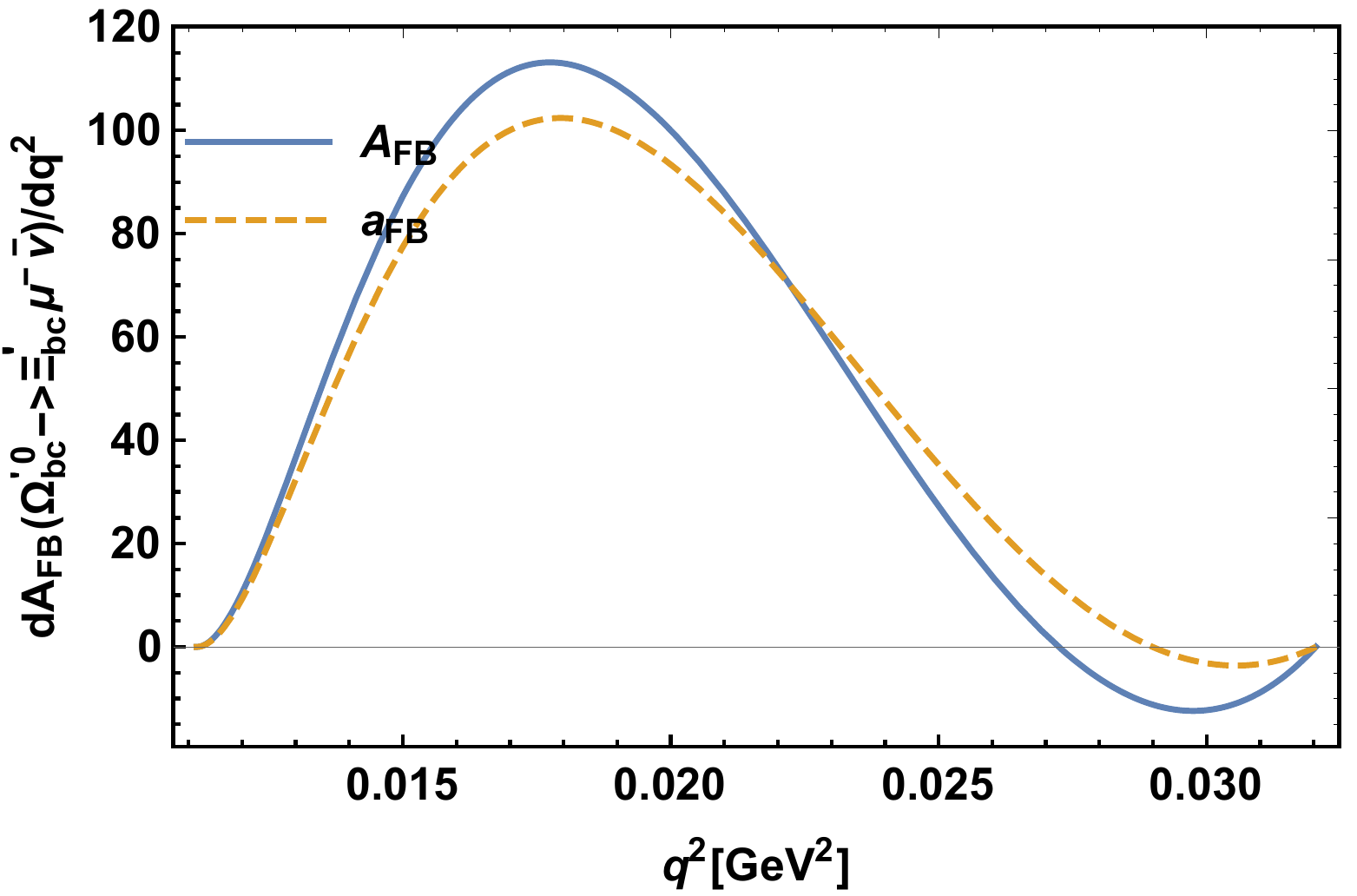}
	\end{minipage}
	\caption{The $dA_{FB}/dq^2$ and $da_{FB}/dq^2$
		of spin-1/2 to spin-1/2 doubly heavy baryon decay processes. $dA_{FB}/dq^2$ is depend on $F(q^2)$ and $da_{FB}/dq^2$ is depend on $F(0)$.}
	\label{fd6}
\end{figure*}



\begin{figure*}[htbp!]
	\begin{minipage}[t]{0.4\linewidth}
		\centering
		\includegraphics[width=1\columnwidth]{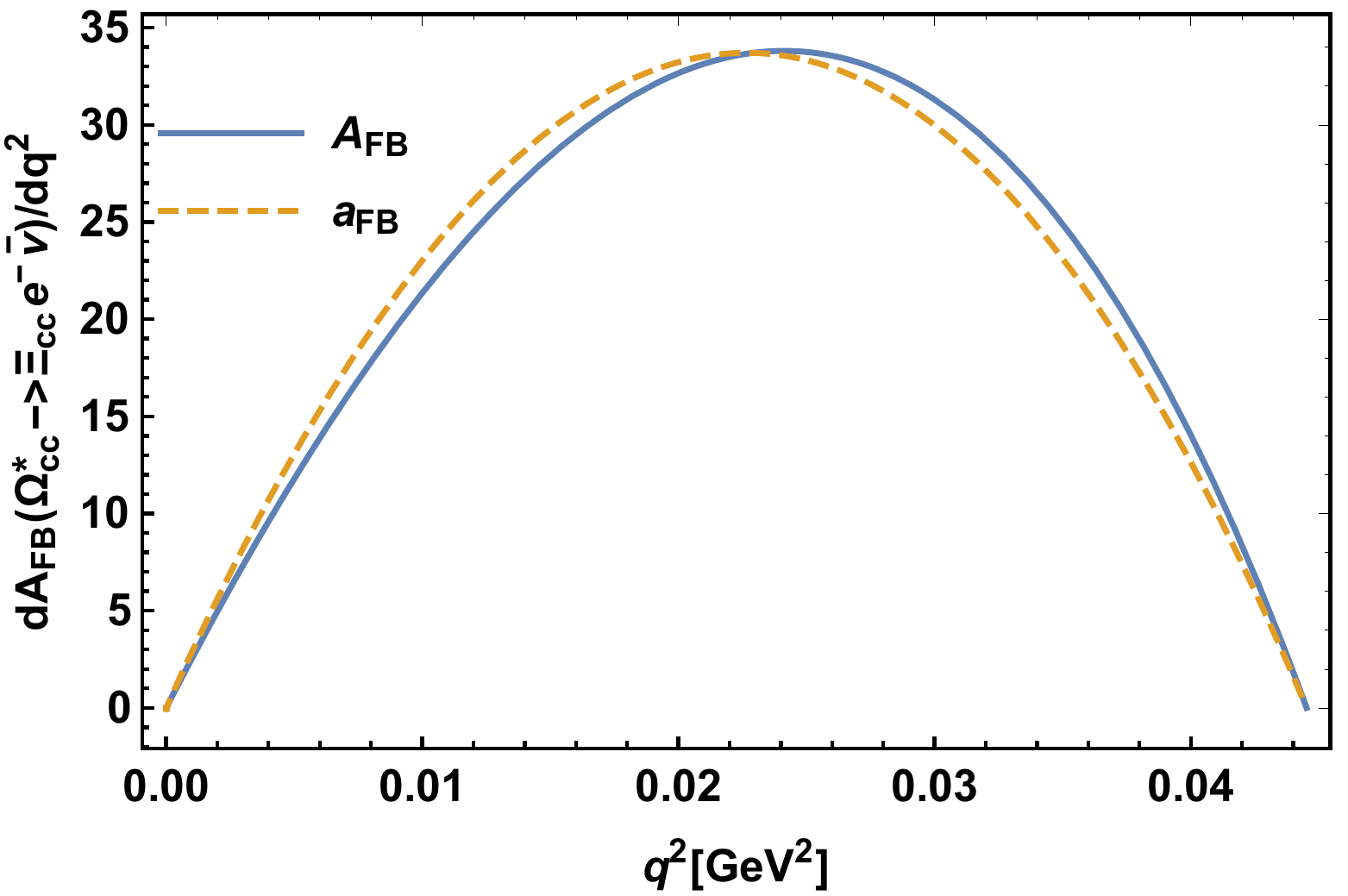}
	\end{minipage}
	\begin{minipage}[t]{0.4\linewidth}
		\centering
		\includegraphics[width=1\columnwidth]{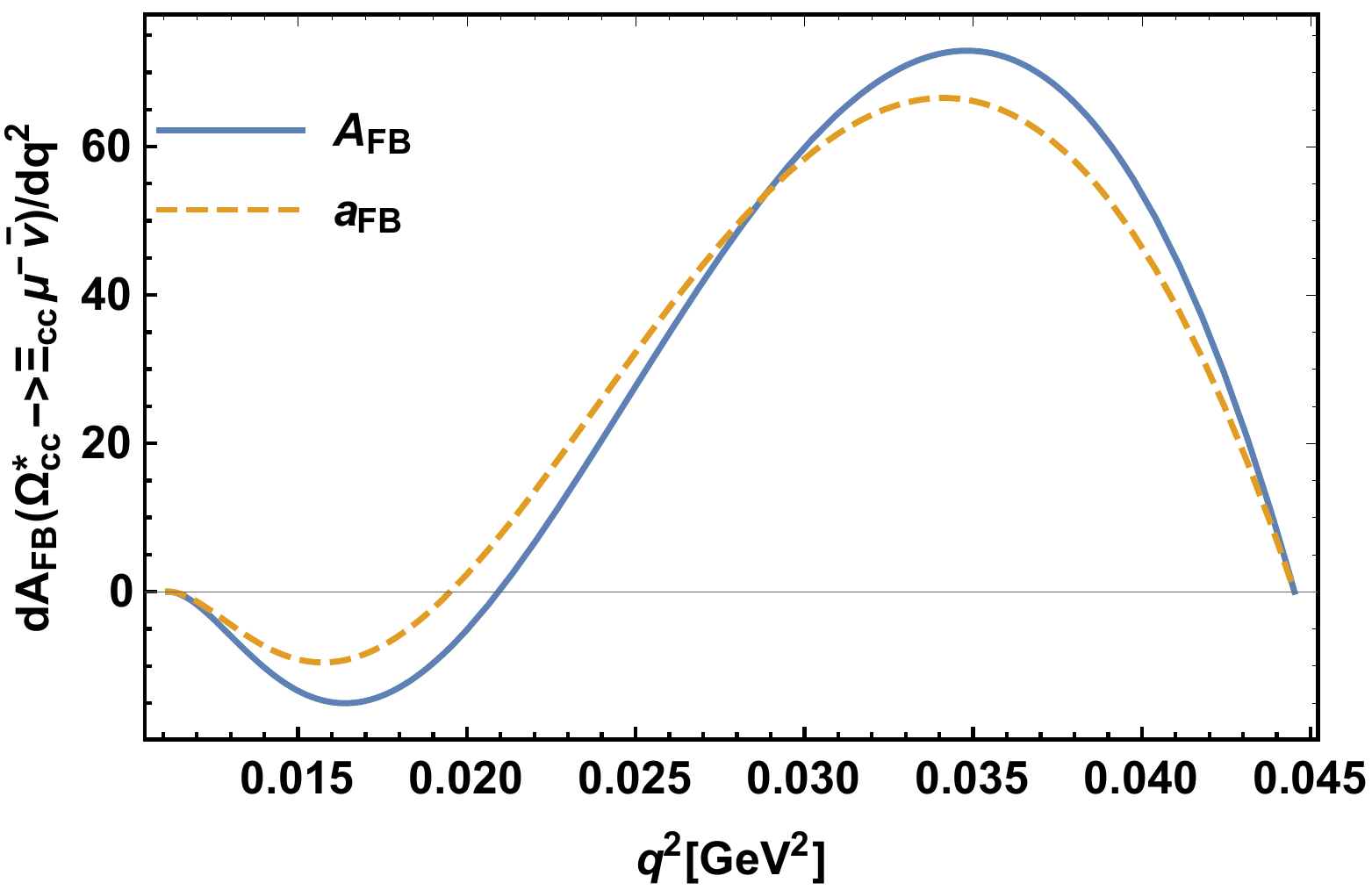}
	\end{minipage}
	\begin{minipage}[t]{0.4\linewidth}
		\centering
		\includegraphics[width=1\columnwidth]{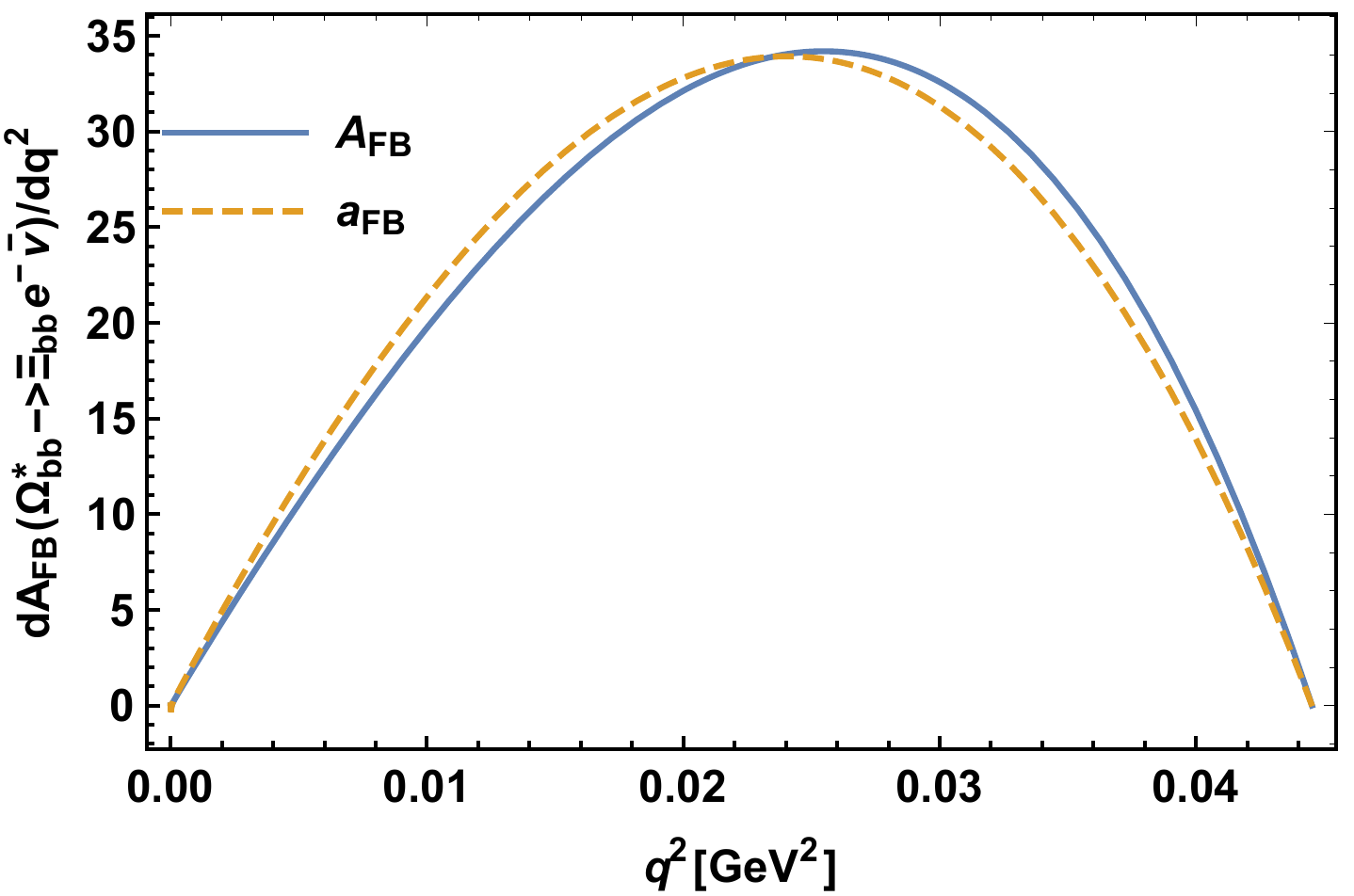}
	\end{minipage}
	\begin{minipage}[t]{0.4\linewidth}
		\centering
		\includegraphics[width=1\columnwidth]{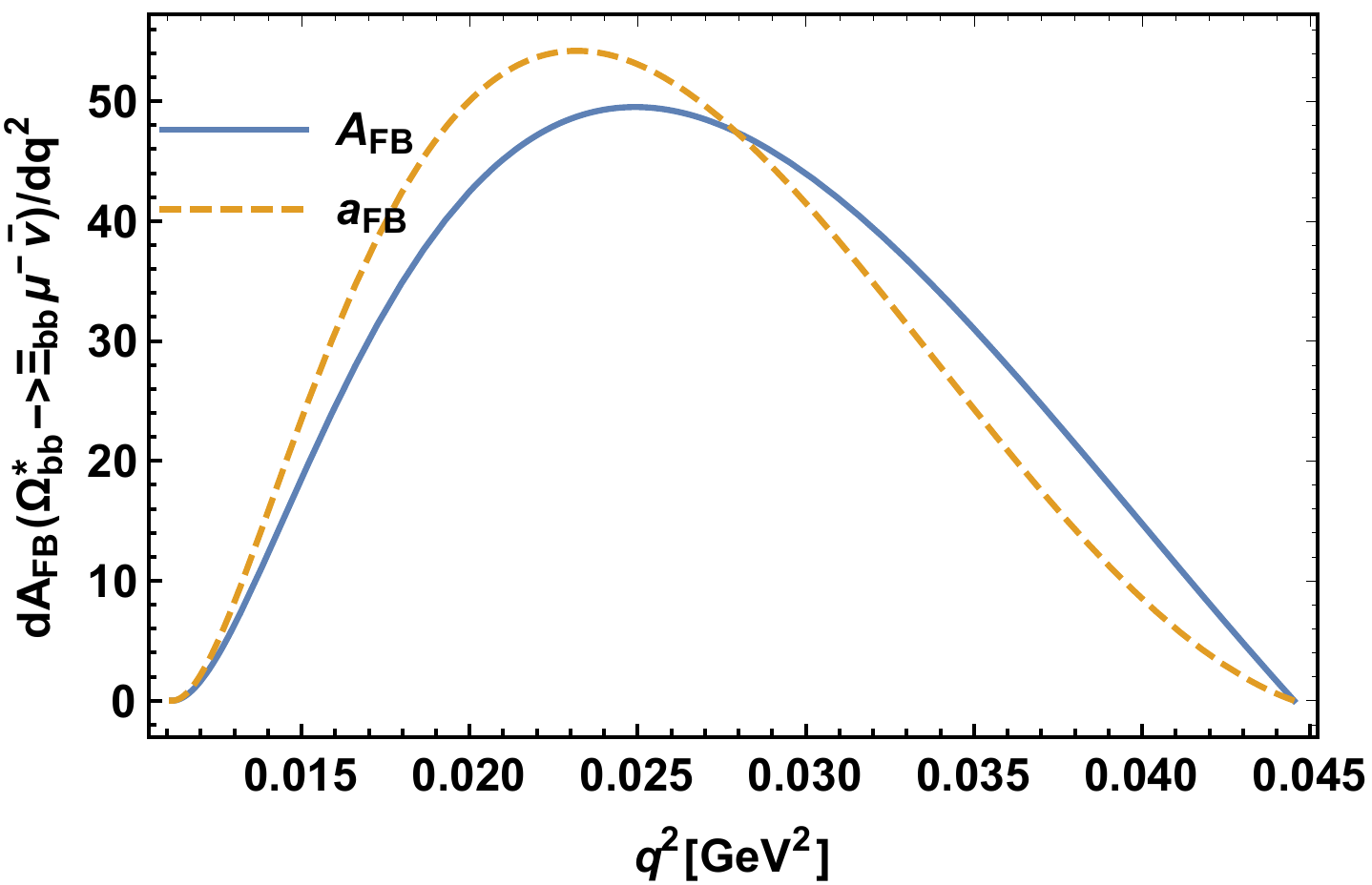}
	\end{minipage}
	\begin{minipage}[t]{0.4\linewidth}
		\centering
		\includegraphics[width=1\columnwidth]{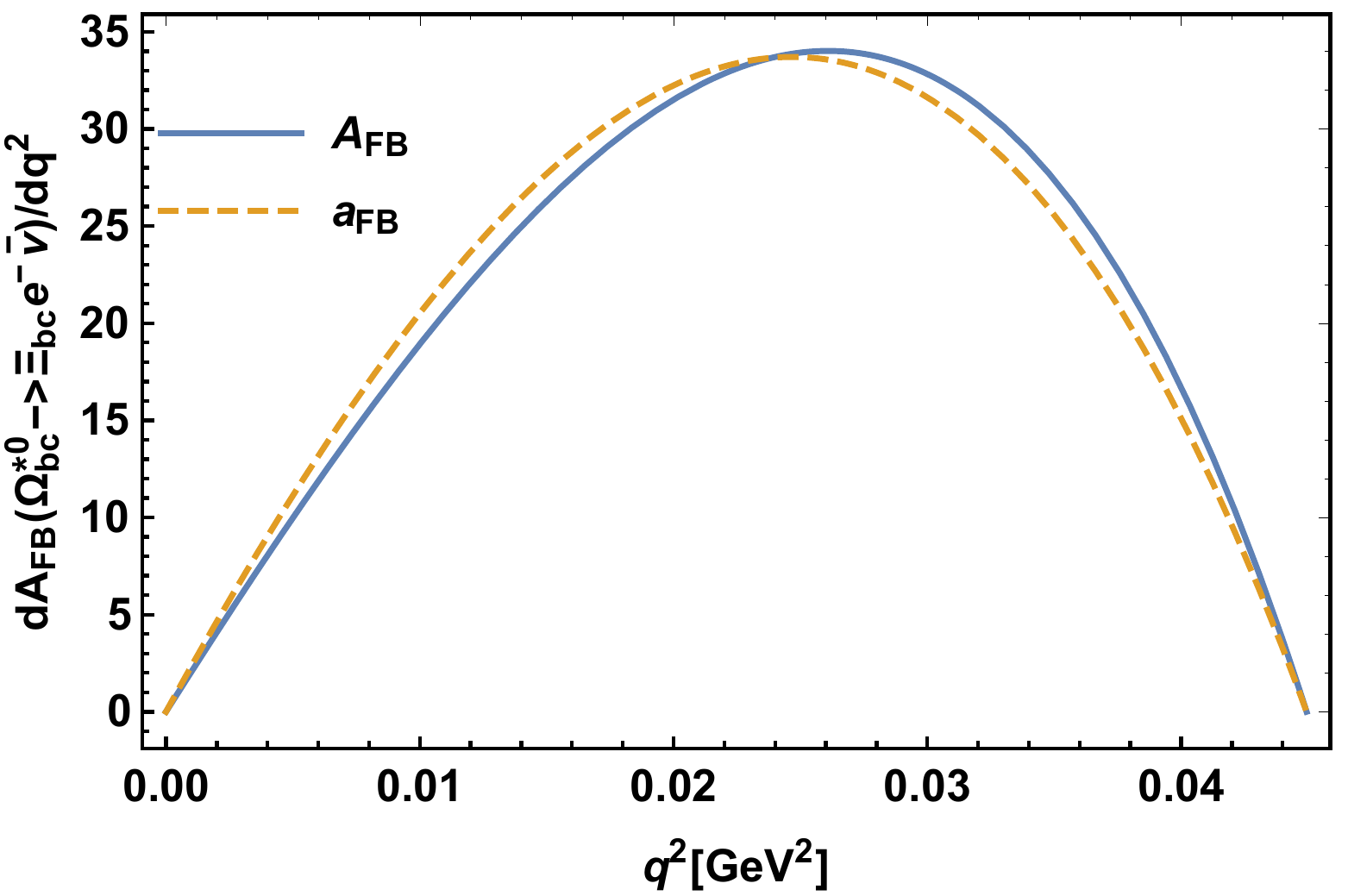}
	\end{minipage}
	\begin{minipage}[t]{0.4\linewidth}
		\centering
		\includegraphics[width=1\columnwidth]{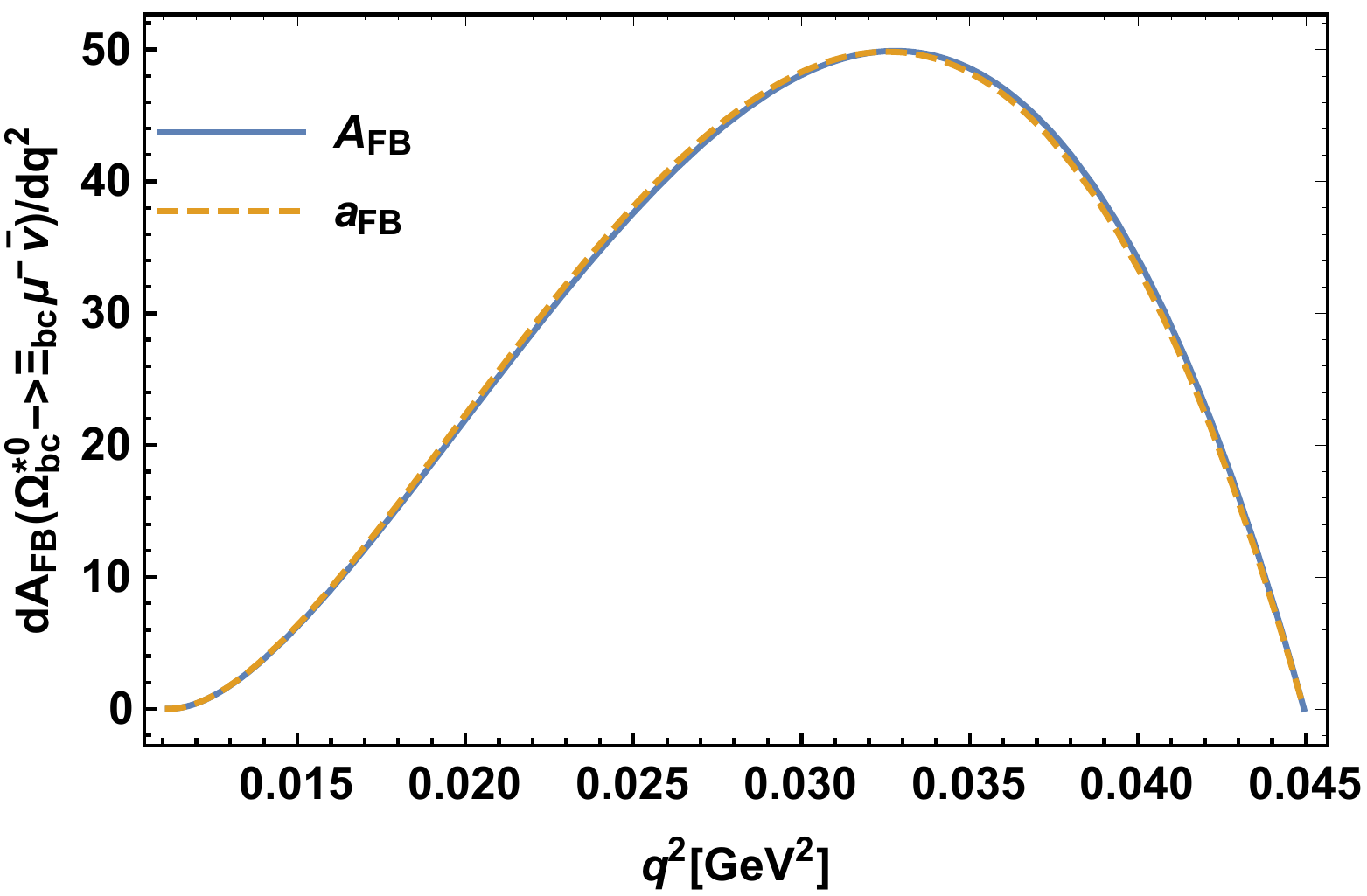}
	\end{minipage}
\caption{The $dA_{FB}/dq^2$ and $da_{FB}/dq^2$
		of spin-3/2 to spin-1/2 doubly heavy baryon decay processes. $dA_{FB}/dq^2$ is depend on $F(q^2)$ and $da_{FB}/dq^2$ is depend on $F(0)$.}
	\label{fd7}
\end{figure*} 


\end{document}